\documentclass[12pt]{article}

\pdfoutput=1

\usepackage{amsfonts,latexsym,eucal,graphicx,epsfig,amssymb,amsmath}
\usepackage{ccaption}
\usepackage[usenames]{color}
\usepackage{mathrsfs}

\usepackage[
      colorlinks=true,
      linkcolor=blue,
      urlcolor=blue,
      filecolor=black,
      citecolor=red,
      pdfstartview=FitV,
      pdftitle={},
        pdfauthor={Dan Brattan, Joan Camps, R. Loganayagam, Mukund Rangamani},
        pdfsubject={},
        pdfkeywords={},
        pdfpagemode=None,
        bookmarksopen=true
      ]{hyperref}
      
\usepackage{rotating}

\makeatletter
\@addtoreset{equation}{section}

\makeatletter
\renewcommand\section{\@startsection {section}{1}{\z@}%
                                   {-3.5ex \@plus -1ex \@minus -.2ex}
                                   {2.3ex \@plus.2ex}%
                                   {\normalfont\large\bfseries}}
\renewcommand\subsection{\@startsection{subsection}{2}{\z@}%
                                     {-3.25ex\@plus -1ex \@minus -.2ex}%
                                     {1.5ex \@plus .2ex}%
                                     {\normalfont\bfseries}}

  \captionnamefont{\bfseries}
  \captiontitlefont{\small\sffamily}
  \captiondelim{: }
  \hangcaption

\def\baselinestretch{1.2}
\def\sec#1{\S\ref{#1}}
\def\fig#1{Fig.\,\ref{#1}}
\def\req#1{(\ref{#1})}
\def\App#1{Appendix \ref{#1}}

\definecolor{purple}{rgb}{0.8,0.3,0.5}
\definecolor{green}{rgb}{0.1,0.8,0.2}

\definecolor{orange}{rgb}{0.8,0.5,0.1}

\definecolor{dred}{rgb}{0.3,0.3,0.9}
\def\red#1{\color{dred}{#1}}

\def\AdS#1{AdS$_{#1}$}

\newcommand{\be}{\begin{eqnarray}}
\newcommand{\ee}{\end{eqnarray}}
\newcommand{\beq}{\begin{eqnarray}}
\newcommand{\eeq}{\end{eqnarray}}

\newcommand{\dalm}{\kern1pt\vbox{\hrule height 0.9pt\hbox{\vrule width 0.9pt\hskip 2.5pt\vbox{\vskip 5.5pt}\hskip 3pt\vrule width 0.3pt}\hrule height 0.3pt}\kern1pt}

\title{{\bf CFT dual of the AdS Dirichlet problem:  \\ 
Fluid/Gravity on cut-off surfaces}}

\author{ Daniel Brattan$^a$\footnote{d.k.brattan@durham.ac.uk},  
Joan Camps$^a$\footnote{joan.camps@durham.ac.uk}, R. Loganayagam$^b$\footnote{nayagam@physics.harvard.edu}, Mukund Rangamani$^a$\footnote{mukund.rangamani@durham.ac.uk} \\ \\
\small{\emph{$^{a}$  Centre for Particle Theory \& Department of
Mathematical Sciences}}, \\[-1.5mm]
\small{\emph{Science Laboratories, South Road, Durham DH1 3LE, United Kingdom}}\\
\small{\emph{$^{b}$  Junior Fellow, Harvard Society of Fellows}}, \\[-1.5mm]
\small{\emph{Harvard University, Cambridge, MA 02138, USA}}
}

\begin{document}

\setlength{\baselineskip}{16pt}
\begin{titlepage}
\maketitle
\begin{picture}(0,0)(0,0)
\put(380, 330){DCPT-11/25}
\end{picture}
\vspace{-50pt}

\begin{abstract}
We study the gravitational Dirichlet problem in AdS spacetimes with a view to understanding the boundary CFT interpretation. We define the problem as bulk Einstein's equations with Dirichlet boundary conditions on fixed timelike cut-off hypersurface.  Using the fluid/gravity correspondence, we argue that one can determine non-linear solutions to this problem in the long wavelength regime. On the boundary we find a conformal fluid with Dirichlet constitutive relations, viz., the fluid propagates on a `dynamical' background metric which depends on the local fluid velocities and temperature. This boundary fluid can be re-expressed as an emergent hypersurface fluid which is non-conformal but has the same value of the shear viscosity as the boundary fluid. The hypersurface dynamics arises as a collective effect, wherein effects of the background are transmuted into the fluid degrees of freedom. Furthermore, we demonstrate that this collective fluid is forced to be non-relativistic below a critical cut-off radius in AdS to avoid acausal sound propagation with respect to the hypersurface metric. We further go on to show how one can use this set-up to embed the recent constructions of flat spacetime duals to non-relativistic fluid dynamics into the AdS/CFT correspondence, arguing that a version of the membrane paradigm arises naturally when the boundary fluid lives on a background  Galilean manifold.
\end{abstract}
\thispagestyle{empty}
\setcounter{page}{0}
\end{titlepage}

\renewcommand{\baselinestretch}{1}  
\tableofcontents
\renewcommand{\baselinestretch}{1.2}  

\section{Introduction}
\label{s:intro}

The AdS/CFT correspondence~\cite{Maldacena:1997re} which postulates a remarkable duality between large $N$ quantum field theories and gravitational dynamics, provides a useful theoretical laboratory to address questions underlying the dynamics of these systems. Not only has it proven useful to obtain quantitative information about the dynamics of strongly coupled field theories, but it also provides a unique perspective into the geometrization of field theoretic concepts.
 
Since the early days of the AdS/CFT correspondence it has been known that the radial direction of the bulk spacetime encodes in some sense the energy scale of the dual field theory \cite{Susskind:1998dq}. While the nature of this map is not terribly precise outside of the simple example of pure \AdS{} geometry (dual to the vacuum state of the field theory), it nevertheless provides valuable intuition about certain basic aspects of effective field theory dynamics \cite{Banks:1998dd,Peet:1998wn}, and has led to the idea of the holographic  renormalisation group \cite{deBoer:1999xf}, which relates the radial `evolution' in AdS to RG flows in field theories. More recently this idea has been exploited to geometrize Wilson's concept of integrating out momentum shells to generate field theory effective actions, in terms of integrating out regions of the bulk geometry which in turn lead to  effective multi-trace boundary conditions on the cut-off surface, a fixed radial slice (in some preferred foliation) in AdS \cite{Heemskerk:2010hk,Faulkner:2010jy}. One of the key features of this holographic Wilsonian approach was the emergence of multi-trace deformations of the field theory even in the planar limit,  consistent with field theory expectation. 

A natural question in this context is what does this RG flow mean for the gravitational equations of motion?  More precisely, consider the problem of integrating out radial geometric shells in Einstein gravity with negative cosmological constant (which  is a consistent truncation of string theory/supergravity). One anticipates based on the standard dictionary which relates the bulk metric to the boundary energy momentum tensor to obtain a scale dependent effective action for the energy momentum tensor, containing arbitrarily high multi-traces of the stress tensor. The reason for the generation of these multi-traces is clear, once one factors in the intrinsic non-linearity of gravity. The basic equations in this context are of course easy to write down; as explained in \cite{Heemskerk:2010hk,Faulkner:2010jy} the flow is driven by the radial ADM Hamiltonian and one can in principle solve the resulting Hamilton-Jacobi like equation for the effective action on the cut-off hypersurface. Despite the conceptual simplicity  of the formulation of Wilsonian RG in terms of geometric effective actions, the point still remains that gravity's intrinsic non-linearity makes explicit solutions hard to come by.   

One can ask whether there is a tractable sector of the gravitational flow equations which leads to new insight. A natural avenue for exploration is suggested by the long-wavelength regime where we restrict attention to fluctuations of low frequency in the field theory directions. As evidenced by the fluid/gravity correspondence \cite{Bhattacharyya:2008jc}  there is an essential simplification in this regime; bulk Einstein's equations can be explicitly solved order by order in a long-wavelength expansion along the boundary.\footnote{See \cite{Rangamani:2009xk,Hubeny:2010wp} for reviews which describe developments in this area and \cite{Bhattacharyya:2008ji,Bhattacharyya:2008mz,Bhattacharyya:2008kq} for generalizations that will be of great use in the following.}  As such one should be able to use this framework in conjunction with the fluid dynamical expansion to derive an effective action for the low frequency degrees of freedom which live on a cut-off surface in the interior of the AdS spacetime.\footnote{It is natural to expect that such a construction might provide a holographic derivation of the
hydrodynamic deconstruction described in \cite{Nickel:2010pr}.}  Rather than tackle this problem directly we will take a slightly different tack in this paper, one which we believe clarifies some aspects of evolution in the radial direction and its possible connection to RG flows. One of our main conclusions will be that imposing rigid cut-offs in AdS is more naturally viewed in terms of perturbing the CFT by some non-local deformation or equivalently by introducing explicit state-dependent
sources in the boundary theory.

A second motivation is the recent work  \cite{Bredberg:2010ky,Bredberg:2011jq} which derives an explicit map between solutions of   vacuum Einstein equations (with no cosmological constant) and those of  incompressible  Navier-Stokes equations, thereby making direct contact with some of the ideas of the black hole membrane paradigm in asymptotically flat spacetime \cite{Damour:1978cg, Thorne:1986iy}. This problem, which has been further generalized in \cite{Compere:2011dx}, is the zero cosmological constant analog of the problem we consider (see also \cite{Eling:2009pb,Eling:2009sj} for another approach and \cite{Cai:2011xv} for related work). The idea is to consider a fixed timelike hypersurface with Dirichlet data enforcing a flat metric on the slice. Given these boundary conditions one wants to solve vacuum Einstein equations so as to obtain a solution which has a regular future horizon.\footnote{In the asymptotically flat spacetime one also has to specify initial data for radial evolution on the past null infinity ${\mathscr I}^-$. The boundary conditions chosen in \cite{Bredberg:2011jq} are such that no disturbance propagates into the bulk spacetime from ${\mathscr I}^-$.} By explicit construction which involves long wavelength fluctuations around flat space in a Rindler patch the authors of \cite{Bredberg:2011jq,Compere:2011dx} construct solutions to vacuum Einstein's equations order by order in a perturbation expansion in gradients along the hypersurface directions. The resulting geometry has a regular Rindler horizon, and one obtains a regular solution to Einstein's equations contingent on the fact that dynamics of the induced stress tensor on the hypersurface satisfies the incompressible Navier-Stokes equations. 

While this development is fascinating, one is hampered from a first principles understanding of the physics from a holographic viewpoint, owing to the rather poorly understood concepts of flat space holography. Moreover, given the connection between fluid dynamics (albeit relativistic and conformal) and Einstein's equations with negative cosmological constant as described by the aforementioned fluid/gravity correspondence \cite{Bhattacharyya:2008jc}(and its non-relativistic extension in \cite{Bhattacharyya:2008kq} ), it is interesting to ask whether the construction in \cite{Bredberg:2011jq}  can be obtained as a limit of the fluid/gravity map. If this is possible,  one can then look for the field theoretic interpretation of the flat space problem.

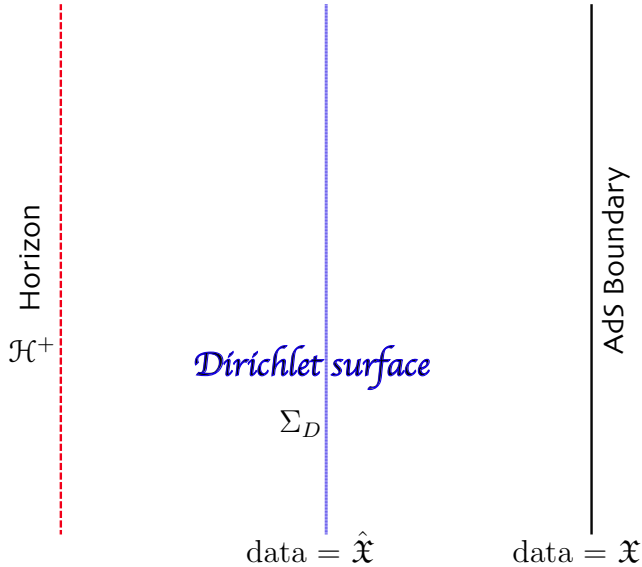
\begin{figure}[h!]
 \begin{center}
 \includegraphics[scale=0.5]{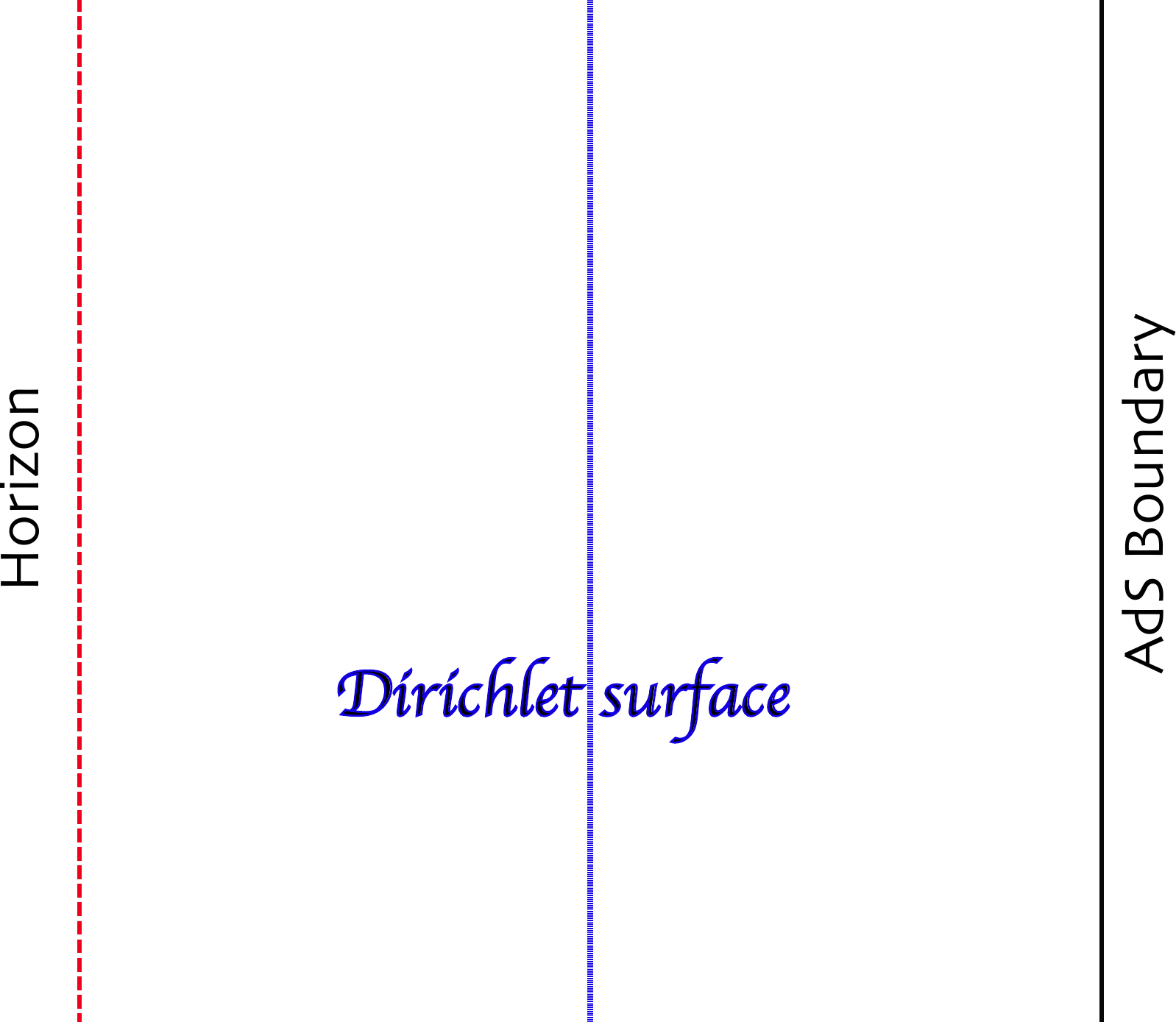}
 \begin{picture}(0,0)
\setlength{\unitlength}{1cm}
\put(-4.8,1.36){$\Sigma_D$}
\put(-8.4,2.3){${\mathcal H}^+$}
\put(-5.24,-0,36){data = $\hat{\mathfrak X}$}
\put(-1.7,-0.36){data = ${\mathfrak X}$}
\end{picture}
\end{center}
\caption{Schematic representation of the Dirichlet problem we consider in this paper. The Dirichlet surface is taken to be at some value $r = r_D$  where we impose boundary conditions on the fields. The solutions will further be constrained by requiring that they be regular on any putative horizon ${\mathcal H}^+$ (shown in the figure) or the origin. The question we are after is what is the boundary image of this Dirichlet data?}
\label{f:setup}
\end{figure}

Motivated by these issues, we consider a region of the AdS spacetime bounded by a timelike hypersurface $\Sigma_D$ at some radial position, say $r = r_D$ in the supergravity limit of AdS/CFT. We are interested in solving for the bulk dynamics
where will give ourselves the freedom to specify boundary conditions on $\Sigma_D$. The second boundary condition (which is necessary to zero-in onto a unique solution) will be specified by demanding regularity in the interior of the spacetime. We have schematically depicted the set-up in \fig{f:setup}. In the large $N$ limit, the specification of the problem thus is tantamount to solving classical partial differential equations (PDEs) in AdS with a Dirichlet boundary condition imposed on various fields at the hypersurface $\Sigma_D$. The question we would like to know the answer to is simply: ``What is the problem that we are solving in the CFT language?'' 

As we have reviewed above, various results exist in literature that suggest that solving such a Dirichlet problem is analogous to some  kind of RG from the CFT point of view. Despite the strongly suggestive nature of this holographic RG point of view, it is also not very clear what kind of an RG is one  speaking of within a CFT. A-priori,  for one, it does not seem like the RG flow that arises from cutting off a CFT a la Wilson is the correct way to dualize the Dirichlet problem. Hence, our question - what is the CFT dual of a bulk Dirichlet problem?\footnote{To avoid confusion we wish to emphasize that we will refer to the bulk Dirichlet problem as one defined in the preceding paragraph. While this reduces to the standard Dirichlet problem on the boundary of AdS when we take the surface $\Sigma_D$ to infinity, we will soon see that the bulk Dirichlet problem induces different boundary conditions at infinity when the surface $\Sigma_D$  is retained at a finite position and it therefore pays to maintain the distinction.}

As a warm-up we first consider the bulk Dirichlet problem for linear PDEs, using the simple setting of a Klein-Gordon field propagating in a cut-off AdS spacetime. In this case it is not hard to see that one is  deforming the field theory by a non-local double-trace operator, whose precise form, we argue, can be extracted by suitable convolution of appropriate bulk-to-boundary propagators.

We then turn to the issue of setting up the problem in a gravitational setting, outlining it in general before moving on to the tractable setting of the fluid/gravity regime. In the long wavelength regime we will argue that the bulk Dirichlet problem reduces to a particular forcing of the  fluid on the boundary of the asymptotically AdS spacetime. The fluid/gravity correspondence was generalized to fluids propagating on curved backgrounds with slowly varying curvatures in \cite{Bhattacharyya:2008ji} and the most general solutions which will prove to be of interest to us were presented in \cite{Bhattacharyya:2008mz}. Using these results it transpires that we can immediately write down the solution to the bulk Dirichlet problem in the long wavelength regime. 

The logic is the following: we wish to prescribe on the hypersurface $r =r_D$ a Lorentzian metric which we denote as ${\hat g}_{\mu\nu}$. This is arbitrary subject to the requirement that its curvatures be slowly varying so that we can treat it with the fluid/gravity perturbation scheme. We then solve Einstein's equations demanding regularity in the interior of the spacetime. Using standard intuition from the AdS/CFT correspondence it can be argued that the seed geometry which we need to set up the gradient expansion should simply be a black hole geometry which has a regular future event horizon, which furthermore satisfies the prescribed Dirichlet boundary condition.\footnote{The implicit idea behind the  uniqueness of the seed geometry here is based on the notion that equilibrium dynamics in the field theory at finite temperature is governed by a black hole, i.e., we are always in the `deconfined phase' in the field theory on the boundary.} It is not hard to see that such a seed solution is obtained by  simply performing a coordinate transformation of the well known planar Schwarzschild-AdS black hole.

But this is precisely the set-up of \cite{Bhattacharyya:2008ji,Bhattacharyya:2008mz}, the only difference being the fact that in these works the Dirichlet data is imposed on the boundary at $r =\infty$. Let's call this boundary metric $g_{\mu\nu}$, which is also by definition slowly varying etc.. The solution to the asymptotic Dirichlet problem is characterized by the boundary metric $g_{\mu\nu}$, a distinguished velocity field $u^\mu$ (which is unit normalized) and a scalar function $b$ (determining the temperature or equivalently the local energy density). Let us denote these variables collectively as ${\mathfrak X}$. The boundary Brown-York stress tensor (up to counter-terms) takes the fluid dynamical form and is built out of the data contained in ${\mathfrak X}$.  

Now given the space of solutions to the asymptotic Dirichlet problem, we can reparametrize that
space of solutions appropriately to obtain the solutions of the new bulk Dirichlet problem. The only condition we have to satisfy is that the induced metric on $\Sigma_D$ in the solutions obtained this way\footnote{Note that in order to get something interesting, it is important that we have a solution space allowing for non-trivial metrics at infinity (which we do thanks to \cite{Bhattacharyya:2008mz}). We are not simply slicing a single spacetime but working with a family of geometries characterized by the boundary data ${\mathfrak X}$ which is being adjusted so as to agree with the Dirichlet boundary conditions.} be equal to ${\hat g}_{\mu\nu}$. Furthermore, we can extract the stress tensor on $\Sigma_D$\footnote{There is an issue of counter-terms that one can use when working at finite radial coordinate in an asymptotically AdS spacetime. We will take the conservative view that the relevant counter-terms are the same as those necessary asymptotically.}. We will
argue that there is a corresponding velocity field ${\hat u}^\mu$ (normalized with respect to the hypersurface metric) and a scalar function (the hypersurface temperature), which parameterize the stress tensor of the hypersurface, which not surprisingly takes the fluid dynamical form. The main novelty is that the stress tensor does not however 
correspond to that of a conformal fluid. The introduction of an explicit scale by way of the Dirichlet surface's location engenders a non-vanishing trace, which curiously evolves in a highly suggestive manner under change of cut-off surface position, see \req{trerd}.

Calling the totality of the data on the hypersurface ${\hat {\mathfrak X}}$ we further show that within the gradient expansion there is a one-to-one correspondence between  the hypersurface data and the boundary data; $\varphi_D: {\mathfrak X} \to {\hat {\mathfrak  X}}$ is bijective. This then has the advantage that we can immediately understand the boundary dual of the bulk Dirichlet problem as a conformal fluid which is placed on a\footnote{We use the word dynamical to characterize the boundary metric in the following sense: the metric on the boundary depends on the dynamical degrees of freedom of the system, viz., $u_\mu$ and $T$, as in \req{bdyggu}. This is a pre-specified constitutive relation for the boundary metric in terms
of the fluid variables, not unlike the constitutive relation for the energy momentum tensor. We will call such a constitutive relation coming from bulk Dirichlet problem as the Dirichlet constitutive relation. The fluid at the 
boundary, therefore, sees a dynamic metric background  but no new degrees of freedom are introduced and hence for example, there are no boundary Einstein's equations that need to be solved.}  `dynamical background' whose metric depends on the same set of variables that characterizes the fluid itself (in addition to the prescribed hypersurface metric). So from the boundary viewpoint there is a complete mixing between intrinsic and extrinsic data, which is the long-wavelength non-linear analog of the double trace deformation seen for the scalar toy model. Moreover, this solution allows us to see that the dynamics of the fluid on the Dirichlet surface, as given by the conservation equation on $\Sigma_D$, `emerges' as collective dynamics of the boundary CFT. In particular, the boundary fluid lives on a `dynamical background', and the effects of the background can be suitably subsumed into the fluid description. This suggests that the correct way to think about the hypersurface physics is in terms of a `dressed fluid' living on an inert geometry. Thus, the effective description of a fluid on this dynamical background is geometrically encapsulated in terms of the Dirichlet hypersurface dynamics.

Examining the resulting dynamics on $\Sigma_D$ we find that the hypersurface or effective fluid suffers from a possible pathology for $r_D$ smaller than some critical value $r_{D,snd}$. At $r_{D,snd}$  the sound mode of the effective fluid starts to propagate outside the inert background $\Sigma_D$'s light-cone. We suggest that in the CFT, this effect is due to the extreme forcing of the fluid on the boundary by the `dynamical' metric, and moreover propose that one can obtain sensible dynamics by projecting out the  sound mode. This involves looking at the fluid at a scaling  limit  and this can be formalized as taking the incompressible non-relativistic limit of the fluid on the hypersurface in a manner entirely analogous to the scaling limit described in for generic relativistic fluids in \cite{Bhattacharyya:2008kq,Fouxon:2008tb}.\footnote{We shall refer to scaling limit as the BMW limit after the authors of \cite{Bhattacharyya:2008kq}, despite the fact that we are focussing on sub-sonic excitations and that it has been well documented in classic textbooks, cf., \cite{Landau:1965pi}.}
 
Having understood the Dirichlet problem for generic $\Sigma_D$ away from the horizon ${\mathcal H}^+$, we then proceed to push this surface deeper into the spacetime and ask what happens as we approach the horizon. In this regime  $\Sigma_D$ dynamics continues to be described by incompressible Navier-Stokes equations in the limit, though with some slight differences from the BMW limit mentioned above. Zooming in onto the region between $\Sigma_D$ and the horizon, we provide an embedding of the construction of  \cite{Bredberg:2011jq,Compere:2011dx} into the fluid/gravity correspondence \cite{Bhattacharyya:2008jc}. Further, we demonstrate that  in this limit both the bulk metric in the region between $\Sigma_D$ and the boundary, and the boundary metric degenerate from metrics on a Lorentzian manifold to Newton-Cartan like structures. This raises interesting questions about the natural emergence of the Galilean structures in the AdS/CFT correspondence which we postpone for future work. 

The plan of this paper is as follows: In \sec{s:dscalar} we first address the Dirichlet problem for a scalar field propagating in \AdS{d+1} using this linear problem to build intuition. In \sec{s:dgrav} we pose the bulk Dirichlet problem for gravity in \AdS{} spacetime and solve it in the long wavelength approximation borrowing heavily on the results from the fluid/gravity correspondence. The remainder of the paper is then devoted to understanding the physics of our construction in various regimes: \sec{s:emergence} demonstrates how the Dirichlet surface dynamics, as governed by the conservation equation, arises from the boundary physics. Aided by this we argue that the Dirichlet dynamics is probably pathological past a critical radius and propose a non-relativistic scaling of the resulting fluid a la BMW in \sec{s:dgravnr} to cure this possible pathology. Finally, in \sec{s:nh} we study the near-horizon Dirichlet problem and make contact with the recent work on the flat space Dirichlet problem (and its connection with Navier-Stokes equations). We end with a discussion in \sec{s:discuss}. Various appendices contain useful technical results. In particular, to aid the reader we provide a comprehensive glossary of our conventions and key formulae in \App{app:notation}. This is followed by a complete `Dirichlet dictionary' relating hypersurface variables to boundary variables in \App{A:dirdict} for ready reference.

\noindent
{\em Note added:} While this work being completed we received \cite{Kuperstein:2011fn} which has partial overlap with the results presented in \sec{s:dgrav}. These authors also attempt to solve for bulk geometries with prescribed boundary conditions on $\Sigma_D$ in the long wavelength regime and interpret their results in terms of a RG flow of fluid dynamics.

\noindent
{\em Note added in v2:} In the first version of the paper the non-relativistic metrics quoted in \sec{s:dgravnr} and 
in \App{A:bmw} were incorrect; the metrics as presented do not solve the bulk Einstein's equations to the desired order. These are now corrected in the current version. However, the full set of terms that we need to include in order to see the Naiver-Stokes equation on the boundary is quite large. Hence in the main text we only report the results for the case where the non-relativistic fluid moves on a Ricci flat spatial manifold in \sec{s:dgravnr} and present the general results in a new appendix \App{s:bmwR}. We note that the results of \sec{A:bmw} also correct the expressions originally derived in \cite{Bhattacharyya:2008kq}.

\section{Dirichlet problem for probe fields}
\label{s:dscalar}

To set the stage for the discussion let us consider setting up the bulk Dirichlet problem for linear PDEs in an asymptotically \AdS{d+1} spacetime. As a canonical example we will consider the dynamics of a probe scalar field $\Phi(r,x^\mu)$ of mass $m$. Generalizations to other linear wave equations such as the free Maxwell equation are straightforward. 

We will let this scalar field propagate on a background asymptotically \AdS{d+1} geometry with spatio-temporal translational symmetries so that the background metric  can be brought to the form
\begin{equation}
ds^2 =r^2\, g_{\mu\nu}(r) \, dx^\mu \, dx^\nu + \frac{g_{rr}(r)}{r^2}  \, dr^2 
\label{bggen}
\end{equation}	
The boundary of the spacetime is at $r \to \infty$ and we will assume that the boundary metric is  the Minkowski metric on ${\mathbb R}^{d-1,1}$ for simplicity, so that asymptotically $g_{\mu\nu} \to \eta_{\mu\nu}$ and $g_{rr} \to 1$ (as $r\to \infty$).

The dynamical equation of motion for the scalar is the free Klein-Gordon equation which can be written as an ODE in the radial direction for the Fourier modes $\Phi_{k}(r)$ of  $\Phi(r,x^\mu)$ 
\begin{equation}
\Phi(r,x^\mu)=\int \frac{d^dk}{(2\pi)^d} \,e^{i\, k\cdot x} \, \Phi_k(r)\ , 
\end{equation}
and takes the form
\begin{equation}
 \frac{1}{\sqrt{-g\, g_{rr}}}\, \partial_r \, \left(\sqrt{-g\, g^{rr}} \,  \partial_r \Phi_{k}(r) \right) - \left(g_{\mu\nu} \,k^\mu\,k^\nu +m^2 \right) \Phi_{k} = 0 
\end{equation}	
As a second order equation we need to specify two boundary conditions. We are going to restrict attention to the finite part of the geometry as illustrated in \fig{f:setup} and impose Dirichlet boundary conditions for the field at some  hypersurface $\Sigma_D$ at $r = r_D$. The second boundary condition in general can
take the form of a regularity boundary condition in the interior of the spacetime. 
If we were working in a spacetime with a horizon this would demand that the mode functions 
of interest are purely ingoing at the future horizon. 

The question we wish to pose is the following: usually in an asymptotically \AdS{d+1} spacetime we know that the solution to the scalar wave equation above has two linearly independent modes with power-law fall-off characterized by the source $J_\phi$ and vev $\phi$ of the dual boundary operator ${\cal O}_\Phi$ (which we recall is a conformal primary). We wish to ask what is the characterization of the boundary data as a functional of the Dirichlet hypersurface data. In this simple linear problem it is easy to see that there is a one-one map between the two sets of data. Essentially we are asking how to tune $J_\phi$ and $\phi$ so that the value of the scalar field on the Dirichlet hypersurface at $r=r_D$ takes on its given value. 
 
While it is possible to derive a formal answer to the above question, it is useful to first visit the simple setting of pure \AdS{d+1} spacetime where we have the luxury of being able to solve the scalar wave-equation explicitly to see an explicit answer to the question.

\subsection{Probe scalar in \AdS{d+1}}
\label{s:adss}

We specialize our consideration to the pure \AdS{d+1} geometry where $g_{\mu\nu} = \eta_{\mu\nu} $ and $g_{rr} = 1$ and one has enhanced Lorentz symmetry on the constant $r$ slices. The wave equation simplifies to 
\begin{equation}
\frac{1}{r^{d-1} }\, \frac{d}{dr} \, \left(r^{d+1} \, \frac{d}{dr} \Phi_{k}(r) \right) - \left(k^2 +m^2 \right) \Phi_{ k} = 0 
\label{mkgads}
\end{equation}	
This is well known to have solutions in terms of Bessel functions, but we will proceed to examine the behavior in a gradient expansion to set the stage for the real problem of interest later.

\subsubsection{The $k=0$ case}
The translationally invariant solution of the massive Klein-Gordon equation \req{mkgads} in the bulk is\footnote{We will for the moment refrain from imposing any IR boundary condition so as to be able to see the general structure. After all in pure AdS imposing regularity at the Poincar\'e horizon would kill the vev which has to vanish in the vacuum. }
\begin{equation} \Phi(r) = \frac{\phi}{(2\nu)\,r^\Delta} +  r^{\Delta-d} J_\phi 
\end{equation}
where the dual primary has a scaling dimension $\Delta$ obeying
$\Delta(\Delta-d)=m^2$ along with a source $J_\phi$  and a normalized vev\footnote{Note that the vev is $(16\pi\, G_{d+1})^{-1} \, \phi$ where $G_{d+1}$ is the gravitational constant in AdS$_{d+1}$.}
 $\phi$  defined via
\begin{equation} 
J_\phi \equiv \left[r^{d-\Delta}\, \Phi(r)\right]_{r\to\infty}
\end{equation}
\begin{equation}
\phi \equiv \left[-r^{2\nu}\times r\partial_r \left( r^{d-\Delta}\Phi \right)\right]_{r\to\infty} = \left[-r^\Delta\left( r\partial_r \Phi -(\Delta-d)\Phi\right)\right]_{r\to\infty}
\end{equation}
with 
\begin{equation}
 \nu \equiv \Delta - \frac{d}{2} = \sqrt{\frac{d^2}{4} + m^2} 
\end{equation}	
for convenience. For simplicity, we will assume $\Delta > \frac{d}{2} $ and choose $m$ such that $\nu \notin {\mathbb Z}$ to avoid complications with logarithms. Extension to $\Delta \in [\frac{d}{2}-1, \frac{d}{2}]$ with the lower end of the interval saturating the unitary bound is possible with the added complication of taking proper account of the necessary boundary terms.

We will begin by rewriting this solution in terms of the quantities on the Dirichlet surface which we denote with a hat to distinguish them from the boundary data:
\begin{equation} 
\hat{J}_\phi \equiv \left[r^{d-\Delta}\,\Phi(r)\right]_{r\to r_D}=J_\phi+ \frac{\phi}{(2\nu)\,r_D^{2\nu}} \ ,
\label{j0}
\end{equation}
\begin{equation} 
\hat{\phi} \equiv \left[-r^{2\nu}\times r\partial_r \left( r^{d-\Delta}\Phi \right)\right]_{r\to r_D} =\phi \ .
\label{ph0}
\end{equation}

Since the transformation between the data on the boundary $\{J_\phi,\phi\}$ and that on the hypersurface $\{\hat{J}_\phi, \hat{\phi}\}$ is linear it is a simple matter to write the bulk solution in terms of the hypersurface variables. One simply has
\begin{equation} 
\Phi(r) =\frac{\hat{\phi}}{(2\nu)\, r^\Delta} +  r^{\Delta-d}  \left(\hat{J}_\phi- \frac{\hat{\phi}}{(2\nu)r_D^{2\nu}}\right) . 
\label{pDsol}
\end{equation}
This is the answer we seek and all that remains is to interpret this result.

It is now easy to notice that the imposition of  the Dirichlet condition on a hypersurface inside the bulk is equivalent to making the boundary source a specific function of the vev.  From \req{pDsol} we can read off  the specific deformation of
the boundary CFT action to be given by
\begin{equation}  \delta \mathcal{L}_{CFT} = -\frac{1}{16\pi \,G_{d+1}\, } \, \left(\hat{J}_\phi\,\hat{\phi}- \frac{1}{2(2\nu)\, r_D^{2\nu}}\, \hat{\phi}^2 \right) \propto \hat{J}_\phi \, {\cal O}_\Phi - \frac{(16\pi \,G_{d+1})}{2(2\nu)\, r_D^{2\nu}}\, {\cal O}_\Phi^2
\end{equation}
which happens to be an irrelevant double-trace deformation \cite{Witten:2001ua,Berkooz:2002ug} of the boundary CFT. Hence, at least in this simple setup the dual of the Dirichlet problem is to make the source of a primary ${\cal O}_\Phi$  a particular joint function of the vev of the primary in the given state and  another fixed (state-independent) auxiliary source.

\subsubsection{The $k\neq0$ case : Derivative expansion up to $k^2$}

Having seen the result for the translationally invariant case $k=0$, we now proceed with $k \neq 0$. It is well known that general solution to the wave equation \req{mkgads} is given in terms of Bessel functions which we parameterize as\footnote{We are working here with space-like momenta having $k^2>0$. Results for time-like momenta 
follow from replacing $k$ by $i\,k$ and using the Bessel relations
\begin{equation*}
\begin{split}
\frac{2(ix)^{\nu}}{\Gamma(\nu)}\,K_{\nu}(2ix) &= -\frac{\pi}{2}\,\frac{2x^{\nu}}{\Gamma(\nu)}Y_{\nu}(2x)
+ i \sin\left[(-\nu)\pi\right]\Gamma(-\nu+1) x^{\nu}J_{\nu}(2x)\\
\frac{\Gamma(\nu)}{2(ix)^{\nu}}\, I_{\nu}(2ix) &= \frac{\Gamma(\nu)}{2x^{\nu}}J_{\nu}(2x)\\
\end{split}
\end{equation*} 
}  
\begin{equation}
\Phi_k(r) = \frac{\phi_k}{r^\Delta}\times\frac{\Gamma(\nu)}{2(k/2r)^{\nu}}I_{\nu}(k/r) +  r^{\Delta-d} (J_\phi)_k  \times \frac{2(k/2r)^{\nu}}{\Gamma(\nu)}K_{\nu}(k/r)
\end{equation}
Note that our previous result for $k=0$ follows from just keeping the leading $x^0$ terms in the expansions
\begin{equation}
\begin{split}
\frac{2x^{\nu}}{\Gamma(\nu)}\;K_{\nu}(2x) &= \sum_{j=0}^{\infty} \frac{\Gamma(\nu-j)}{\Gamma(\nu)} \frac{(-x^2)^j}{j!} +x^{2\nu}\sum_{j=0}^{\infty} \frac{\Gamma(-\nu-j)}{\Gamma(\nu)} \frac{(-x^2)^j}{j!}\\
\frac{\Gamma(\nu)}{2x^{\nu}}I_{\nu}(2x) &=\sum_{j=0}^{\infty} \frac{\Gamma(\nu)}{(2\nu+2j)\Gamma(\nu+j)} \frac{x^{2j}}{j!}
\end{split}
\label{eq:expn}
\end{equation}
For a general $k$, we can repeat the analysis of the previous section. While this
can be done generally at all orders in $k$ with some work, for simplicity
we will resort to  derivative expansion keeping terms upto order $k^2$. Not only will this allow us to see some of the structures emerging explicitly, but it also sets the stage for our gravitational computation in  later sections.

Using the expansion \req{eq:expn} above, we have 
\begin{equation}
\begin{split}
\Phi_k(r) &= \frac{\phi_k}{(2\nu)\, r^\Delta}\left(1 +\frac{2\nu}{(2\nu+2)^2} \frac{k^2}{2r^2}+\ldots \right) +   r^{\Delta-d} (J_\phi)_k \left( 1-\frac{1}{(2\nu-2)} \frac{k^2}{2r^2}+\ldots\right.\\
&\qquad\qquad \quad \left. +
\frac{\Gamma(-\nu)}{\Gamma(\nu)}\left(\frac{k}{2r}\right)^{2\nu}\left\{1 +\frac{1}{(2\nu+2)} \frac{k^2}{2r^2}+\ldots\right\}\right)
\end{split}
\end{equation}
with the ellipses representing order $k^4$ terms and higher. 

The source at the intermediate surface is as before easily determined 
\begin{equation}
\begin{split}
(\hat{J}_\phi)_k &\equiv \left[r^{d-\Delta}\Phi_k(r)\right]_{r\to r_D}\\
&=  (J_\phi)_k \left(1-\frac{1}{(2\nu-2)} \frac{k^2}{2\,r_D^2}+\ldots + \frac{\Gamma(-\nu)}{\Gamma(\nu)}\left(\frac{k}{2r_D}\right)^{2\nu}\left\{1 +\frac{1}{(2\nu+2)} \frac{k^2}{2\,r_D^2}+\ldots\right\}\right)\\
&\qquad \qquad +\frac{\phi_k}{(2\nu)\, r_D^{2\nu}}\left(1 +\frac{2\nu}{(2\nu+2)^2} \frac{k^2}{2\,r_D^2}+\ldots \right) 
\end{split}
\label{j2}
\end{equation}
while the normalized vev of the primary to this order in derivative expansion can be determined after subtracting an appropriate counter-term as\footnote{If we keep $k^4$ terms and higher, one needs to subtract appropriate counter-terms at that order. These counter-terms are determined by requiring that $\hat{\phi}_k$ is finite as $r_D\to\infty$. The explicit expressions for counter-terms to any required order can be determined using the expansions in \eqref{eq:expn} (see for e.g., \cite{Papadimitriou:2010as}). }
\begin{equation}
\begin{split}
\hat{\phi}_k &=\left[-r^{2\nu}\times r\partial_r \left( r^{d-\Delta}\Phi_k \right)+\frac{r^\Delta}{2\nu-2}\frac{k^2}{r^2}\Phi_k +\ldots \right]_{r\to r_D}\\
&=\phi_k\left(1 +\frac{2\nu-2}{(2\nu)^2} \frac{k^2}{2\,r_D^2}+\ldots \right) +\frac{4(J_\phi)_k}{(2\nu-2)} \times\frac{\Gamma(-\nu)}{\Gamma(\nu+2)}\left(\frac{k}{2}\right)^{2\nu}+\ldots  \\
\end{split}
\label{ph2}
\end{equation}

To solve the Dirichlet problem,we need to solve for $\phi_k,(J_\phi)_k $ in terms of the hatted variables from \req{j2} and \req{ph2} which can be inverted to get 
\begin{equation}
\begin{split}
(J_\phi)_k &=   \frac{(\hat{J}_\phi)_k}{\mathfrak{D}}\left(1 +\frac{2\nu-2}{(2\nu)^2} \frac{k^2}{2r_D^2}+\ldots \right)  -\frac{\hat{\phi}_k}{\mathfrak{D}\, (2\nu) \, r_D^{2\nu}}\left(1 +\frac{2\nu}{(2\nu+2)^2} \frac{k^2}{2\,r_D^2}+\ldots \right) \\
\phi_k &=\frac{\hat{\phi}_k}{\mathfrak{D}} \left( 1-\frac{1}{(2\nu-2)} \frac{k^2}{2\,r_D^2}+\ldots +
\frac{\Gamma(-\nu)}{\Gamma(\nu)}\left(\frac{k}{2\,r_D}\right)^{2\nu}\left\{1 +\frac{1}{(2\nu+2)} \frac{k^2}{2\,r_D^2}+\ldots\right\}\right)\\
&\qquad \qquad -\frac{4(\hat{J}_\phi)_k}{\mathfrak{D}(2\nu-2)} \times\frac{\Gamma(-\nu)}{\Gamma(\nu+2)}\left(\frac{k}{2}\right)^{2\nu}+\ldots  \\
\end{split}
\end{equation}
where the momentum dependent coefficient ${\mathfrak D}$ is 
\begin{equation}
\begin{split}
\mathfrak{D}&\equiv 1-\frac{4(2\nu-1)}{(2\nu)^2(2\nu-2)} \frac{k^2}{2\,r_D^2}+
\frac{\Gamma(-\nu)}{\Gamma(\nu)}\left(\frac{k}{2\,r_D}\right)^{2\nu}\left\{1-\frac{1}{\nu^2(\nu-1)} \right.\\
&\left.\qquad \qquad \qquad+\; \frac{(2 \nu +1)^4-4 (2 \nu +2)^2+7}{(2 \nu)^2
   (2 \nu +2)^3}\;\frac{k^2}{r_D^2}+\ldots\right\}\\
\end{split}
\end{equation}

As we saw in the $k=0$ case, we have yet again determined a state dependent source on the boundary for the primary operator ${\cal O}_\Phi$.  The key feature to note from the above analysis, is that the expression for the boundary source $J_\phi$ is non-analytic in $k$ and hence non-local when Fourier-transformed back to position space. Hence, we see that in general we have a map between the non-local double trace deformation on the boundary and the Dirichlet data on $\Sigma_D$ (similar non-local double-trace deformations were explored earlier in \cite{Marolf:2007in}).

\subsection{A general proposal for linear systems}

From the analysis of the free scalar wave equation in \AdS{d+1} the picture is rather clear. In the CFT, in general one can make the source a non-local functional of the vev of the primary operator. Usually such a function can be fed into the  holographic dictionary via a `state-dependent' boundary condition, which whilst  somewhat unnatural from a field theory
is a perfectly sensible boundary condition to consider. For some special classes of functionals, this state-dependent boundary condition has a very simple bulk interpretation as a Dirichlet boundary condition imposed on an intermediate surface, implying that we can trade the non-locality of the boundary sources into local behavior at some lower radius.

We just have one further  question to answer before we declare victory: how do we in practice determine this special set of sources in various holographic setups? For the general backgrounds we can formally write the solution to the wave equations in terms of integrals over the Dirichlet data convolved with suitable `Dirichlet bulk to boundary propagators', ${\cal K}_\text{source}$ and ${\cal K}_\text{vev}$. The former propagates the information contained in $\hat{J}_\phi$ to the boundary source, while the latter allows determination of the contribution from the vev $\hat{\phi}$ on $\Sigma_D$, i.e., formally
\begin{equation}
J_\phi(x)  = \int d^dx' \, \left\{{\cal K}_\text{source}(r_D, x; x') \, \hat{J}_\phi(x') + {\cal K}_\text{vev}(r_D, x; x') \, \hat{\phi}(x')\right\}
\end{equation}	
Note that implicit in our definition of these Dirichlet bulk to boundary propagators is the information of the boundary condition in the interior of the geometry and the necessary counter-terms. While it is possible to work this out in more specific geometries, such as a Schwarzschild-AdS$_{d+1}$ spacetime to see the interplay of these IR boundary conditions, we will leave this toy problem for now, and proceed to analyze the more interesting case of gravitational dynamics in \AdS{d+1} wherein we do have to face-up with non-linearities of the equations of motion.\footnote{In fact, even for pure \AdS{d+1} it is interesting to ask what the deformation on the boundary is when we move $\Sigma_D$ close to the Poincar\'e horizon; in this limit it seems natural to expect that the asymptotically we obtain a  a Neumann boundary condition. We thank Don Marolf for emphasizing this to us.}

Before proceeding to the gravitational setting, however, let us make a few pertinent observations relevant to the motivation mentioned at the beginning of \sec{s:intro}. The result we have obtained is quite intuitive; demanding that our fields take on the desired value at $\Sigma_D$ entails a linear relation between the two pieces of data at infinity, thereby leading to the observation about the source depending on the vev. We also see that despite some superficial resemblance to the Wilsonian RG flow where too one encounters multi-trace operators there is a crucial distinction in the physics. In the formulation of \cite{Heemskerk:2010hk,Faulkner:2010jy} one finds that 
for fixed asymptotic data, upon integrating out the region of the geometry between the boundary and a cut-off surface (which we can for simplicity take to be $\Sigma_D$ for the sake of discussion) one obtains an effective action for a cut-off field theory living on $\Sigma_D$ with scale dependent sources. These are irrelevant double traces (which are the only terms generated in a Gaussian theory which the linear models under discussion are), and one obtains the $\beta$-functions for the double trace couplings along the flow. In the present context however what we have is a situation wherein we are forced to engineer a specific double trace deformation on the boundary so as to ensure that we satisfy the Dirichlet boundary conditions on $\Sigma_D$. This is conceptually different from usual notions of RG, where one does not conventionally consider state dependent boundary conditions in the UV. However, there is a sense in which renormalisation of sources takes place which will become quite clear when we look at the gravitational problem.

\section{The Dirichlet problem for gravity}
\label{s:dgrav}

Having understood the boundary meaning of the Dirichlet problem for probe fields in a fixed background, we now turn to the situation where we consider dynamical gravity in the bulk. While we could consider other matter degrees of freedom in the bulk whose backreaction we now have to take into account, we choose for simplicity to restrict attention to the dynamics in the pure gravity sector which, as is well known, is a consistent truncation of the supergravity equations of motion. From a field theory perspective, we are going to work in the planar limit and focus on the dynamics of a single operator, the stress tensor and its source, the CFT metric $g_{\mu\nu}$.

\subsection{Setting up the general Dirichlet problem}
\label{s:setdg}

First of all one should ask what does it mean to consider the Dirichlet problem at a fixed hypersurface in the bulk when gravity is dynamical. We will take the view that the location of the hypersurface $\Sigma_D$ is specified by a scalar function on the bulk manifold ${\cal M}_{d+1}$. We want to determine the  metric on this spacetime by solving the dynamical equations of motion subject to the boundary condition on the prescribed hypersurface.   To wit, we demand that ${\cal M}_{d+1}$ be endowed with  a Lorentzian metric ${\cal G}_{MN}$ which solves Einstein's equations with a negative cosmological constant. The equations of motion are  (in units where $R_{AdS} =1$)\footnote{We will use upper-case Latin indices for the bulk spacetime indices, reserving lower-case Greek indices for hypersurface or boundary indices. See \App{app:notation} for a  list of conventions.}
\begin{equation}
E_{MN} = {\cal R}_{MN} + d \, {\cal G}_{MN} = 0 
\label{eins}
\end{equation}	
We will adapt coordinates $X^A = \{r ,x^\mu\}$ to the hypersurface $\Sigma_D$ and take this distinguished surface to be at $r = r_D$ with intrinsic coordinates $x^\mu$.\footnote{It might be useful to view this scalar function as physically being specified either as the level set of the red-shift factor, or by introducing a dynamical scalar field.}  We impose the boundary condition 
\begin{equation}
{\cal G}_{MN} \, dX^M \, dX^N  \big|_{r\to r_D}  =   \;r_D^2\, \hat{g}_{\mu\nu}(x) \, dx^\mu\, dx^\nu
\label{indgh}
\end{equation}	
where ${\hat g}_{\mu\nu}(x)$ is the Dirichlet data we wish to specify. To complete the specification of the problem we should impose some boundary condition in the interior of the spacetime, which we will canonically take to be a regularity condition. The scaling by $r_D^2$ above whilst unconventional from the bulk perspective, is more natural in the AdS/CFT context for it makes it easy to compare with the case where we push the hypersurface to the boundary. 

The general problem as stated above is quite hard. For one it is not clear that for generic choices of Dirichlet data one obtains a solution compatible with regularity in the interior of the spacetime. While a local solution in an open neighbourhood of $\Sigma_D$ can presumably be obtained by adapting a Fefferman-Graham like expansion, one is unlikely to be able to gain much insight using this procedure. Moreover, given that the map between $\Sigma_D$ and the boundary is expected to be non-local (borrowing intuition from the linear problem) one wonders whether there are causality issues as well. In particular, for generic state dependent boundary conditions causality is murky -- does the source adjust itself acausally to obtain the appropriate response? Likewise on $\Sigma_D$ there is a concern that signals can propagate outside the light-cone of the metric $\hat{g}_{\mu\nu}$ (which they could for instance do through the bulk); does this imply a corresponding pathology for the boundary physics as well?\footnote{We thank Veronika Hubeny for extensive discussions on these issues.} In short, the well-posedness of the Dirichlet problem is a-priori unclear. Nevertheless, we will ignore all these subtleties for
now and forge ahead. 

After solving  Einstein's equations with the boundary conditions we have set-up, we can extract the Brown-York stress tensor on $\Sigma_D$, denoted $\hat{T}_{\mu\nu}$, using the standard set of boundary counter-terms  \cite{Henningson:1998gx,Balasubramanian:1999re} 
\begin{equation}
\hat{T}_{\mu\nu}=- \frac{r_D^d}{8\pi G_{d+1}} \left(\hat{K}_{\mu\nu}-\hat{K} \,\hat{g}_{\mu\nu}+(d-1)\, \hat{g}_{\mu\nu} +\ldots\right) .
\label{hypst}
\end{equation}	
$r_D^2\,\hat{K}_{\mu\nu}$ is the extrinsic curvature of the hypersurface $\Sigma_D$ and $\hat{K}$ is its trace, defined as usual in terms of the normal to the surface. 

Note that we have written the answer in terms of the intrinsic metric on the hypersurface which is related to the induced metric from the bulk up to a  rescaling by $r_D^2$. The homogeneous scaling of the stress tensor under allows us to fix an overall $r_D$ dependent pre-factor. The hypersurface stress tensor \req{hypst} is of course covariantly conserved (the Gauss-Codacci constraint), i.e., 
\begin{equation}
\hat{\nabla}_\mu{\hat{T}^\mu}{}_{\nu}=0\,.
\label{hcons}
\end{equation}	

Given $\{\hat{g}_{\mu\nu}, \hat{T}_{\mu\nu} \}$ we can ask what are the corresponding boundary conditions on the boundary that lead to the same geometry. Based on our scalar problem we can conclude that the boundary source $g_{\mu\nu}$ and stress tensor $T_{\mu\nu}$  are in general non-local functionals of the Dirichlet data. We would like to characterize the map between these two sets of data\footnote{\label{alert} At this point it is worthwhile to get a technical point out of the way. We will have two metric structures in the story henceforth, the hypersurface metric $\hat{g}_{\mu\nu}$ and a boundary metric $g_{\mu\nu}$. To avoid confusion, we will write all equations intrinsic to the hypersurface or to the boundary consistent with the respective metric structures. In practice this simply means that one raises/lowers indices of the equations with respect to the appropriate metric. These have to be handled with care, but by judicious use of the two metrics one can relate other components if required.}
\begin{equation}
\varphi_D: \{ \hat{g}_{\mu\nu}, \hat{T}_{\mu\nu} \} \to  \{g_{\mu\nu}, T_{\mu\nu} \} \ .
\end{equation}	

\begin{figure}[h!]
 \begin{center}
 \includegraphics[scale=0.5]{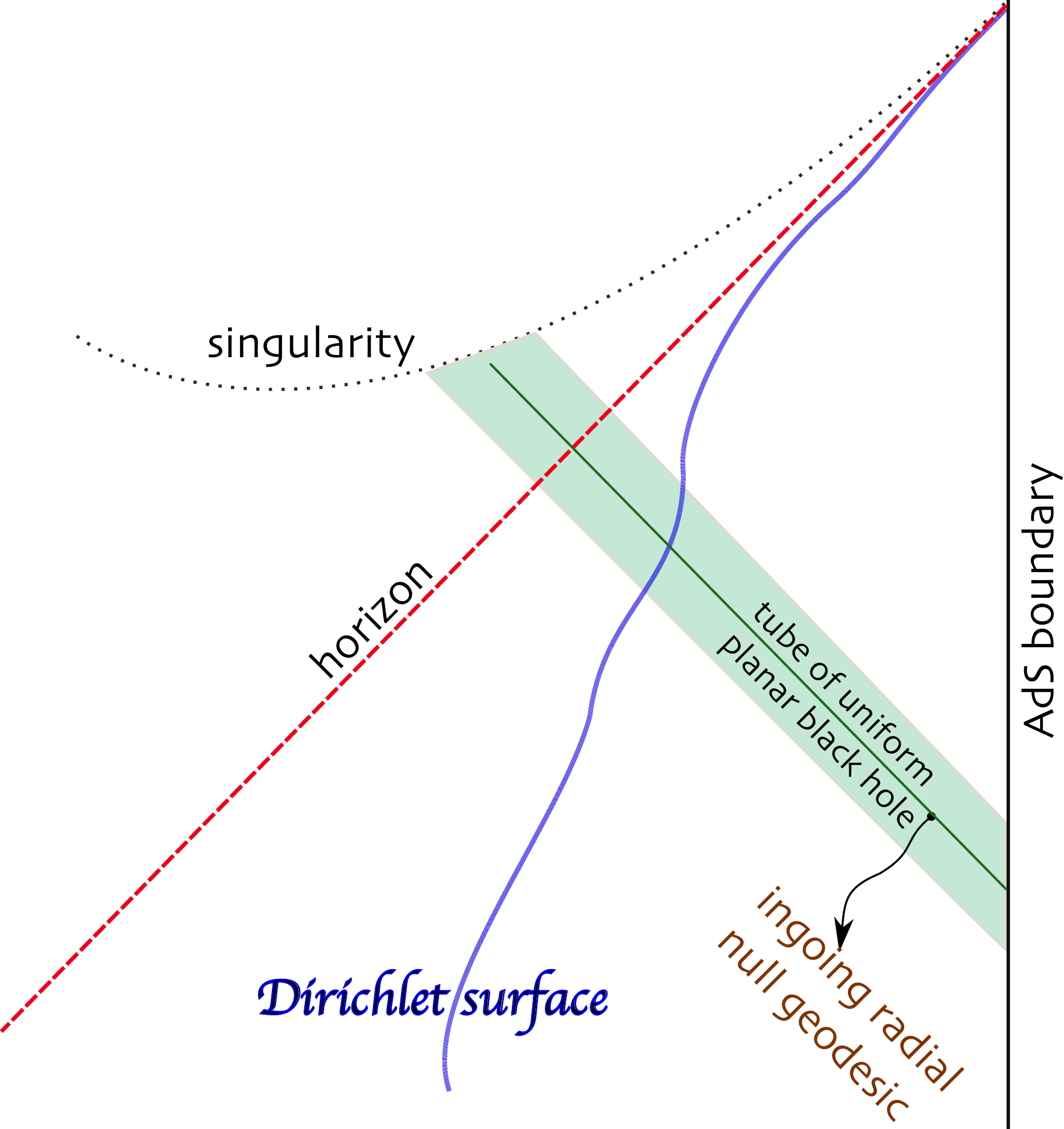}
  \begin{picture}(0,0)
\setlength{\unitlength}{1cm}
\put(-4.8,2){$\Sigma_D$}
\put(-6.5,-0.5){ $\hat{\mathfrak X}=\{\hat{g}_{\mu\nu}, \hat{u}^\mu, \hat{T}\} \quad \stackrel{\varphi_D}{\longmapsto}\quad{\mathfrak X}=\{g_{\mu\nu}, u^\mu, T\}$}
\end{picture}
\end{center}
\caption{Schematic representation of the gravitational Dirichlet problem in the fluid/gravity regime. The causal structure of the fluid/gravity spacetimes is illustrated emphasizing the tubewise approximation; in each tube the geometry resembles that of a uniformly boosted Schwarzschild-AdS$_{d+1}$ black hole. Suitable choices of the Dirichlet surface allow us to find the map between the boundary data ${\mathfrak X}$ and the Dirichlet hypersurface data $\hat{\mathfrak X}$ within each tube, rendering the problem tractable. }
\label{f:tubes}
\end{figure}
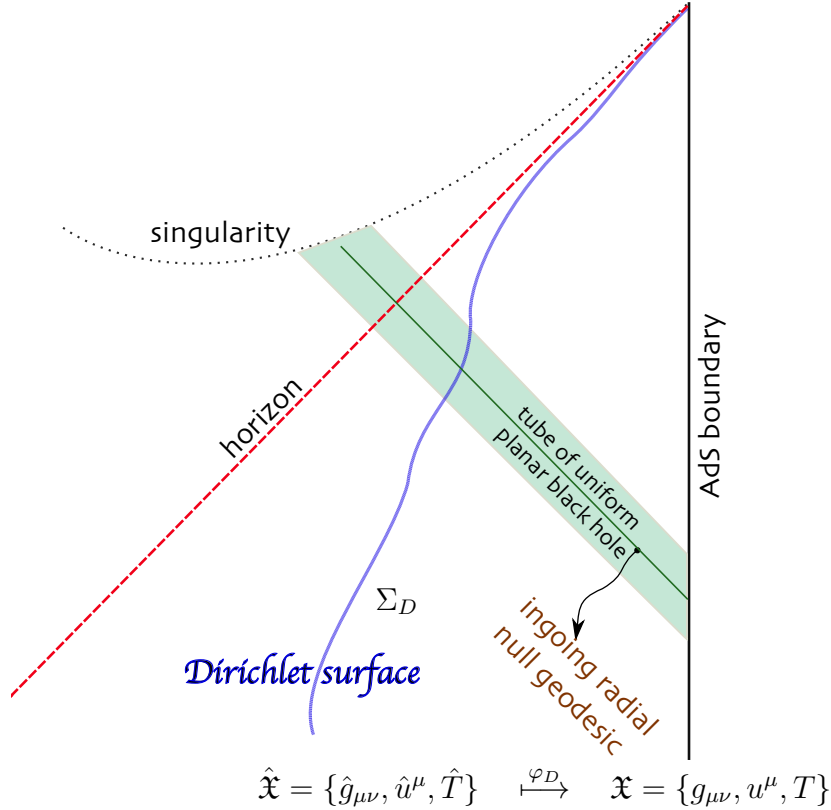

While this problem is in general difficult, there is one context in which we can not only solve for the boundary data in terms of the hypersurface variables, but we can also investigate the issues raised above in precise terms. This is the long-wavelength hydrodynamical regime along the hypersurface as in the fluid/gravity correspondence \cite{Bhattacharyya:2008jc,Bhattacharyya:2008ji,Bhattacharyya:2008mz}, wherein gravitational duals to arbitrary fluid flows on the boundary were constructed order by order in a gradient expansion.  The reason this is possible, as explained in these works,  is that the bulk spacetime in this long-wavelength regime is well approximated `tubewise' by the boosted planar Schwarzschild-\AdS{d+1} solution, see \fig{f:tubes}. As a result one finds that  Einstein's become ultra-local in the $x^\mu$ directions leading one to effective ODEs to determine the radial profiles. The tubes in question are centered around radially ingoing null geodesics, which can be used to translate information from any bulk hypersurface of interest to the boundary (see \cite{Bhattacharyya:2008xc,Bhattacharyya:2008mz} for a discussion of the causal structure). Given this, it is actually easy to solve the problem of finding the map $\varphi_D$ and we now describe the construction in the rest of this section.

\subsection{Dirichlet problem and the Fluid/Gravity correspondence}
\label{s:fg1}

To set the stage for our discussion, let us recall that the fluid/gravity map constructs in a gradient expansion, regular solutions to the bulk Einstein's equations \req{eins} which are dual to arbitrary fluid flows on the boundary of the asymptotically \AdS{d+1} spacetime. For a boundary metric $g_{\mu\nu}(x)$ which is slowing varying, the boundary stress tensor in this context is a not an arbitrary symmetric traceless two tensor, but constrained to take the hydrodynamical form. It is parameterized by $d$ independent parameters - a velocity field $u_\mu(x)$ (unit normalized so that  $g_{\mu\nu}\, u^\mu\,u^\nu =-1$) and a scalar function $b(x)$ which parameterizes the temperature. The bulk metric ${\cal G}_{MN}$ is determined in terms of the data ${\mathfrak X}  = \{g_{\mu\nu}(x), u_\mu(x), b(x)\}$.

We wish to implement the same procedure, but starting with analogous data $\hat{{\mathfrak X}}  = \{\hat{g}_{\mu\nu}(x), \hat{u}_\mu(x), b(x)\}$ on the Dirichlet hypersurface $\Sigma_D$. But given the ultra-locality inherent 
in the long-wavelength regime and the fact that \cite{Bhattacharyya:2008mz} have solved the problem
for arbitrary boundary metrics (corresponding to fluids on arbitrary slowly varying curved backgrounds),
we don't need to solve any equations. The solution space of the bulk Dirichlet problem coincides with the 
solution space found in \cite{Bhattacharyya:2008mz} and the problem at hand is readily solved by
slicing this solution space appropriately. With this aim, we now review the solutions constructed 
in \cite{Bhattacharyya:2008mz}.

\subsubsection{Review of fluid/gravity}
\label{s:fgrev}
The general solutions of the bulk equations of motion \req{eins} in the fluid/gravity regime take the form
\begin{equation} 
ds^2 ={\cal G}_{MN} \,dX^M \,dX^N = - 2 \,   {\mathfrak u}_\mu(x) \, dx^\mu \,\left( dr + r\,{\mathfrak V}_\nu(r,x)\,\,dx^\nu\right)+ r^2\,{\mathfrak G}_{\mu \nu}(r,x) \, dx^\mu\, dx^\nu \ , 
\label{formmetw}
\end{equation}
where the fields ${\mathfrak V}_\mu$ and ${\mathfrak G}_{\mu\nu}$ are functions of $r$ and $x^\mu$ which admit an expansion in the boundary derivatives and are known to second order in the gradients. For our purposes it will suffice to consider the first order metric where\footnote{In the fluid/gravity literature one chooses to maintain ${\mathfrak u}_\mu = u_\mu$ to all orders in the gradient expansion for simplicity. We will  generalize this suitably in the rest of the discussion to simplify our formulae.} 
\begin{eqnarray}
\mathfrak{u}_\mu &=& u_\mu\ ,\qquad \mathfrak{V}_\mu =  \mathcal{A}_\nu+\frac{r}{2}\, f(br)\, u_\nu \\
\mathfrak{G}_{\mu\nu} &=&  P_{\mu\nu} +2b \,F(br)\  \sigma_{\mu\nu} 
\end{eqnarray}	
with the functions
\begin{equation}
f(x) \equiv 1 - \frac{1}{x^d} \ , \qquad F(x)\equiv \int_{x}^{\infty}\frac{y^{d-1}-1}{y(y^{d}-1)}dy\,. 
\label{fFdef}
\end{equation}	
${\cal A}_\mu$ is the Weyl covariant connection introduced in \cite{Loganayagam:2008is} 
which is expressed in terms of the acceleration and the expansion of the velocity $u_\mu$ 
\begin{equation}
\mathcal{A}_\mu\equiv u^\lambda\nabla_\lambda u_\mu-\frac{\nabla_\lambda u^\lambda}{d-1}u_\mu =  a_\mu - \frac{\theta}{d-1} \, u_\mu ,
\end{equation}	
while $\sigma_{\mu\nu}$ is shear strain rate tensor of $u_\mu$: 
\begin{equation}
\sigma_{\mu\nu}\equiv{P_\mu}^\alpha {P^\beta}_\nu\,
\left[\nabla_{(\alpha} u_{\beta)}-g_{\alpha\beta}\frac{\nabla_\lambda u^\lambda}{d-1}\right],\quad\textrm{with}\quad P_{\mu\nu}\equiv g_{\mu\nu}+u_\mu u_\nu .
\end{equation}	

So the bulk metric to first order in derivatives explicitly takes the form:
\begin{equation}
ds^2=-2 \,u_\mu \,dx^\mu \left( dr + r\ \mathcal{A}_\nu dx^\nu \right) 
+  r^2 \left[ g_{\mu\nu}  +\frac{u_\mu u_\nu}{(br)^d}+2b \,F(br)\  \sigma_{\mu\nu}\right] dx^\mu dx^\nu + \ldots
\label{metricsimp:eq}
\end{equation}
and we have refrained from explicitly denoting the $x^\mu$ dependence of ${\mathfrak X}$ and the ellipses denote second order and higher gradient terms. The corresponding co-metric (the metric on the cotangent bundle/the inverse metric) is  given by 
\begin{equation}\begin{split}
\mathcal{G}^{AB}&\partial_A\otimes\partial_B \\
&= \left[r^2 \, f(br)-\frac{2\,r\,\theta}{d-1}\right]\partial_r\otimes\partial_r+2\left[u^\mu -r^{-1} a^{\mu}\right]\partial_\mu\otimes_s\partial_r\\
&\qquad +r^{-2}\left[P^{\mu\nu}   -2b \,F(br)\  \sigma^{\mu\nu}\right]\partial_\mu\otimes\partial_\nu\\
\end{split}\end{equation}
The stress tensor on the boundary is that of a viscous relativistic fluid:
\begin{equation}
T_{\mu\nu} = p\, g_{\mu\nu} + (\varepsilon  + p) \, u_\mu\,u_\nu - 2 \, \eta\, \sigma_{\mu\nu} + \ldots
\label{Tbdy}
\end{equation}	
with thermodynamic state variables
\begin{equation}
p = \frac{1}{d-1}\,  \varepsilon  = \frac{1}{16\pi\, G_{d+1} } \, \frac{1}{b^d} \,  
\label{epbdy}
\end{equation}	
and shear viscosity
\begin{equation}
\eta=\frac{1}{16\pi \,G_{d+1}} \;\frac{1}{b^{d-1}}\,.
\label{etab}
\end{equation}	
The expression in \req{metricsimp:eq} should be thought of as a way to generate solutions of Einstein equations
when provided with hydrodynamic configurations that solve the ideal fluid equations derived form the $T_{\mu\nu}$
above, i.e., when provided with $u^\mu,b$ that satisfy $\nabla_\mu T^{\mu\nu} = 0$ to first order in gradients. 
Our aim is to reformulate this set of solutions as solutions  to the bulk Dirichlet problem.

\subsubsection{Dirichlet data from fluid/gravity solutions}
\label{s:dfgsol}

Given the solution \req{metricsimp:eq} to the bulk equations of motion, we will begin by simply slicing it at a given radial position $r=r_D$ and extract the intrinsic metric on the Dirichlet surface\footnote{We would like to thank Sayantani Bhattacharyya for collaboration on  some of the  ideas in this section.}.

The advantage of working with the Weyl covariant form of the bulk metric is that one can simultaneously deal with Dirichlet surfaces specified by slowly varying functions $r=\rho(x)$. These can always be brought to the form $r=r_D$ by working in a suitable boundary Weyl-frame (local rescaling by a conformal factor $\log(1-\rho(x)/r_D)$ will do the trick). 
The Weyl connection $\mathcal{A}$ eats $\rho(x)$ so that
\[ \mathcal{A}_{\mu}=\rho^{-1} \left(\nabla_\mu+ a_\mu - \frac{\theta}{d-1}\, u_\mu \right) \rho\]
with this understanding all our formulae hold for arbitrary $\rho(x)$.

At fixed $r=r_D$ the hypersurface metric reads (recalling \req{indgh}) to first order
\begin{equation}
\hat{g}_{\mu\nu}= g_{\mu\nu}   + \frac{u_\mu u_\nu}{(b\,r_D)^d}+2b  \,F(br_D)\  \sigma_{\mu\nu} -\frac{2}{r_D}\, u_{(\mu} \mathcal{A}_{\nu)} + \ldots
\end{equation}
While this relation was obtained by slicing a known solution with prescribed asymptotic boundary conditions, it has the nice feature of having solved the equations of motion (to the desired order) and moreover satisfies the regularity condition in the interior. We will now turn the logic around and imagine $\hat{g}_{\mu\nu}$ to be specified at $\Sigma_D$ and view the equation above as specifying the boundary  intrinsic metric in terms of the hypersurface metric. Hence, the above equation
giving $g_{\mu\nu}$ in terms of $\hat{g}_{\mu\nu}$ and $u_\mu$ is the Dirichlet constitutive relation we 
seek -- when such a relation is imposed on the boundary data, the intrinsic hypersurface metric is 
automatically fixed to $\hat{g}_{\mu\nu}$. 

To complete  the specification of the  map $\varphi_D$ we need to further eliminate the boundary velocity field in 
favour of a vector field on the hypersurface. To do this we need to examine the hypersurface stress tensor 
and parameterize it appropriately.  The stress tensor at $r_D$ is easily obtained from \req{hypst} to be 
\begin{equation}
\hat{T}_{\mu\nu} =  \hat{p}\, \hat{g}_{\mu\nu} + \frac{1}{\hat{\alpha}^2}\, (\hat{\varepsilon}+\hat{p})\, u_\mu u_\nu - 2\, \hat{\alpha}\, \eta \, \sigma_{\mu\nu} +\frac{2}{r_D} \, (\hat{\varepsilon}+\hat{p})\, u_{(\mu} \mathcal{A}_{\nu)} +\ldots
\label{eq:rDquantities}
\end{equation}
with\footnote{The expressions for the hypersurface energy density and pressure $\hat{\varepsilon}$ and $\hat{p}$ can be re-written in terms of the hypersurface temperature $\hat{T}$ (see \req{hst2})
which is the more natural quantity on $\Sigma_D$. However, it is convenient for practical reasons to leave these definitions in terms of $b$.}
\begin{equation}
\hat{\varepsilon}\equiv\frac{(d-1)}{8\pi\, G_{d+1}}\; \frac{\hat{\alpha}}{\hat{\alpha}+1}\; \frac{1}{b^d} \,,\quad\quad
\hat{\varepsilon} + \hat{p} \equiv\frac{d }{16\pi \,G_{d+1}}\,  \frac{\hat{\alpha}}{b^d}\,,
\label{eqofstate:eq}
\end{equation}
with $\eta$ as given before in \req{etab} and we have defined a hypersurface scalar
\begin{equation}
\hat{\alpha}\equiv \frac{1}{\sqrt{f(b\, r_D)}} =\frac{1}{\sqrt{1-(b\,r_D)^{-d}}}\,.
\label{alhatdef}
\end{equation}
The stress tensor is tantalizingly similar to that of a viscous fluid, but as yet, we cannot interpret $\eta$ as the shear viscosity since we have not expressed  $\hat{T}_{\mu\nu}$, in terms of hypersurface variables.

This is however easy to remedy.  Define $\hat{u}_\mu$ to be the unit normalized (with respect to  $\hat{g}_{\mu\nu}$ of course) timelike eigenvector of $\hat{T}_{\mu\nu}$. A simple computation shows that 
\begin{equation}
\hat{u}_\mu\equiv \frac{u_\mu}{\hat{\alpha}} + \frac{\hat{\alpha}}{r_D} \mathcal{A}_\mu\,.
\end{equation}
We want to express the hypersurface stress tensor in terms of $\hat{u}_\mu$ and its gradients with respect to the $\hat{g}_{\mu\nu}$ compatible connection $\hat{\nabla}_\mu$. This can be done by the standard computation of the difference of $\nabla_\mu - \hat{\nabla}_\mu$. We outline the calculation in \App{app:CovDeriv} and simply quote the relevant result here: 
\begin{equation}
\hat{\sigma}_{\mu\nu}=\hat{\alpha} \, \sigma_{\mu\nu}\,,\quad\quad \mathcal{A}_\nu =\hat{\mathcal{A}}_\nu - \frac{\frac{d}{2}(\hat{\alpha}^2-1)}{1+\frac{d}{2}(\hat{\alpha}^2-1)}\;  \hat{a}_\nu\,.
\end{equation}
Armed with this data we can write now the stress tensor at $r=r_D$ as
\begin{equation}
\hat{T}_{\mu\nu} = \hat{p}\, \hat{g}_{\mu\nu} + (\hat{\varepsilon}+\hat{p})\, \hat{u}_\mu \,\hat{u}_\nu - 2\, \eta\, \hat{\sigma}_{\mu\nu}+\ldots
\end{equation}
We see that the result indeed is a stress tensor of a relativistic  fluid with  energy density $\hat{\varepsilon}$, pressure $\hat{p}$, given in \req{eqofstate:eq} and the same value of shear viscosity as the boundary theory $\eta$, \req{etab}.   The dynamical content of this system is still the conservation equation 
\begin{equation}
\hat{\nabla}_\mu{\hat{T}^\mu}{}_{\nu}=0\,.
\label{hypcons}
\end{equation}	
which follows from realizing that this is the `momentum constraint' equation for the radial slicing of Einstein's equations (or if one prefers the Gauss-Codacci constraint on the hypersurface).  Its equation of state is given in \eqref{eqofstate:eq} and we will momentarily evaluate the trace of $\hat{T}_{\mu\nu}$ in \eqref{eq:traceDirichlet}.

Further note that the stress tensor has no contribution associated with the expansion of the fluid, i.e., the bulk viscosity vanishes identically on the hypersurface. Nevertheless, the fluid is not a conformal fluid, for the trace of the stress tensor is non-vanishing. 
\begin{equation}
\hat{T}^\mu{}_\mu = \hat{T}_{\mu\nu}\, \hat{g}^{\mu\nu}= -  \hat{\varepsilon} + (d-1)\hat{p} =\frac{d(d-1)}{16\pi \, G_{d+1}}\ \frac{\hat{\alpha}-1}{\hat{\alpha}+1}\ \frac{\hat{\alpha}}{b^d} \ .
\label{eq:traceDirichlet}
\end{equation}
This is not entirely surprising for we have introduced an explicit scale $r_D$ into the problem, and as required for consistency the trace vanishes in the limit $r_D \to \infty$ as $\hat{\alpha} \to 1$. More curious is the fact that rate of change of the trace with the $\Sigma_D$'s radial location is simple:
\begin{equation}
\hat{T}^\mu{}_\mu =-r_D\frac{d\hat{\varepsilon}}{dr_D}\,
\label{trerd}
\end{equation}	
The evolution of the trace is highly suggestive for \req{trerd} can be interpreted as saying the the trace is generated by the variation of the local energy density with respect to some scale. This kind of a relation probably hints at some 
kind of non-linear realization of scale invariance. Again this is reminiscent of the holographic RG ideas and it would be interesting to flesh this out in greater detail.

Having the notion of the hypersurface velocity field $\hat{u}_\mu$ we can now proceed to write the boundary metric in terms of hypersurface data. Inverting the relation for the velocities to obtain (see \App{app:CovDeriv})
\begin{equation}
u_\mu = \hat{\alpha}\, \hat{u}_\mu - \frac{\hat{\alpha}^2}{r_D} \,\left( \hat{\mathcal{A}}_\nu-\frac{\frac{d}{2}(\hat{\alpha}^2-1)}{1+\frac{d}{2}(\hat{\alpha}^2-1)}\; \hat{a}_\nu\right)
\label{buhu1}
\end{equation}	
one can show that
\begin{equation}
g_{\mu\nu} = \hat{g}_{\mu\nu} -(\hat{\alpha}^2-1) \, \hat{u}_\mu \, \hat{u}_\nu  + \frac{2\,\hat{\alpha}^2}{r_D} \,  \left[ \hat{u}_{(\mu} \hat{\mathcal{A}}_{\nu)}-\frac{\frac{d}{2}(\hat{\alpha}^2-1)}{1+\frac{d}{2}(\hat{\alpha}^2-1)}\;  \hat{u}_{(\mu}\hat{a}_{\nu)} \right] - \frac{2b}{\hat{\alpha}}\,  F(br_D) \hat{\sigma}_{\mu\nu}
\label{bghg1}
\end{equation}	
The equations \req{buhu1} and \req{bghg1} together specify the map $\varphi_D$ we seek in the long-wavelength regime. Note that the hypersurface and the boundary data are determined by the same scalar function $b(x)$, which however enters non-trivially through $\hat{\alpha}$ in the determination of dynamics on $\Sigma_D$. 

Note that the light-cones of $g_{\mu\nu}$ are enlarged by a factor determined by $\hat{\alpha}$ relative to that of $\hat{g}_{\mu\nu}$. This is the first signal that there is some interesting interplay between boundary causal structures and fixing boundary conditions on $\Sigma_D$; we will address this issue in some detail in \sec{s:emergence}. But first we finish the solution to the Dirichlet problem as stated and write down the bulk metric in the long wavelength regime.

\subsection{Bulk metric in terms of Dirichlet data}
\label{s:dbulk}

Given the map in \req{buhu1} and \req{bghg1} we are in a position to re-write the bulk metric \req{metricsimp:eq} in terms of $\Sigma_D$ data $\hat{{\mathfrak X}}$ alone. Substituting the transformations we can write the final result as in \req{formmetw} with 
\begin{equation}\begin{split}
\mathfrak{u}_\mu &= u_\mu = \hat{\alpha}\, \hat{u}_\mu - \frac{\hat{\alpha}^2}{r_D}\left[\frac{\hat{a}_\mu}{\left[1+\frac{d}{2}(\hat{\alpha}^2-1)\right]}-\frac{\hat{\theta}}{d-1}\hat{u}_\mu\right] \\
\mathfrak{V}_\mu &=  \mathcal{A}_\mu+\frac{r}{2}\, f(br)\, u_\mu \\
&= \hat{\xi}\left[\frac{\hat{a}_\mu}{\left[1+\frac{d}{2}(\hat{\alpha}^2-1)\right]}-\frac{\hat{\theta}}{d-1}\hat{u}_\mu\right] +\frac{r}{2}\, f(br) \,\hat{\alpha}\, \hat{u}_\mu\\
\mathfrak{G}_{\mu\nu} &=  P_{\mu\nu} +2\,b \,F(br)\  \sigma_{\mu\nu} = \hat{P}_{\mu\nu} +2\,b \,\hat{F}(br)\  \hat{\sigma}_{\mu\nu}\\
\label{dirbulk1}
\end{split}\end{equation}	
and we have defined
\begin{equation} 
\hat{\xi} \equiv 1-\frac{1}{2}\,\frac{r}{r_D}\;\frac{f(br)}{f(br_D)}
= 1-\frac{\hat{\alpha}^2}{2}\, \frac{r}{r_D}\, f(br)
\label{xidef}
\end{equation}
and
\begin{equation}
\hat{F}(br) \equiv \frac{1}{\hat{\alpha}}\left(F(br)-F(br_D)\right) = \frac{1}{\hat{\alpha}} \;  \int_{br}^{br_D}\; \frac{y^{d-1}-1}{y(y^{d}-1)}dy\,.
\end{equation}
The factors of $\hat{u}_\mu$ are distributed between ${\mathfrak G}_{\mu\nu}$ and ${\mathfrak V}_\nu$ by the requirement that the former be transverse to ${\mathfrak u}_\mu$. 

This bulk metric ${\cal G}_{MN}$ solves the gravitational Dirichlet problem in the long-wavelength regime. We have thus solved the problem posed at the beginning of this section completely in terms in this regime aided by the ultra-locality of the gradient expansion. We summarize the complete map $\varphi_D$ and the resultant dictionary between CFT variables and the 
hypersurface variables in the \App{A:dirdict}. As a consistency check note that the results agree when we send the surface $\Sigma_D$ to the boundary with those derived in \cite{Bhattacharyya:2008mz}; one simply sets $r_D=\infty$ and $\hat{\alpha} =1$. In the remainder of the paper we will explore the relation between the dynamics on $\Sigma_D$ and 
that on the boundary, with an aim towards getting better intuition for various
issues raised in \sec{s:intro} and \sec{s:setdg}.

\section{Emergence of Dirichlet dynamics in the CFT$_d$}
\label{s:emergence}

Having obtained a solution to the gravitational Dirichlet problem in the long-wavelength regime, we 
now turn to analyze the underlying physics of the system.  From the viewpoint of the bulk, the dynamics of the system under
study has two equivalent descriptions -- one in terms of hypersurface (hatted)
variables $\hat{\nabla}_\mu\hat{T}^{\mu\nu}=0$, and another in terms 
of the un-hatted boundary variables $\nabla_\mu T^{\mu\nu}=0$. While the former
is more natural in the bulk (since it is $\hat{g}_{\mu\nu}$ which 
is fixed in the bulk for our Dirichlet boundary conditions on $\Sigma_D$), only the latter has a straightforward
interpretation in the CFT$_d$. Hence, it is interesting to ask how the hypersurface description should be interpreted within the CFT.

From the CFT point of view, the fluid is living in a metric background
with a Dirichlet constitutive relation (rewriting \req{bghg1} in terms of $u_\mu$)
\begin{equation}
\begin{split}
g_{\mu\nu}&=\hat{g}_{\mu\nu}    - \left(1-\frac{1}{\hat{\alpha}^2}+\frac{2\,\theta}{(d-1)\,r_D}\right)u_\mu u_\nu+\frac{2}{r_D}\, u_{(\mu} a_{\nu)} -2\,b  F(br_D)\  \sigma_{\mu\nu} +\ldots\\
g^{\mu\nu}&= \hat{g}^{\mu\nu}-\left(1-\hat{\alpha}^2-\frac{2\,\hat{\alpha}^4\,\theta}{(d-1)\,r_D}\right)u^\mu u^\nu \ +2\,b F(br_D)\  \sigma^{\mu\nu} -\frac{2\,\hat{\alpha}^2}{r_D}\, u^{(\mu} a^{\nu)}+ \ldots  \\
\end{split}
\label{bdyggu}
\end{equation}
where $\hat{\alpha}^2$ is defined in \req{alhatdef}.\footnote{We alert the reader again to footnote \ref{alert} - the expressions in \req{bdyggu} should be dealt with care as the l.h.s. and r.h.s contain contributions from quantities defined with respect to different metric structures.} The function $\hat{\alpha}$ increases with
increase in the local temperature (decreases with increase in the local $b$). These expressions
basically tell us how the ambient spacetime background the CFT$_d$ lives on responds to the motion of the  CFT fluid.

Let us first understand the physical content of the piece in the constitutive relation
with zero-derivatives. This can mostly be done heuristically which we will do first 
and then confirm it with an explicit calculation. The zero-derivative piece of 
the boundary metric $g_{\mu\nu}$ is made of an inert piece $\hat{g}_{\mu\nu}$
which does not respond to the fluid motion along with an additional piece 
proportional to $u_\mu u_\nu$. The presence of a term proportional to $u_\mu u_\nu$  
means the boundary metric effectively has a correction in its $dt^2$
piece in the local fluid rest-frame (which is responsible for opening up of the light-cone in the boundary). This kind of correction as is well known in 
general relativity just represents a gravitational potential well.
Note that this potential well travels along with the fluid and hence it is tempting to
think that there is a way to describe the collective packet of fluid and the local graviton cloud 
that it carries along in terms of a `dressed' fluid.

To guide our intuition, let us draw analogies with another familiar physical situation
where the background responds to the system locally via this kind of a relation. One
analogous situation is that of a charge carrier moving in a polarizable medium.
The polarizability of the medium defines the constitutive relation of the medium
in exact analogy with the constitutive relations for the metric above. We know that
in the case of the charge carrier moving in a polarizable medium often the
polarizability can be taken in to account by {\em shifting} the dispersion
of the charge carrier and pretending that this `dressed' charge carrier
is essentially moving through an inert medium. 

What we want to argue out in this section is the fact that a similar `dressing'
phenomenon happens in the case of the CFT$_d$ fluid  - we want to rewrite 
the problem of a fluid with $T^{\mu\nu}$ moving in the `polarizable' $g_{\mu\nu}$
into the problem where a dressed fluid with $T^{\mu\nu}_{\text{dressed}}$ moving
in the an inert spacetime $g_{\mu\nu,\text{inert}}$. It is clear that the inert
background is just the non-dynamical part of the metric, i.e., 
$g_{\mu\nu,\text{inert}} =\hat{g}_{\mu\nu}$. We would like to claim that 
$T^{\mu\nu}_{\text{dressed}} = \hat{T}^{\mu\nu} $. This then would be 
a complete physical picture of how the dynamics on a Dirichlet  hypersurface
in the bulk emerges directly from the boundary description. 

\subsection{Conservation equations at the boundary and on the Dirichlet surface}
\label{s:conseq}
Let us now implement the dressing picture heuristically described above at the level of the equations to derive 
the conservation equations on the Dirichlet surface from those on the boundary.
Given an arbitrary energy-momentum tensor 
\begin{equation}
T^{\mu\nu} = \varepsilon \,u^\mu u^\nu + p \,P^{\mu\nu} +\pi^{\mu\nu}
\label{piTdef}
\end{equation}
with $\pi^{\mu\nu}$ capturing the dissipative terms involving at least one gradient of the velocity field or thermodynamic state variables, we have
\begin{equation}\begin{split}
\nabla_{\nu}T^{\mu\nu} &= u^\mu\left[ u^\nu \nabla_\nu \varepsilon  +  (\varepsilon+ p) \nabla_\nu u^\nu\right] + (\varepsilon+p)\, a^\mu+ P^{\mu\nu}\nabla_{\nu}\, p  +\nabla_{\nu}\, \pi^{\mu\nu}\\
&= u^\mu\left[ \frac{s}{c_{snd}^2} u^\nu\nabla_\nu T + T\, s \, \theta-u_\alpha\nabla_{\beta}\pi^{\alpha\beta}\right]  + T\,s \,a^\mu+ s\, P^{\mu\nu}\nabla_{\nu} T  +P^{\mu\nu}\nabla_{\lambda}\pi_\nu^{\lambda}\\
\end{split}
\label{conseom}
\end{equation}
where we have introduced 
\[ c_{snd}^2\equiv \frac{dp}{d\varepsilon}= s\frac{dT}{d\varepsilon} \ , \]
and used the Euler relation $\varepsilon + p = s\, T$, with $s$ being the entropy density of the fluid.  Since the part proportional to $u^\mu$ and the part transverse to $u^\mu$ should separately vanish, we get 
\begin{equation}\begin{split}\label{eq:dT}
s\left[\partial_\mu + a_\mu-c_{snd}^2 \,u_\mu \,\theta \right] T+P_\mu{}^{\nu}\nabla_{\lambda}\pi_\nu{}^{\lambda}-c_{snd}^2 \,u_\mu \, \pi^{\alpha\beta}\nabla_\alpha u_\beta &=0
\end{split}\end{equation}

Similarly, $\hat{\nabla}_{\nu}\hat{T}^{\mu\nu}=0$ is equivalent to the equation
\begin{equation}\begin{split}\label{eq:HatdT}
\hat{s}\left[\partial_\mu + \hat{a}_\mu-\hat{c}_{snd}^2 \, \hat{u}_\mu \,\hat{\theta} \,\right] \hat{T}+\hat{P}_\mu{}^{\nu}\hat{\nabla}_{\lambda}\hat{\pi}_\nu{}^{\lambda}-\hat{c}_{snd}^2\, \hat{u}_\mu \,\hat{\pi}^{\alpha\beta}\hat{\nabla}_\alpha \hat{u}_\beta &=0
\end{split}\end{equation}
Using the relations 
\begin{equation}
\hat{s}=s \ , \qquad \text{and} \;\;  \hat{T}=T\, \hat{\alpha} ,
\label{hbst}
\end{equation}	
which are derived in \sec{s:csq} we can write
\begin{equation}\begin{split}
\hat{s}\left[\partial_\mu + \hat{a}_\mu-\hat{c}_{snd}^2\, \hat{u}_\mu\, \hat{\theta} \right] \hat{T}
&= s \left[\partial_\mu + \hat{a}_\mu-\hat{c}_{snd}^2 \,\hat{u}_\mu \,\hat{\theta} \right] T\hat{\alpha}\\
&= \hat{\alpha} \,s \left[\left(1+\frac{d\,\ln \hat{\alpha}}{d\, \ln T}\right)\partial_\mu +  \hat{a}_\mu-\hat{c}_{snd}^2 \,\hat{u}_\mu \,\hat{\theta} \right] T \\
&= \hat{\alpha} \left(1+\frac{d\ln \hat{\alpha}}{d\, \ln T}\right) \,s \left[\partial_\mu +  \frac{\hat{a}_\mu-\hat{c}_{snd}^2\, \hat{u}_\mu \, \hat{\theta}}{\left(1+\frac{d\, \ln \hat{\alpha}}{d\, \ln T}\right)} \right] T \\
\end{split}\end{equation}
so that \eqref{eq:HatdT} becomes
\begin{equation}\begin{split}\label{eq:HatdT2}
s \left[\partial_\mu +  \frac{\hat{a}_\mu-\hat{c}_{snd}^2\, \hat{u}_\mu \,\hat{\theta}}{\left(1+\frac{d\,\ln \hat{\alpha}}{d\,\ln\ T}\right)} \right] T+\frac{\hat{P}_\mu{}^{\nu}\hat{\nabla}_{\lambda}\hat{\pi}_\nu{}^{\lambda}-\hat{c}_{snd}^2\, \hat{u}_\mu\, \hat{\pi}^{\alpha\beta}\hat{\nabla}_\alpha \hat{u}_\beta}{\hat{\alpha} \left(1+\frac{d\,\ln \hat{\alpha}}{d\,\ln\ T}\right)} &=0
\end{split}\end{equation}

For the equation \eqref{eq:HatdT2} to describe the same dynamical system  as the equation \eqref{eq:dT}, it is necessary
and sufficient that
\begin{equation}\begin{split}\label{eq:matchEqn}
&\hat{a}_\mu-\hat{c}_{snd}^2 \,\hat{u}_\mu \,\hat{\theta} +\frac{\hat{P}_\mu{}^{\nu}\hat{\nabla}_{\lambda}\hat{\pi}_\nu{}^{\lambda}-\hat{c}_{snd}^2 \,\hat{u}_\mu\, \hat{\pi}^{\alpha\beta}\hat{\nabla}_\alpha \hat{u}_\beta}{\hat{T}\,\hat{s} } \\
&\quad\stackrel{?}{=}\left(1+\frac{d\,\ln \hat{\alpha}}{d\,\ln\ T}\right)\left[a_\mu-c_{snd}^2 \,u_\mu \,\theta +\frac{P_\mu{}^{\nu}\nabla_{\lambda}\pi_\nu{}^{\lambda}-c_{snd}^2 \,u_\mu\, \pi^{\alpha\beta}\nabla_\alpha u_\beta}{T\,s}\right]
\end{split}\end{equation}

We can show that this is indeed true at the  first derivative level by using the conversion formulae (rewriting \req{buhu1})
\begin{equation}
\begin{split}
{u}_\mu &= \left(1+\frac{\hat{\alpha}\,\hat{\theta}}{r_D(d-1)}\right)\hat{\alpha}\,\hat{u}_\mu - \frac{\hat{\alpha}^2}{r_D} \frac{\hat{a}_\mu}{\left[1+\frac{d}{2}(\hat{\alpha}^2-1)\right]}\\
\frac{d\ln \hat{\alpha}}{d\ln\ T}&= \frac{d}{2}(\hat{\alpha}^2-1), \quad {\theta}=\frac{1}{\hat{\alpha}}\hat{\theta}  \\
\hat{a}_\nu &= \left(1+\frac{d}{2}(\hat{\alpha}^2-1)\right) a_\nu
 ,\quad {\theta}{u}_\mu=\hat{\theta}\hat{u}_\mu \\
\hat{c}^2_{snd} &= \left(1+\frac{d}{2}(\hat{\alpha}^2-1)\right) c^2_{snd} .
\end{split}
\end{equation}
Hence, till this order, we have proved that $\hat{T}^{\mu\nu}$ is indeed the dressed energy-momentum
tensor that we were looking for.

It should be instructive to extend this analysis to 
higher orders in derivatives. In particular, it would be interesting to pin down the
specific property of the Dirichlet constitutive relation which leads to the fact
that the dressed viscosity $\hat{\eta}$ is same as the bare value $\eta$ and furthermore understand why the hypersurface fluid has no bulk viscosity. This would complement the analysis of \cite{Iqbal:2008by} who demonstrated the absence of corrections to the shear viscosity by considering a flow equation in the linearized regime between the boundary and the horizon.  

\subsection{Causality and relativistic fluids on the Dirichlet hypersurface}
\label{s:csq}

Having established a clear connection between the dynamics of the dressed fluid on the Dirichlet surface 
and that of fluid on a `dynamical' boundary metric, we now turn to examining the 
properties of the fluid motion. It seems a priori that all is well in 
the long wavelength regime with regards to the issues raised at the beginning of 
section \sec{s:dgrav} viz. the issue of locality and causality of the Dirichlet 
problem in \AdS{d+1}. However, this is probably a bit too quick; while 
it is true that we have local dynamical equations given by the 
conservation of the hypersurface stress tensor \req{hypcons}, we 
have not established firmly that these equations arise from a 
sensible thermodynamic system. We now proceed to address this issue.

The energy momentum tensor of the dressed fluid on the hypersurface 
is characterized by an energy density $\hat{\varepsilon}$ and 
a pressure $\hat{p}$ which are given in \req{eqofstate:eq}. 
In particular, the pressure of the fluid is
\begin{equation}
\hat{p} \equiv \frac{\left[1+\frac{d}{2}(\hat{\alpha}-1)\right]}{8\pi G_{d+1}b^d}\frac{\hat{\alpha}}{\hat{\alpha}+1} =  \frac{2\hat{\alpha}}{\hat{\alpha}+1}\left[1+\frac{d}{2}(\hat{\alpha}-1)\right] p
\label{prhyper}
\end{equation}
We also note that $\hat{\varepsilon} = \frac{2\,\hat{\alpha}}{\hat{\alpha}+1} \, \varepsilon$ which is useful in what follows.

Using the thermodynamic relations
\begin{equation}
\frac{d\hat{s}}{\hat{s}}= \frac{d\hat{\varepsilon}}{\hat{\varepsilon}+\hat{p}} ,\quad \frac{d\hat{T}}{\hat{T}}= \frac{d\hat{p}}{\hat{\varepsilon}+\hat{p}} \quad\text{and}\quad \hat{\varepsilon}+\hat{p} = \hat{T} \hat{s} 
\end{equation}
we get  the entropy density and the temperature of this fluid as
\begin{equation}
\hat{s}= \frac{1}{4 G_{d+1}}\, \frac{1}{b^{d-1}}=s ,\quad\text{and} \quad \hat{T}=\frac{d}{4\pi b}\hat{\alpha}=\hat{\alpha}\, T  
\label{hst2}
\end{equation}
as quoted above in \req{hbst}. The first of these relations follows from the fact that the entropy of the fluids on the asymptotic boundary as well as on the Dirichlet surface are given in terms of the area of the horizon which is unchanged by the solution. To determine the temperature on the hypersurface one has to account for the fact that the surface is in the interior of the spacetime. In the planar Schwarzschild-AdS$_{d+1}$ solution we get deviation  from the Hawking temperature (which is temperature in the CFT) via a red-shift factor $\hat{\alpha}$. Conversely, given the above relations \req{hst2}, the expressions for $\hat{\varepsilon}$ and $\hat{p}$ can be deduced using $d\hat{\varepsilon}=\hat{T}d\hat{s}$ and $d\hat{p}=\hat{s}d\hat{T}$.

The speed of sound mode in this system is given by
\begin{equation}\label{spsnd:eq}
\begin{split}
\hat{c}_{snd}^2 &\equiv \frac{\partial\hat{p}}{\partial\hat{\varepsilon}}= \frac{1}{d-1}\left[1+\frac{d}{2}(\hat{\alpha}^2-1)\right] = {c}_{snd}^2 \left[1+\frac{d}{2}(\hat{\alpha}^2-1)\right]
\end{split}
\end{equation}
This exceeds the speed of light as measured by $\hat{g}_{\mu\nu}$, i.e., we get superluminal sound propagation, for $\hat{\alpha}>\hat{\alpha}_{snd}$ where
$\hat{\alpha}_{snd}\equiv \sqrt{3-\frac{4}{d}}$. This corresponds to 
\begin{equation}\label{r_snd:eq}
\begin{split}
b\ r_{D,snd} &\equiv \left(\frac{\hat{\alpha}_{snd}^2}{\hat{\alpha}_{snd}^2-1}\right)^{1/d}\\
& =\left(\frac{3-\frac{4}{d}}{2-\frac{4}{d}}\right)^{1/d}\\
& \approx 1+ \frac{1}{d}\ln (3/2) + O(d^{-2}) \\
\end{split}
\end{equation}

One can intuitively understand this result from the viewpoint of the boundary fluid. As we noted earlier the 
boundary fluid is subject to a gravitational potential well. Should one locally increase the strength of
this well then the fluid would get sufficiently accelerated, perhaps leading to a pathology. This 
is manifest in the picture of the dressed fluid moving on an inert background achieved 
by translating over to the Dirichlet surface. In particular, this gets reflected in the
fact that the effective pressure $\hat{p}$ felt by the dressed fluid increases 
relative to its energy density $\hat{\varepsilon}$ thus driving the dressed fluid into
a regime where the dominant energy condition is violated. Such violations of the dominant
energy condition are known to be susceptible to superluminal sound modes\footnote{For 
an early discussion see \cite{Bludman:1968zz} where similar issues for fluids models
driven to high pressure regimes are discussed.} as we observed above.

How much should one be worried by this apparent acausal behavior where the dressed sound mode travels superluminally with respect to the inert part of the boundary metric?
After all, as is easily verified the mode with dispersion $\omega \sim \hat{c}_{snd}\, k$
propagates within the local light-cone of the `dynamical' boundary metric $g_{\mu\nu}$.
This is achieved by the phenomenon we had already alluded to towards the end of the
section \sec{s:dfgsol}: as we move our Dirichlet surface into the AdS, the Dirichlet
constitutive relation  ensures that the boundary light-cone opens up (see \req{bghg1})
thus ensuring that the dressed sound mode is not superluminal when measured 
with respect to $g_{\mu\nu}$. Of course, pending a detailed analysis of the initial value
problem posed by this dynamical system (and other possible global issues), 
one cannot assert that the  boundary physics is sensible from above observations alone.

We would however like to suggest that viewing the hypersurface fluid as 
an autonomous dynamical system, a superluminal sound mode probably indicates 
a pathology.  As described in \cite{Adams:2006sv} one should anticipate 
that the corresponding initial value problem\footnote{While the initial 
value problem for ideal, or even viscous fluids is ill-posed as the 
conservation equations are parabolic, we here want to drive home the
point that the pathology we want to encounter happens for the sounds modes 
that are usually non-problematic.} for the hypersurface fluid might be ill-posed.
Is this the way the long wavelength problem is telling us that the dual of 
the  generic bulk Dirichlet problem in the CFT is ill-posed? Is it possible 
that the bulk Dirichlet problem in gravity is pathological the moment
$r_D$ is finite and the fact that the effective dynamics of the fluid 
of the CFT remains sensible up to a critical radius $r_{D,snd}$ is just a
long wavelength artifact? Clearly this issue deserves further investigation.

We will now take an alternate approach which sidesteps these deep questions -- 
given that  our issue is with the superluminal sound mode on $\Sigma_D$ 
(for $r_D<r_{D,snd}$), is it possible to project this offending mode out of 
our dynamics  hence avoiding the entire issue?  We will now argue that
fortunately  the answer is yes  -- there is indeed a way to project out
the sound mode, retaining sensible physics at least 
within the long wavelength regime. 

The way to do this is to move to the incompressible non-relativistic 
regime of fluid/gravity correspondence first studied in \cite{Bhattacharyya:2008kq}.
We will now implement their construction in our setting allowing us to 
obtain sensible dynamics for the Dirichlet problem. We will postpone
some of the more general questions raised above to the 
discussion section \sec{s:discuss}.

\section{The Dirichlet problem for gravity with non-relativistic fluids}
\label{s:dgravnr}

The discussion so far has concentrated on mapping relativistic fluid dynamics on a curved 
background $\Sigma_D$ to a corresponding problem on the boundary where we have a relativistic
fluid on a `dynamical metric'. Moreover in the previous section, we cited the 
possible problems with the sound mode as a motivation to project it out 
consistently so as to get a  clearly sensible dynamical system 
with no possible issues with the initial value problem etc.. Our goal
now is to describe how this can be done consistently within our 
gradient expansion inspired by the non-relativistic incompressible 
scaling limit of \cite{Bhattacharyya:2008kq,Fouxon:2008tb}. As we
will see this limit has the added advantage that it naturally allows 
us to make contact with the metric derived in \cite{Cai:2011xv}.

The idea as explained beautifully in \cite{Bhattacharyya:2008kq}
is the following: every relativistic fluid has a scaling limit 
where we freeze out the propagating sound mode, which drives 
the fluid into a non-relativistic regime, while simultaneously
 making it incompressible. This BMW scaling can 
essentially be derived by the requirement  that 
one retains the non-linearities of the conservation equation 
(at least at first order). The resulting conservation equation is 
the classic incompressible non-relativistic Navier-Stokes equations.
Using the fluid/gravity map  \cite{Bhattacharyya:2008kq} constructed
a gravitational dual of this system. 

The BMW scaling involves two ingredients. Firstly, the velocities and 
the temperatures of the fluid are taken to be slowly varying functions
of a specific kind, with spatial and temporal gradients having different
scaling dimensions (heuristically $\partial_t \sim \partial_x^2$). 
Secondly, the amplitude of the spatial velocity and the temperature
fluctuation (about some constant equilibrium value) are also taken to 
be small and are of the same order as $\partial_x$ and $\partial_t$ 
respectively. It is convenient to introduce a large parameter $\aleph$
(which is inverse of the small parameter $\epsilon$ in 
\cite{Bhattacharyya:2008kq})  in terms of which the
above statements can be written as  $\partial_t\sim \aleph^{-1}$,
$\partial_x\sim \aleph^{-2}$ etc. We will denote the 
corresponding parameter for the hypersurface fluid by the
hatted $\hat{\aleph}$.

Under the large $\aleph$ limit it is possible to show that 
the relativistic conservation equations map straightforwardly into
the incompressible Navier-Stokes equations. We review this scaling
in \App{A:bmw} for the convenience of the reader and proceed
in the main text to directly implement an analogous scaling 
on the hypersurface $\Sigma_D$ using a large parameter  $\hat{\aleph}$. 

\subsection{Non-relativistic fluids on the Dirichlet hypersurface}
\label{s:nrhyp1}

To keep the computation sufficiently general we will take the metric on the 
Dirichlet surface to have non-vanishing curvature. Furthermore, it is useful
 as in \cite{Bhattacharyya:2008kq} to allow for the background metric to be 
decomposed to slowly varying parts of different orders so as to recover 
non-relativistic fluids which are forced on $\Sigma_D$. One reason 
for doing so is that we are going to obtain  a boundary metric, via the 
map $\varphi_D$ described in \sec{s:dfgsol}, which naturally contains such terms.
Hence it pays to be more general to see the mixing of various contributions at the boundary.

To start off let us consider on the hypersurface a metric $\hat{g}_{\mu\nu}$ of the form:
\begin{equation}
\hat{g}_{\mu\nu} = \hat{g}^{(0)}_{\mu\nu} + \hat{h}_{\mu\nu} 
\label{Dgans}
\end{equation}	
with
\begin{equation}
\hat{g}^{(0)}_{\mu\nu} = -dt^2 + \hat{g}^{(0)}_{ij}(x) \, dx^i \, dx^j
\label{Dg0}
\end{equation}	
where $\hat{g}^{(0)}_{ij}$ are slowly varying functions of $x^i$ and %
with $\hat{h}_{\mu\nu}$ are the metric perturbations which we take it as 
\begin{equation}
\hat{h}_{\mu\nu} \,dx^\mu\,dx^\nu= 2\, \hat{\aleph}^{-1}\, \hat{k}^*_{i}\,   dt\, dx^i +  \hat{\aleph}^{-2}\, \left(\hat{h}^*_{tt}\, dt^2 + \hat{h}^*_{ij} \, dx^i \, dx^j \right)
\label{Dhans}
\end{equation}	
To keep things simple it turns out to be useful to work with the background spatial metric $g^{(0)}_{ij}$ being Ricci flat i.e., $R^{(0)}_{ij} =0$. This turns out to simplify the analysis considerably for a host of terms dependent on the curvature of $g^{(0)}_{ij}$ drop out -- a more comprehensive analysis for general backgrounds is presented in \App{s:bmwR}. We indicate the various corrections that arise from the curvature terms  at appropriate stages in the main text.

All the functions which have a $*$ subscript or superscript (which we freely interchange to keep formulae clear) are of a specific functional form with anisotropic scaling of their spatial and temporal gradients.
\begin{equation}
\hat{{\cal Y}}_*(t,x^i) : {\mathbb R}^{d-1,1} \mapsto {\mathbb R}\ ,  \;\; \text{such that}
\;\; \{ \partial_t \hat{{\cal Y}}_*(t,x^i), \hat{\nabla}^{(0)}_i \hat{{\cal Y}}_*(t,x^i)\}  \sim \{{\cal O}(\hat{\aleph}^{-2}) ,{\cal O}(\hat{\aleph}^{-1})\} 
\end{equation}	
where $\hat{\aleph}$ is a counting parameter introduced to implement the BMW scaling (on the boundary $\aleph^{-1} = \epsilon_\text{BMW}$ as discussed in \App{A:bmw}).

Following \cite{Bhattacharyya:2008kq}, we parameterize the velocity field as
\begin{equation}
\hat{u}^{\mu} = \hat{u}^t \left(1, \hat{\aleph}^{-1}\, \hat{v}_*^i \right) 
\end{equation}	
where the function $\hat{u}^t$ is determined by requiring that $\hat{g}_{\mu\nu}\hat{u}^{\mu}\hat{u}^{\nu}=-1$.
This gives the full velocity field in a large $\hat{\aleph}$ expansion as \footnote{Note that most of 
these expressions are readily obtained by just `hatting' the formulae in \App{A:bmw} .}
\begin{equation}
\begin{split}
\hat{u}^t &=1 + \frac{\hat{\aleph}^{-2}}{2} \left( \hat{h}^*_{tt} + 2 \,\hat{k}^*_{j}\, \hat{v}^{j}_{*} + \hat{g}^{(0)}_{jk} \, \hat{v}^{j}_{*}\,  \hat{v}^{k}_{*} \right)+ {\cal O}(\hat{\aleph}^{-4})\\
\hat{u}^i  &= \hat{\aleph}^{-1} \, \hat{v}_*^i + \frac{\hat{\aleph}^{-3}}{2} \left( \hat{h}^*_{tt} + 2 \,\hat{k}^*_{j}\, \hat{v}^{j}_{*} + \hat{g}^{(0)}_{jk} \, \hat{v}^{j}_{*}\,  \hat{v}^{k}_{*} \right)\, \hat{v}_*^i + {\cal O}(\hat{\aleph}^{-4})\\
{\hat{u}}_t &= -1 - \frac{1}{2}\,  {\hat{\aleph}}^{-2} \, \left(- {h}^*_{tt} +  {\hat{g}}^{(0)}_{jk} \, {\hat{v}}^j_* \,  {\hat{v}}^k_* \right)+ {\cal O}(\hat{\aleph}^{-4})\\
{\hat{u}}_i &=  {\hat{\aleph}}^{-1}\, \left( {\hat{v}}^*_i +  {k}^*_i \right)+ \hat{\aleph}^{-3}\left[{h}^*_{ij}\hat{v}^{j}_{*}+ \frac{1}{2} \left( \hat{h}^*_{tt} + 2 \,\hat{k}^*_{j}\, \hat{v}^{j}_{*} + \hat{g}^{(0)}_{jk} \, \hat{v}^{j}_{*}\,  \hat{v}^{k}_{*} \right)\, \left( {\hat{v}}^*_i +  {k}^*_i \right) \right] + {\cal O}(\hat{\aleph}^{-4})
\end{split}
\end{equation}	
and the velocity gradients are given by
\begin{eqnarray}
\hat{\theta} &=& {\cal O}(\hat{\aleph}^{-4}) \nonumber \\
\hat{\mathcal{A}}_{\mu} dx^{\mu} &=& \hat{a}_{\mu} dx^{\mu}= \hat{\aleph}^{-3} \left[ \partial_{t} \hat{v}_{i}^{*} 
+ \hat{v}_{*}^{j} \hat{\nabla}^{(0)}_{j} \hat{v}_{i}^{*} - \hat{f}_i^*  \right] dx^{i}  + {\cal O}(\hat{\aleph}^{-4})  \nonumber \\
\hat{\sigma}_{\mu \nu} dx^{\mu} dx^{\nu} &=&\hat{\aleph}^{-2}\, \hat{\nabla}^{(0)}_{(i} \hat{v}^{*}_{j)} \,dx^{i} dx^{j} 
-  2\hat{\aleph}^{-3} \, \hat{v}^{j}_{*} \,\hat{\nabla}^{(0)}_{(i} \hat{v}^{*}_{j)} \,dx^{i} dt + {\cal O}(\hat{\aleph}^{-4})  
\label{Duder}	
  \end{eqnarray}
where $\hat{\nabla}^{(0)}_{\mu}$ is the covariant derivative compatible with $\hat{g}^{(0)}(x^{i})$ and we 
have freely raised and lowered the spatial indices with $\hat{g}^{(0)}_{ij}$ for brevity. 
Further, $\hat{f}^i$ is a forcing function determined as a functional of $\hat{h}_{\mu\nu}$ data
\begin{equation}
\hat{f}_{i} =  \frac{1}{2} \partial_i  \hat{h}^*_{tt}- \partial_t \hat{k}^*_i + \hat{q}^{*}_{\ ij} \, \hat{v}_*^j \,.
\label{hypforce}
\end{equation}	
and $\hat{q}^*_{ij} =  \hat{\nabla}^{(0)}_i \hat{k}^*_j -\hat{\nabla}^{(0)}_j \hat{k}^*_i$ . In deriving these expressions we have used the fact that to leading order in the $\aleph \to \infty$ expansion, the velocity field $v^i_*$ is divergenceless (see below). 
 
We take the scaling in $b$ to be of the form
\begin{equation}
b = b_0 + \hat{\aleph}^{-2}\, \delta b_* \ .
\label{Dbexp}
\end{equation}	
Using \req{alhatdef}
\begin{eqnarray}
\hat{\alpha} &=& \hat{\alpha}_0 + \hat{\aleph}^{-2} \, \left(\frac{d}{2}\,  \hat{\alpha}_0 \left( 1 - \hat{\alpha}_0^2 \right) \,\frac{\delta b_*}{b_0} \right)
+{\cal O}(\hat{\aleph}^{-4})  \ , \qquad   \hat{\alpha}_0 \equiv \frac{1}{\sqrt{f \left(b_0\, r_{D}\right)}}
\end{eqnarray}
we can evaluate the non-relativistic pressure per mass density and kinematic viscosity:
 \begin{eqnarray}
\hat{p}_{*} &=&\frac{\delta \hat{p}}{\hat{\varepsilon}_0+\hat{p}_0} =  - \hat{\aleph}^{-2} \, \left(1 + \frac{d}{2} \left(\hat{\alpha}^2_{(0)} -1\right) \right)  \frac{\delta b_*}{b_0} +{\cal O}(\hat{\aleph}^{-4}) \nonumber \\
\hat{ \nu}_{0} &=& \frac{\eta_0}{\hat{\varepsilon}_0+ \hat{p}_0}=  \frac{b_0}{d \, \hat{\alpha}_0}+ {\cal O}(\hat{\aleph}^{-2}) 
\label{bmwpnu}	   
\end{eqnarray}
with $\hat{\rho}_0\equiv \hat{\varepsilon}_0+ \hat{p}_0 = \frac{d \,\hat{\alpha}_0}{16 \pi G_{d+1}\, b_0^d}$ playing the role of the non-relativistic mass density.

Given these data\footnotemark\ we can show that the  conservation equations \req{hypcons}  reduce to the incompressible Navier-Stokes equations:\footnotemark
\begin{eqnarray}
&& \hat{\nabla}^{(0)}_i \, v^i_* = 0  \nonumber \\
&& \hat{\nabla}^{(0)}_i \hat{p}_* + \partial_t \hat{v}^*_i + \hat{v}_*^j \, \hat{\nabla}^{(0)}_j \hat{v}^*_i - 2\, \hat{\nu}_0\, \hat{\nabla}^{(0)^j} \left(\hat{\nabla}^{(0)}_{(i} \hat{v}^{*}_{j)}\right) = \hat{f}_i
\label{hypns}
\end{eqnarray}	

\footnotetext{Note that BMW limit does not involve any scaling of the metric on $\Sigma_D$ which is taken to be a fixed Lorentzian  structure. Only the hydrodynamic fields (velocity, pressure etc.) are rescaled and their gradients are constrained to scale in  a specific way. Thus the dressed fluid  while non-relativistic still lives on a Lorentzian geometry. Further in this limit, the fluid dynamics has  certain Galilean symmetries which are actually enlarged into a enhanced symmetry algebra \cite{Gusyatnikova:1989nx,Bhattacharyya:2008kq,Bagchi:2009my}.}

\footnotetext{If the spatial geometry on which the non-relativistic fluid moves is not Ricci-flat then there is an additional term in Naiver-Stokes equations due to the background curvature, see \eqref{nsbdy1}.}

We now proceed to derive the expressions entering into the bulk metric 
and the map $\varphi_D$ given in \sec{s:dfgsol}.

\subsection{Bulk metric in terms of Dirichlet data}
\label{s:nrhyp2}

Armed with the results from \sec{s:nrhyp1} we can proceed to use those of \sec{s:dbulk}
to construct the bulk metric corresponding to the non-relativistic fluid on the Dirichlet
hypersurface $\Sigma_D$. In principle Einstein equations need to be solved in the 
new gradient expansion to obtain the non-relativistic solutions. As argued by the 
authors of \cite{Bhattacharyya:2008kq}, this can be done via an algorithm very similar
to the algorithm used to find the metric dual of the relativistic fluid. The main
difference is the anisotropic scaling of space with respect to time and the fact that
the bulk metric is no-more ultra-local in space but is still ultra-local in time.
We can again proceed from the space of  non-relativistic solutions with asymptotic
boundary conditions that we present in \App{A:bmw} and reparametrize it in terms
of Dirichlet data.

But we will take here instead an easier route and directly derive it from the 
the bulk relativistic metric \req{dirbulk1} written in terms 
of Dirichlet data. We should be careful though -- given the difference in 
derivative counting between the relativistic scaling and the non-relativistic
scaling, it is in principle possible that a higher order term according to the
relativistic counting contributes at a lower order according to the 
non-relativistic counting. In order to obtain the metric accurate to the order where the Navier-Stokes equations can be seen, ${\cal O}(\hat{\aleph}^{-3})$, we need to have the certain terms in the relativistic metric accurate to third order in gradients.\footnote{This point was initially missed by both our analysis and that of \cite{Bhattacharyya:2008kq}. It was originally thought that it would be sufficient to obtain the non-relativistic metric from just the first order relativistic metric. This unfortunately is not true and as a result the non-relativistic metrics quoted in v1 of this paper and in \cite{Bhattacharyya:2008kq} only solve Einstein's equations to first order in the non-relativistic gradient expansion. The correct form of the general expressions are now collected in \App{s:bmwR}.} 

 If we however restrict to the case of Ricci flat spatial metric on the hypersurface $\Sigma_D$, then we can obtain the non-relativistic metric from the second order relativistic fluid/gravity metric obtained in \cite{Bhattacharyya:2008mz}. There are three terms we need to account for which give rise to non-relativistic contributions proportional to $\nabla^{(0)}_j\nabla^{(0) j} \hat{v}^*_i \equiv \nabla_{(0)}^2 \hat{v}^*_i$ and $ \nabla^{(0)}_j \hat{q}_*^{j}{}_i$. These involve new radial functions which we collect below after presenting the bulk metrics highlighting the terms that were missed in the original analyses. General expressions including  spatial curvatures can be found in  \App{s:bmwR}. 

Since in the scaling limit $\hat{\aleph} \gg 1$ one has from \req{Duder} that the $\hat{a}_\mu = \hat{\mathcal{A}}_\mu$ to leading order things simplify considerably. Using the formulae in the last subsection and including the terms from the second order metric\footnote{Note that the highlighted terms involve Weyl invariant Ricci  ${\hat{\mathcal R}}_{\mu\nu}$  and Schouten ${\hat{\mathcal S}}_{\mu\nu}$ curvature tensors which are defined in \App{A:bmw}.} (highlighted) it is  easy to show that the bulk metric \req{dirbulk1} becomes\footnote{Note that we have retained certain terms at ${\cal O}(\hat{\aleph}^{-3})$ which are actually not necessary to solve the equation of motion \req{eins} at this order (eg., the velocity cubed term). This is to facilitate ease of comparison of our results when  we undertake the near-horizon analysis in \sec{s:nh}, with those in the existing literature.} 

\begin{equation}
\label{hypbmwf1}
\begin{split}
ds^2 &= 
	-2 \hat{\alpha}\, \hat{u}_\mu \,dx^\mu dr 
+\frac{2\hat{\alpha}^2}{r_D}\left[\frac{\hat{a}_\mu}{\left[1+\frac{d}{2}(\hat{\alpha}^2-1)\right]}\right]dx^\mu dr 	-2\,r\,\hat{\alpha}\, (2\, \hat{\xi}-1) 
	\, \frac{\hat{u}_{(\mu}\,\hat{a}_{\nu)} }{1+\frac{d}{2}\, (\hat{\alpha}^2 -1)} 
\, dx^\mu \,dx^\nu\\
&\qquad 
	+\; r^2\left[\hat{g}_{\mu\nu} + \left(1-\hat{\alpha}^2\, f(br) \right) 
	\hat{u}_\mu \,\hat{u}_\nu + 2b\,\hat{F}(br)\, \hat{\sigma}_{\mu\nu}\right]
	 dx^\mu \,dx^\nu  \\
&\qquad 
	\red{- 4b^2\kappa_L\hat{\alpha}\hat{P}_{\mu}^{\lambda}\hat{\mathcal{D}}_{\alpha}
	{\hat{\sigma}^{\alpha}}_{\lambda}\,dx^\mu\,dr}\\
&\qquad 
	 \red{-2\,\frac{\hat{\alpha}^3}{r_D^2}\hat{\mathcal{S}}_{\mu\lambda}\hat{u}^\lambda\,dx^\mu\,dr+\,2
	 \frac{\hat{\alpha}^3} 
	 {r_D^2(d-2)}\left[1+\frac{2}{d\hat{\alpha}(\hat{\alpha}+1)}\right]
	 \hat{\mathcal{R}}_{\mu\lambda}\hat{u}^\lambda\,dx^\mu\,dr}\\
	&\qquad 
	\red{+\; 2\,(br)^2\left[\hat{M}_1(br)\, \hat{u}_{(\mu}\hat{\mathcal{S}}_{\nu)\lambda}\hat{u}^\lambda -
	\hat{M}_2(br)\, \hat{u}_{(\mu}\hat{\mathcal{R}}_{\nu)\lambda}\hat{u}^\lambda +2\, \hat{L}_1(br)\,
	\hat{u}_{(\mu}\hat{P}_{\nu)}^{\lambda}\hat{\mathcal{D}}_{\alpha}{\hat{\sigma}^{\alpha}}_{\lambda} 
	\right]dx^\mu dx^\nu} \\
&= ds_0^2 + \hat{\aleph}^{-1} ds_1^2 + \hat{\aleph}^{-2} ds_2^2 + \hat{\aleph}^{-3} ds_3^2 + {\cal O}(\hat{\aleph}^{-4}) \\
\end{split}
\end{equation}
with
\begin{equation}
\label{hypbmwf1a}
\begin{split}
ds_0^2 &= 
	2\,\hat{\alpha}_0\ dt\ dr + r^2\left(-\hat{\alpha}_0^2 \,f_0 \,dt^2 
	+ \hat{g}^{(0)}_{ij}\,dx^i dx^j\right)\\
ds_1^2 &= 
	-2\, \hat{\alpha}_0\left( \hat{v}^*_i +  \hat{k}^*_i \right)\ dx^i\ dr 
	+ 2\, r^2 \left[\hat{k}^*_i -\left(1-\hat{\alpha}_0^2\, f_0\right)
	\left( \hat{v}^*_i +  \hat{k}^*_i \right)\right] dx^i\, dt \\
ds_2^2 &= 
	2\, \hat{\alpha}_0 \left[- \frac{1}{2}\hat{h}^*_{tt} 
	+ \frac{1}{2}\, \hat{g}^{(0)}_{jk} \, \hat{v}^j_* \,  \hat{v}^k_* 
	+\hat{p}_* \frac{\frac{d}{2}\, (\hat{\alpha}_0^2 -1)}{1+\frac{d}{2}\, (\hat{\alpha}_0^2 -1)} 	\right]dt\ dr 
	+ r^2\left[\hat{h}^*_{tt}\, dt^2 + \hat{h}^*_{ij} \, dx^i \, dx^j \right]\\
&\quad 
	+r^2 \left(1-\hat{\alpha}_0^2\, f_0\right) \left[\left(- \hat{h}^*_{tt} 
	+ \hat{g}^{(0)}_{jk} \, \hat{v}^j_* \,  \hat{v}^k_*
	+\hat{p}_* \frac{d\hat{\alpha}_0^2 }{1+\frac{d}{2}\, (\hat{\alpha}_0^2 -1)}
	 \right)dt^2 \right.\\
&\qquad \left. \qquad 
	+ \left( \hat{v}^*_i +  \hat{k}^*_i \right) \left( \hat{v}^*_j 
	+  \hat{k}^*_j \right)dx^i dx^j \right]
	+2\,r^2\,b_0\, \hat{F}_0\,\hat{\nabla}^{(0)}_{(i} \hat{v}^{*}_{j)} \,dx^{i} dx^{j}\\
ds_3^2 &= 
	-2\,\hat{\alpha}_0\left[\hat{h}^*_{ij}\hat{v}^{j}_{*}+ 
	 \left( \frac{1}{2}\hat{h}^*_{tt} +  \,\hat{k}^*_{j}\, \hat{v}^{j}_{*} 
	 +\frac{1}{2} \hat{g}^{(0)}_{jk} \, \hat{v}^{j}_{*}\,  \hat{v}^{k}_{*} 
	 +\hat{p}_* \frac{\frac{d}{2}\, (\hat{\alpha}_0^2 -1)}{1+\frac{d}{2}\, (\hat{\alpha}_0^2 -1)}	\right)\, \left( \hat{v}^*_i +  \hat{k}^*_i \right) \right] dx^i dr \\
&\quad  
         +\frac{2\hat{\alpha}_0^2}{r_D\left(1+\frac{d}{2}(\hat{\alpha}_0^2-1)\right)}\left[ \partial_{t} \hat{v}_{i}^{*}+\hat{v}_{*}^{j}  \hat{\nabla}^{(0)}_{j}  \hat{v}_{i}^{*} 
	-\hat{f}_i^* \right] dx^{i}dr\\
&\quad 
	+ 2r\,\frac{\hat{\alpha}_0(2\,\hat{\xi}_0-1)}{1+\frac{d}{2}\, (\hat{\alpha}_0^2 -1)}
	\left[ \partial_{t} \hat{v}_{i}^{*}+\hat{v}_{*}^{j}  \hat{\nabla}^{(0)}_{j}  \hat{v}_{i}^{*} 
	-\hat{f}_i^* \right] dx^{i}dt\\
&\quad 
	-2\,r^2 \left(1-\hat{\alpha}_0^2 \,f_0\right)\left[\hat{h}^*_{ij}\hat{v}^{j}_{*}
	+ \left(\hat{k}^*_{j}\, \hat{v}^{j}_{*} + \hat{g}^{(0)}_{jk} \, \hat{v}^{j}_{*}\,  \hat{v}^{k}_{*}
	+\hat{p}_* \frac{d\hat{\alpha}_0^2 }{1+\frac{d}{2}\, (\hat{\alpha}_0^2 -1)} \right)\, 
	\left( \hat{v}^*_i +  \hat{k}^*_i \right) \right] dx^i dt \\
&\qquad 
	-4\, r^2\, b_0\, \hat{F}_0\, \hat{v}_{*}^j\hat{\nabla}^{(0)}_{(i} \hat{v}^{*}_{j)} \,dx^{i} dt \; 
\red{-\;2\, b_0^2\, r^2\, \hat{L}_1	\nabla^2_{(0)} v_i^* \, dt \, dx^i 
-2\,b_0^2\, \hat{\kappa}_L\,\hat{\alpha}_0\,  \hat{\nabla}^2_{(0)}v^*_i\, dx^i\, dr} \\
&\qquad
\red{+ \;\hat{M}_0 \,\hat{\nabla}^j_{(0)}\hat{q}^*_{ij}\, dx^i \, dt+ 2\, \frac{\hat{\alpha}_0^2}{r_D^2} \,  \frac{\hat{\nabla}^j_{(0)}\hat{q}^*_{ij}}{d\,(d-2)\, (\hat{\alpha}_0 +1) } \,
 dx^i\, dr }
\end{split}
\end{equation}
where 
\begin{equation}
\begin{split}
\hat{\alpha}_0 &\equiv (1-(b_0 r_d)^{-d})^{-1/2}\ ,\quad f_0 \equiv 1-(b_0 r)^{-d}\ ,\\
\hat{\xi}_0 &\equiv 1- \frac{r}{2r_D}\hat{\alpha}_0^2\, f_0\\
\hat{p}_* &\equiv -\frac{\delta b_*}{b_0}\left\{1+\frac{d}{2}\, (\hat{\alpha}^2 -1)\right\} \\
\hat{F}_0 &\equiv \frac{1}{\hat{\alpha}_0} \int_{b_0 r}^{b_0 r_D}\frac{y^{d-1}-1}{y(y^{d}-1)}dy\\ 
\hat{f}_i^* &\equiv  \frac{1}{2}\partial_i \hat{h}^*_{tt} - \partial_t \hat{k}^*_i +\left[\hat{\nabla}^{(0)}_i \hat{k}^*_j - \hat{\nabla}^{(0)}_j \hat{k}^*_i\right] {v}_*^j = \frac{1}{2}\partial_i\hat{h}^*_{tt} - \partial_t\hat{k}^*_i +\hat{q}^*_{ij} {v}_*^j \\
\hat{L}_1 &\equiv \frac{L(br)}{(br)^d}-\frac{L(br_D)}{(br_D)^d}+\hat{\kappa}_L \left[1-\hat{\alpha}^2\, f_0\right]\\
\hat{\kappa}_L &\equiv \frac{1}{d}\left[
\xi(\xi^d-1)\frac{d}{d\xi}\left[\xi^{-d}L(\xi)\right]+
\frac{1}{\xi\left[1+\frac{d}{2}(\hat{\alpha}^2-1)\right]}+\frac{1}{\xi^2(d-2)}\right]_{\xi=br_D} \\
\hat{M}_0 &\equiv \frac{1}{(d-2)}\left[-\hat{\alpha}_0^2\left(1-\frac{r^2}{r_D^2}\right)+\frac{2}{d}\frac{\hat{\alpha}_0}{1+\hat{\alpha}_0}\left(1-\hat{\alpha}_0^2 f_0\right)\frac{r^2}{r_D^2}\right]
\end{split}
\end{equation}
The function $L(x)$ which enters into the above formulae is given as 
\begin{equation}
L(br) = \int_{br}^\infty\xi^{d-1}d\xi\int_{\xi}^\infty dy\ \frac{y-1}{y^3(y^d
-1)} 
\label{}
\end{equation}	
while the other functions $\hat{M}_1$ and $\hat{M}_2$ can be found in \eqref{allhatfunx}. 

 In the final results we have highlighted the terms appearing in $ds_3^2$ that were missed in the first version of the paper and are necessary in order to solve Einstein's equations \req{eins} to ${\cal O}(\hat{\aleph}^{-3})$. Having realized the necessity of these terms it 
 a-posteriori became clear that the BMW scaling metric introduced in \cite{Bhattacharyya:2008kq} to describe the bulk dual of  boundary non-relativistic fluids  also receives corrections. The corrected form of this metric including effects of boundary spatial curvature is now presented in \App{A:bmw}. The general expressions for the Dirichlet non-relativistic fluid are collected in \App{s:bmwR} as they involve many more terms compared to what was originally reported in v1 of this paper. 

As required this metric reduces to the metric derived by the BMW scaling on the boundary \cite{Bhattacharyya:2008kq} that we 
review in \App{A:bmw} in our conventions; we simply set $r_D=\infty,\hat{\alpha}_0 = 1$ in the above metric. As remarked above, it compares with the metric given in \cite{Bhattacharyya:2008kq} up to the new terms mentioned above.

Inspired by the gravity duals of fluids on cut-off hypersurfaces in flat space \cite{Bredberg:2011jq,Compere:2011dx}, in \cite{Cai:2011xv} the result for the bulk dual to a non-relativistic fluid on a Dirichlet hypersurface in \AdS{} has been recently derived. The results there are presented for a fluid on a flat background and are entirely contained within our framework. 
To facilitate ease of comparison with their results we now specify our computation to the case where the metric on $\Sigma_D$ is flat and furthermore switch off the forcing. Taking $\hat{g}_{\mu\nu} = \eta_{\mu\nu}$ which amounts to setting  $\hat{g}^{(0)}_{ij} = \delta_{ij}$ and $\hat{k}_i^* = \hat{h}_{tt}^* = \hat{h}^*_{ij} =0$ in the above one can show that the bulk metric reduces to the form \req{hypbmwf1} with
\begin{equation}\label{hypbmwf}
\begin{split}
ds_0^2 &= 
	2\,\hat{\alpha}_0\ dt\ dr + r^2\left(-\hat{\alpha}_0^2\, f_0\, dt^2 
	+ \delta_{ij}dx^i dx^j\right)\\
ds_1^2 &= 
	-2\, \hat{\alpha}_0\, \hat{v}^*_i \ dx^i\ dr 
	- 2\, r^2 \,\left(1-\hat{\alpha}_0^2 \,f_0\right)\hat{v}^*_i dx^i dt \\
ds_2^2 &=
	 2 \,\hat{\alpha}_0 \left[ \frac{1}{2}  \, \hat{v}^2_* 
	 +\hat{p}_* \frac{\frac{d}{2}\, (\hat{\alpha}_0^2 -1)}{1+\frac{d}{2}\, 
	 (\hat{\alpha}_0^2 -1)} \right]dt\ dr \\
&\quad 
	+r^2 \left(1-\hat{\alpha}_0^2 \,f_0\right) \left[\left( \hat{v}^2_*
	+\hat{p}_* \frac{d\hat{\alpha}_0^2 }{1+\frac{d}{2}\, (\hat{\alpha}_0^2 -1)} \right)dt^2
	+  \hat{v}^*_i \hat{v}^*_j dx^i dx^j \right]\\
&\quad
	+2\,r^2\,b_0\,\hat{F}_0\,\hat{\nabla}^{(0)}_{(i} \hat{v}^{*}_{j)} \,dx^{i} dx^{j}\\
ds_3^2 &= 
	-2\,\hat{\alpha}_0\, \left( \frac{1}{2} \hat{v}^{2}_{*} 
	+\hat{p}_* \frac{\frac{d}{2}\, (\hat{\alpha}_0^2 -1)}{1+\frac{d}{2}\,
	 (\hat{\alpha}_0^2 -1)}\right) \hat{v}^*_i\, dx^i\, dr 
         +\frac{2\hat{\alpha}_0^2}{r_D\left(1+\frac{d}{2}(\hat{\alpha}_0^2-1)\right)}\left[ \partial_{t} \hat{v}_{i}^{*}+\hat{v}_{*}^{j}  \partial_j  \hat{v}_{i}^{*} 
	\right] dx^{i}dr\\
&\quad	 + 2\,r\,\frac{\hat{\alpha}_0(2\hat{\xi}_0-1)}{1+\frac{d}{2}\, (\hat{\alpha}_0^2 -1)}
	\left[ \partial_{t} \hat{v}_{i}^{*}+\hat{v}_{*}^{j}  \partial_{j}  \hat{v}_{i}^{*} \right] dx^{i}\,dt\\
&\quad 
	-2\,r^2 \left(1-\hat{\alpha}_0^2\, f_0\right) \left(\hat{v}^{2}_{*}\ 
	+\hat{p}_* \frac{d\hat{\alpha}^2 }{1+\frac{d}{2}\, (\hat{\alpha}^2 -1)} \right)
	\hat{v}^*_i \,dx^i \,dt -4\,r^2\,b_0\, \hat{F}_0\,\hat{v}_{*}^j\hat{\nabla}^{(0)}_{(i} \hat{v}^{*}_{j)} \,dx^{i} dt \\
&\quad 
\red{-\;2\, b_0^2\, r^2\, \hat{L}_1	\nabla^2_{(0)} v_i^* \, dt \, dx^i 
-2\,b_0^2\, \hat{\kappa}_L\,\hat{\alpha}_0\,  \hat{\nabla}^2_{(0)}v^*_i\, dx^i\, dr}  \,,
\end{split}
\end{equation}
which agrees with that derived in \cite{Cai:2011xv} once one accounts for some differences in convention. In particular,  one has to rescale the time coordinate to absorb the factor of $\hat{\alpha}_0$,  in addition to  redefining $dt \rightarrow \hat{\alpha}_0^{-1}\, dt$ along with $\hat{v}^*_i \rightarrow \frac{\hat{\alpha}_0}{r_D^2} \, \hat{v}^*_i$ and $\hat{v}_*^i \rightarrow \hat{\alpha}_0\, \hat{v}^*_i$. The inhomogeneity in the scaling of the spatial velocities results from the fact that we define our hypersurface metric to be the induced metric rescaled by a factor of $r_D^{-2}$, while \cite{Cai:2011xv} works with the induced metric.  Also, our derivation here allows us to go to ${\cal O}(\hat{\aleph}^{-3})$ which is necessary in order to see the dynamical Navier-Stokes equation on the hypersurface $\Sigma_D$.

\subsection{The boundary data for the non-relativistic Dirichlet fluid}
\label{s:dbdymet}

Apart from the construction of the bulk dual to the non-relativistic fluid on the Dirichlet surface, we would also like to know what the corresponding physics on the boundary is. For instance we see that the boundary velocity field can
be read off from the terms with $dr$ in \req{hypbmwf1} and  is given as 
\begin{equation}\begin{split}
u_t  &=  -\hat{\alpha}_0 - \hat{\aleph}^{-2} \hat{\alpha}_0 \left[- \frac{1}{2}\hat{h}^*_{tt} + \frac{1}{2} \hat{g}^{(0)}_{jk} \, \hat{v}^j_* \,  \hat{v}^k_* +\hat{p}_* \frac{\frac{d}{2}\, (\hat{\alpha}_0^2 -1)}{1+\frac{d}{2}\, (\hat{\alpha}_0^2 -1)} \right]  + {\cal O}(\hat{\aleph}^{-4})  \\
u_i &= \hat{\aleph}^{-1}\, \hat{\alpha}_0\, \left(\hat{v}^*_i + \hat{k}^*_i \right)\\
&\quad +\hat{\aleph}^{-3}\,
\hat{\alpha}_0\left[\hat{h}^*_{ij}\hat{v}^{j}_{*}+  \left( \frac{1}{2}\hat{h}^*_{tt} +  \,\hat{k}^*_{j}\, \hat{v}^{j}_{*} +\frac{1}{2} \hat{g}^{(0)}_{jk} \, \hat{v}^{j}_{*}\,  \hat{v}^{k}_{*} +\hat{p}_* \frac{\frac{d}{2}\, (\hat{\alpha}_0^2 -1)}{1+\frac{d}{2}\, (\hat{\alpha}_0^2 -1)}\right)\, \left( \hat{v}^*_i +  \hat{k}^*_i \right) \right]\\
&\qquad -\hat{\aleph}^{-3}\frac{\hat{\alpha}_0^2}{r_D\left(1+\frac{d}{2}(\hat{\alpha}_0^2-1)\right)}\left[ \partial_{t} \hat{v}_{i}^{*}+\hat{v}_{*}^{j}  \hat{\nabla}^{(0)}_{j}  \hat{v}_{i}^{*}-\hat{f}_i^* \right] \\
& \qquad 
+\hat{\aleph}^{-3}\left[ b_0^2\, \hat{\kappa}_L\,\hat{\alpha}_0\,  \hat{\nabla}^2_{(0)} \hat{v}^*_i  -\frac{\hat{\alpha}_0^2}{r_D^2} \, \frac{\hat{\nabla}^j_{(0)}\hat{q}^*_{ij}}{d\,(d-2)\, (\hat{\alpha}_0 +1) }\right] +  {\cal O}(\hat{\aleph}^{-4}) 
\end{split}\end{equation}	
while the boundary metric can be read off from the large $r$ behavior and is given to be 
\begin{equation}\begin{split}
g_{tt} &= -\hat{\alpha}_0^2 + \hat{\aleph}^{-2} \, \left[\hat{h}^*_{tt}\ +
\left(1-\hat{\alpha}_0^2 \right) \left(- \hat{h}^*_{tt} + \hat{g}^{(0)}_{jk} \, \hat{v}^j_* \,  \hat{v}^k_*+\hat{p}_* \frac{d\hat{\alpha}_0^2 }{1+\frac{d}{2}\, (\hat{\alpha}_0^2 -1)} \right) \right] + {\cal O}(\hat{\aleph}^{-4}) \\
g_{ti} &= \hat{\aleph}^{-1}\left( \hat{k}^*_i + (\hat{\alpha}_0^2-1)\, (\hat{k}_i^* + \hat{v}_i^*) \right) 
- \hat{\aleph}^{-3}\frac{\hat{\alpha}_0^3}{r_D\left(1+\frac{d}{2}\, (\hat{\alpha}_0^2 -1)\right)}
\left[ \partial_{t} \hat{v}_{i}^{*}+\hat{v}_{*}^{j}  \hat{\nabla}^{(0)}_{j}  \hat{v}_{i}^{*} -\hat{f}_i^* \right] \\
&\quad -\hat{\aleph}^{-3} \left(1-\hat{\alpha}_0^2 \right)\left[\hat{h}^*_{ij}\hat{v}^{j}_{*}+ \left(\hat{k}^*_{j}\, \hat{v}^{j}_{*} + \hat{g}^{(0)}_{jk} \, \hat{v}^{j}_{*}\,  \hat{v}^{k}_{*}+\hat{p}_* \frac{d\hat{\alpha}_0^2 }{1+\frac{d}{2}\, (\hat{\alpha}_0^2 -1)} \right)\, \left( \hat{v}^*_i +  \hat{k}^*_i \right) \right]\\
&\quad +\;\hat{\aleph}^{-3} \left[\frac{2 \, b_0}{\hat{\alpha}_0}\, F(b_0\,r_D)\hat{v}_{*}^j\hat{\nabla}^{(0)}_{(i} \hat{v}^{*}_{j)} + \frac{\hat{\alpha}_0^4}{2\,(d-2)\, r_D^2)}\hat{\nabla}^j_{(0)}\hat{q}^*_{ij} -
 \frac{\hat{\alpha}_0^2\,(\hat{\alpha}_0^2-1)}{2\, (d-2)\, r_D^2}\left(1+\frac{2}{d\,\hat{\alpha}_0\,(\hat{\alpha}_0+1)}\right)  \hat{\nabla}^j_{(0)}\hat{q}^*_{ij}\right] \\
 &\quad
+\;\hat{\aleph}^{-3} \, \left[
-\, b_0^2\, \hat{L}_1(\infty) \, \hat{\nabla}^2_{(0)}\hat{v}^*_{i} \right] + \;{\cal O}(\hat{\aleph}^{-4}) \\
g_{ij} &= \hat{g}^{(0)}_{ij} + \hat{\aleph}^{-2} \left(\hat{h}^*_{ij} -(\hat{\alpha}_0^2 -1)\, (\hat{v}^*_i + \hat{k}^*_i) \,(\hat{v}^*_j + \hat{k}^*_j) - \frac{2 \, b_0}{\hat{\alpha}_0}\, F(b_0\,r_D)  \,\hat{\nabla}^{(0)}_{(i} \hat{v}^{*}_{j)} \right) + \;{\cal O}(\hat{\aleph}^{-4})
\end{split}\end{equation}	
with $\hat{L}_1(\infty)$ denoting the asymptotic value of $\hat{L}_1(br)$.

Note that even in the non-relativistic limit the boundary `dynamical' metric $g_{\mu\nu}$'s light-cone is being opened up relative to that of the hypersurface metric. As before this is a consequence of the red-shift effect. We are normalizing here the hypersurface metric to have 
flat Minkowski metric to leading order and this causes a rescaling by an amount $\hat{\alpha}_0^2$ on the boundary. As long as $\hat{\alpha}_0$ is finite this is not an issue for we are just encountering an overall rescaling of the boundary time.

What we have derived here is the AdS analog of the membrane paradigm connection recently proposed in \cite{Bredberg:2011jq,Compere:2011dx}. Recall that the construction described in these papers proceeds by looking at an asymptotically flat geometry with Dirichlet boundary conditions at some timelike hypersurface (the analog of our $\Sigma_D$) and one solves vacuum Einstein's equations. It was shown that the hypersurface dynamics is constrained to obey the incompressible Navier-Stokes equations, just as what we have shown above. However,  the solutions described in this section solve Einstein's equations with a negative cosmological constant and we furthermore have argued that the Dirichlet dynamics is obtained by suitably dressing up of a CFT fluid by allowing it to propagate on a `dynamical' background metric. In our context it is clear that the interpretation of the physics is less in terms of an RG flow, and more along the lines of the medium dependent `dressing up' of the boundary fluid dynamics in contrast  to the  Wilsonian RG perspective put forth as an interpretation of the membrane paradigm originally in \cite{Bredberg:2010ky}.

There is however one regime where our analysis should be able to make some contact with the discussion in \cite{Bredberg:2011jq,Compere:2011dx}; this is the near horizon regime where one expects to encounter a  local Rindler geometry, which is the starting point for analyzing the Dirichlet problem in flat space. In the next section we show how one can embed this construction using the solutions we have described, enabling one thus to explore the AdS version of the membrane paradigm.

\section{The near horizon Dirichlet problem}
\label{s:nh}

So far in our discussion the Dirichlet hypersurface $\Sigma_D$ has been located at some radial position $r_D$ that is finite. We now want to investigate what happens as we push this surface closer towards the horizon, i.e, $\Sigma_D \to {\cal H}^+$ via $r_D \to b^{-1}$. To understand the resulting physics we first have to realize that we are doing something strange:
the horizon is a null surface and has therefore a degenerate metric. $\Sigma_D$ on the other hand is constrained to be a timelike surface with a non-degenerate metric. So it is clear that demanding a well-behaved metric is going to result in an infinite scaling by the red-shift factor $\hat{\alpha}$; the main question is whether one can implement the scaling while  retaining interesting dynamics. We will now proceed to show that there exists a scaling of parameters such that the near-horizon geometry makes sense. Furthermore, we argue that this allows us to embed the flat space constructions of \cite{Bredberg:2011jq} into our AdS set-up.

We outline the construction in a couple of stages to guide intuition: firstly in \sec{s:dhypdyn} we will examine the conservation equation on the hypersurface and  from there infer the scaling of parameters. This is the only sensible thing to do for us, since the entire dynamical system of the boundary CFT fluid has been converted into that of the hypersurface fluid living on an inert background. We then examine the consequences of the scalings we derive in \sec{s:nhdirmet} focussing on the region close to the horizon; this amounts to blowing up the region of spacetime between ${\mathcal H}^+$ and $\Sigma_D$. This blown up region bears close resemblance to the solution of the vacuum Einstein's equations dual to the incompressible Navier-Stokes system discussed in \cite{Bredberg:2011jq}. Indeed one should anticipate this based on the usual intuition that near any non-degenerate horizon one encounters a patch of the Rindler geometry. However, we will also encounter differences owing to the fact at the end of the day we are solving \req{eins} with a non-vanishing cosmological constant.

There is also the further question of what the geometry between the Dirichlet surface and the asymptotic AdS region looks like and moreover what is the boundary physics of our scalings? We argue in \sec{s:nhgal} that the near horizon scaling regime renders the bulk metric viewed as a Lorentzian metric on a spacetime manifold nonsensical. However, it turns out that there is a nice language to describe the geometry in terms Newton-Cartan structures and we show that the bulk co-metric (which in usual Lorentzian geometry is the inverse metric) is well behaved as is the boundary co-metric. The result is quite satisfying from the physical perspective: the near horizon limit demands a drastic modification of the boundary metric, which forces one into a non-relativistic or Galilean regime. Consequently, rather than describing geometry in terms of Lorentzian structure, we are forced to use the less familiar but equally effective geometrization of the idea of a Galilean spacetime, in terms of Newton-Cartan geometry (see  \cite{Misner:1973by} for a nice account of this subject and \cite{Ruede:1996sy} for a more recent review).

\subsection{Scaling of the Dirichlet dynamics in the near horizon region}
\label{s:dhypdyn}

To understand the behavior of the fluid on the Dirichlet surface as we push $\Sigma_D$ closer to the horizon, we first look at the conservation equation in this limit. It turns out that demanding non-trivial dynamics on the Dirichlet surface forces one into a scaling regime of the fluid, effectively making it non-relativistic. However, the scaling we encounter is not quite the BMW scaling \cite{Bhattacharyya:2008kq} discussed earlier in \sec{s:dgravnr} but a slightly modified version of the same. 

\subsubsection{A new scaling regime}
\label{s:nhnewscale}

Ignoring for the moment the fact for $r_D < r_{D,snd}$ we are supposed to be projecting out the sound mode to ensure subluminal propagation on the Dirichlet hypersurface, let us write the relativistic conservation equations and examine them as we zoom in towards the horizon. Consider then the conservation equation on the Dirichlet hypersurface \req{hypcons}; we have analyzed this from a generic viewpoint in \sec{s:conseq} but for now we will focus on the truncated equations to second order by setting $\hat{\pi}^{\mu\nu} = -2\eta\, \hat{\sigma}^{\mu\nu}$ (cf., \req{piTdef}).  Projecting these  conservation equations parallel and transverse to the velocity we get  (using $\eta$ as given in \req{etab})
\begin{eqnarray}
&&   (d-1)\, \hat{u}^\mu \, \frac{ {\hat \nabla}_\mu b}{b} + {\hat \theta}  + \frac{2b}{\hat{\alpha}\ d} {\hat u}_\mu \,{\hat \nabla}_\nu {\hat \sigma}^{\mu\nu} = 0 \nonumber \\
&&  - \,  \, \left(1+\frac{d}{2}\,(\hat{\alpha}^2-1)\right)\, {\hat P}_\mu^{\; \alpha} \frac{{\hat \nabla}_\alpha b}{b} +  {\hat a}_\mu -\frac{2\, b^d}{\hat{\alpha}\ d} \, {\hat P}_{\mu\alpha} \, {\hat \nabla}_\beta \left(\frac{1}{b^{d-1}}\,{\hat \sigma}^{\alpha \beta}\right) =0 
 \label{releom1}
\end{eqnarray}	
We are going to try to analyze these equations in the limit when $\hat{\alpha} \to \infty$.

From \req{releom1} it is clear that if we insist on leaving the hypersurface data independent of $\hat{\alpha}$ then we find that we have contributions at different orders that need to be independently cancelled. The most constraining equation is the $ {\cal O}(\hat{\alpha}^2)$ term from the transverse equation which demands that the spatial gradients of $b$ vanish. Then at $ {\cal O}(1)$ we have to kill the acceleration and have a non-trivial equation from the longitudinal part. At  order $\hat{\alpha}^{-1}$ we would need to kill all terms that show up with the shear viscosity. The upshot is that we are left with vacuous dynamics on the Dirichlet hypersurface should $b$ and $\hat{u}^\mu$ be ${\cal O}(1)$ as $\hat{\alpha} \to \infty$. 

While this sounds a bit strange, a moments pause reveals that this is indeed what one should expect on physical grounds. The horizon of a black hole is a null surface (it is generated by null generators) and in the process of moving the Dirichlet hypersurface to the horizon, we are effectively doing an infinite rescaling (hence $\hat{\alpha} \to \infty$) to bring the Dirichlet metric $\hat{g}_{\mu\nu}$ to be timelike and non-degenerate. Before we do such a rescaling however, we are in the ultra-relativistic regime as far as the horizon goes -- in such a regime it is natural to expect that there is no dynamics, the fluid streams along the null generators and is effectively frozen into a stationary flow. 

It is then clear that in order to obtain non-trivial dynamics on the Dirichlet surface one has to scale the fields $b$ and $\hat{u}^\mu$ in some fashion. The crucial question is whether there is any scaling that retains non-trivial dynamics; operationally demanding that we obtain  an `interesting' non-linear equation. Consider the following scaling:\footnote{This scaling can be compounded with the scaling symmetry of Navier-Stokes equations \cite{Bhattacharyya:2008kq} to change the exponents of $\hat{\varkappa}$ but we refrain from doing so for simplicity (see end of \App{A:bmw}).}
\begin{equation}
b = \hat{\varkappa} \, b_\bullet + \frac{1}{\hat{\varkappa}^{3}}\, \delta b_\star \ , \qquad \hat{u}^\mu = \left( 1 + {\cal O}(\hat{\varkappa}^{-2} ), \;\hat{\varkappa}^{-1} \, v^i_\star\right) \ , \qquad \hat{\alpha} = \hat{\varkappa}\ \hat{\alpha}_\bullet \left( 1 + {\cal O}(\hat{\varkappa}^{-2} )\right)
\label{nhscaling}
\end{equation}	
where the functions with subscript $\star$ have specific functional form with anisotropic scaling of their spatial and temporal gradients. Specifically, 
\begin{equation}
\hat{{\cal Y}}_\star(t,x^i) : {\mathbb R}^{d-1,1} \mapsto {\mathbb R}\ ,  \;\; \text{such that}
\;\; \{ \partial_t \hat{{\cal Y}}_\star(t,x^i), \partial_i \hat{{\cal Y}}_\star(t,x^i)\}  \sim \{{\cal O}(\hat{\varkappa}^{-2}) ,{\cal O}(\hat{\varkappa}^{-1})\}
\end{equation}	
where we assign gradient weight $1$ to the spatial derivatives and $2$ to temporal derivatives.  This is inspired of course by the non-relativistic BMW scaling discussed in \sec{s:nrhyp1} and as discussed there the ${\cal O}(\hat{\varkappa}^{-2})$ part of the velocity field is fixed by normalization.\footnote{Roughly speaking $\hat{\varkappa} \sim \hat{\aleph}$ of \sec{s:nrhyp1}; this is the overall parameter that will organize for us the hierarchy necessary in the near horizon limit.}

However, there is a crucial difference from the BMW scaling; the leading term in the expansion of $b_\bullet$ is growing with  $ \hat{\varkappa}$, which seems to be an issue. Nevertheless, it is easy to check that under this scaling (which admittedly is obtained by demanding a sensible $ {\cal O}(1)$ equation from the longitudinal part) one finds that the equations can be reduced to:
\begin{equation}
\partial_i v^i_\star =0 \ , \qquad -\frac{d\hat{\alpha}_\bullet^2}{2\, b_\bullet} \partial_i \delta b_\star + \partial_t v_i^\star + v^j_\star \partial_j v_i^\star - \frac{b_\bullet}{\hat{\alpha}_\bullet\ d} \nabla^2 v_i^\star = 0
\label{nrnseq1}
\end{equation}	
where we have specified to a flat Minkowski background metric for specificity ($\hat{g}_{\mu\nu} = \eta_{\mu\nu}$) and to make the equations more familiar. These are of course, the unforced (by background) Navier-Stokes equations which we have earlier described  via the BMW scaling in \sec{s:nrhyp1}. The incompressibility condition as in that case arises from the longitudinal conservation equation at ${\cal O}(\hat{\varkappa}^{-2}\,\hat{\alpha})$ while the Navier-Stokes equation itself appears at ${\cal O}(\hat{\varkappa}^{-3} \, \hat{\alpha})$. A similar scaling can be done for the forced Navier-Stokes system, we simply use the BMW scaling for the fluctuating part of the hypersurface metric.

\subsubsection{Deconstructing the near horizon scaling}
\label{s:dcnh}

How do we reconcile this scaling derived here for the relativistic fluid with that derived by BMW \cite{Bhattacharyya:2008kq}? Firstly, let's note that the rationale of our scaling up the background value of $b \sim \hat{\varkappa}\, b_\bullet$ is that  shear term survives the scaling. From the second equation of \req{bmwpnu} we clearly see that this is required in order to retain the shear term in the limit. On the other hand the non-relativistic pressure in the BMW limit is  proportional to $\hat{\alpha}_0^2\, \frac{\delta b_*}{b_0}$ and in the near horizon limit both $b_0 \sim \hat{\varkappa}\, b_\bullet$ and $\hat{\alpha}_0 \sim \hat{\varkappa}$ diverge. The extra scaling down of $\delta b_* \sim \frac{1}{\hat{\varkappa}}\, b_\star$ (over and above the scaling by $\hat{\varkappa}^{-2}$) is necessary to offset this divergence and ensure that the pressure gradient term contributes at the same order as the convective derivative $\partial_t + v^i_*\partial_i $ and the shear. The effective pressure that enters into the equation \req{nrnseq1} is
\begin{equation}
\hat{\varkappa}^{-2}\,\hat{p}_\star =  -\frac{d}{2} \, \hat{\varkappa}^{2}\,\hat{\alpha}_\bullet^2 \frac{\hat{\varkappa}^{-3}\, \delta b_\star}{\hat{\varkappa} \, b_\bullet} 
\end{equation}	
using the scaling of $\hat{\alpha}$ in \req{nhscaling} so that the first term of the Navier-Stokes equation in \req{nrnseq1} is essentially $\partial_i \hat{p}_\star$. So the final equations of motion on the near horizon region for the hypersurface dynamics are simply:
\begin{equation}
\partial_i v^i_\star =0 \ , \qquad  \partial_i  \hat{p}_\star + \partial_t v_i^\star + v^j_\star \partial_j v_i^\star - \, \hat{\nu}_\bullet \nabla^2 v_i^\star = 0
\label{nrnseq2}
\end{equation}	
with kinematic viscosity 
\begin{equation}
\hat{\nu}_\bullet = \frac{b_\bullet}{\hat{\alpha}_\bullet\ d}
\end{equation}	

We have here glossed over the fact that since we scale $b \sim \hat{\varkappa}$ we potentially have a problem. The limit seems to suggest that we are taking the zero temperature limit of the black hole geometry since the Hawking temperature (which is seen on the boundary) is $T \sim b^{-1}$ \req{hst2}. This naively sounds like we are outside the long wavelength regime and as a result should not be using the fluid/gravity map to describe the Dirichlet problem.

A different way to say this from a dynamical equation of motion perspective is that the scalings \req{nhscaling} were derived by demanding that the equations \req{releom1} remained non-trivial, which operationally means that different terms appear to have inhomogeneous weights. Hence by suitable fiddling of amplitudes and derivatives we engineered that terms which originally scaled as $\hat{\alpha}$ to various powers all contribute homogeneously. However, we have not analyzed the higher order contributions to the equations of motion. Is it possible that relativistic two derivative terms in the stress tensor, which show up as the corrections to viscous relativistic hydrodynamics equations of \req{releom1} show up at the same order? Note that this is not a problem for the  case discussed in \cite{Bhattacharyya:2008kq} for they engineered their scalings about a fixed background temperature, and there were no stray factors of $\hat{\aleph}$ (equivalently $\hat{\varkappa}$) in \sec{s:nrhyp1} (in front of $b_0$) to augment higher order contributions. 
 
For us to make the argument that the higher order terms are suppressed requires knowledge of the transport coefficients at second order on the Dirichlet hypersurface. This is in principle calculable by extending the Dirichlet fluid/gravity map of \sec{s:fg1} to one higher order in gradients. However, we will now argue that there is an essential simplification which allows us to make the statement boldly without computation. The rationale is simply that the enhanced scaling of the zero mode part of $b$ is compensated by the suppressing the corresponding  fluctuation term $\delta b$ so as to ensure that $p_\star$ is finite. Since the overall value of $b_0$ scales out in the non-relativistic regime (we will see this clearly from implementing the scalings in the next subsection) it cannot affect the resulting equations. So we conjecture that accounting for higher order corrections as well, one will obtain \req{nrnseq2} as the leading order equations in the $\hat{\varkappa}$ expansion (all further corrections are suppressed by higher powers of $\hat{\varkappa}$).

One physical way to motivate the correction is to first note that while the boundary temperature is being scaled to zero in the limit \req{nhscaling} the hypersurface temperature can indeed be maintained to be finite:\footnote{We thank Sayantani Bhattacharyya for emphasizing this point to us.}
\begin{equation}
\hat{T}_\bullet = \frac{\hat{\alpha}_\bullet\ d}{4\pi \,b_\bullet} 
\end{equation}	
So from the Dirichlet observer's point of view there is still scope for a non-relativistic scaling, and in fact this is how the Dirichlet observer would carry out the BMW regime. What is clear is that due to the extra red-shifting in translating to the boundary, the asymptotic observer is going to have trouble reconciling this near horizon scaling regime in his variables. We will return to this issue in \sec{s:nhgal}.

In summary, the non-relativistic incompressible Navier-Stokes equations  determine the dynamics of the hypersurface fluid as $\Sigma_D$ approaches the horizon; we will now proceed to investigate what this means for the various metrics: first we look at the bulk metric first and then examine what is the effect on the boundary.

\subsection{The bulk metric between $\Sigma_D$ and ${\mathcal H}^+$}
\label{s:nhdirmet}

From the discussion at the end of \sec{s:dcnh} it seems natural that we should try to ask the question as to  whether we can satisfy the constraints of the Dirichlet problem in the region between $\Sigma_D$ and the horizon (for the moment forgetting about the asymptotic region). Clearly we have $\Sigma_D$ approaching the horizon, so we will be forced to consider a double-scaling regime where we zoom in close to the horizon and expand out the spacetime region in-between. Let us temporarily ignore the consequences of this on the equations of motion \req{eins} and formally take the limit of the bulk metric given in \req{formmetw}, \req{dirbulk1}.

To zoom into the region between the horizon and $\Sigma_D$, realizing that $r_D \to \frac{1}{b}$ in the limit, we take:
\begin{equation}
r = \frac{1}{b_\bullet\, \hat{\varkappa}} \left(1+ \frac{\rho}{\hat{\varkappa}^2 b_\bullet\ \hat{\alpha}_\bullet}\right) .
\end{equation}	
from which it follows that 
\[ dr = \frac{1}{\hat{\varkappa}^3\, b_\bullet^2\ \hat{\alpha}_\bullet } \,d\rho \]

Similarly we parametrize the position of our Dirichlet hypersurface via 
\begin{equation}
r_D = \frac{1}{b_\bullet\, \hat{\varkappa}} \left(1+ \frac{\rho_D}{\hat{\varkappa}^2 b_\bullet\ \hat{\alpha}_\bullet}\right) .
\end{equation}
Both $\rho_D$ and $\hat{\alpha}_\bullet$ are a measure of how close we are to the horizon after
one has zoomed into the near horizon region. To relate them, we subtitute $r_D$ into the
expression for $\hat{\alpha}$ and do a large $\hat{\varkappa}$ expansion.  Identifying
the leading term as $\hat{\alpha}_\bullet$, we get
\begin{equation}
\rho_D \equiv\frac{b_\bullet}{\hat{\alpha}_\bullet\ d} = \hat{\nu}_\bullet
\end{equation}	
We see that $\rho_D$ is same as the kinematic viscosity - hence, the $\rho_D\to 0$ limit is
identical to the inviscid limit in hydrodynamics.

Implementing this change of variables and the scaling \req{nhscaling} in the bulk metric we obtain the near-horizon metric. In fact, the fastest way to derive the metric in \req{hypnrscale} is to utilize the fact that the near horizon limit is essentially the non-relativistic limit together with a particular form of the pressure fluctuations; essentially we substitute $\hat{\alpha}_0 = \hat{\varkappa} \, \hat{\alpha}_\bullet$, $b_0 = \hat{\varkappa}\, b_\bullet$, $\hat{p}_* = \hat{p}_\star$ into \req{hypbmwf} along with the identification $\hat{\aleph} = \hat{\varkappa}$ to ensure that we have the correct scalings. 
We first calculate the near horizon expansions of various functions appearing in the metric
\begin{equation}\begin{split}
\hat{\alpha} &= \hat{\alpha}_\bullet \, \hat{\varkappa} 
\left[1 + \hat{\varkappa}^{-2}  \left( \hat{p}_\star  + \frac{d+1}{4d\hat{\alpha}_\bullet^2}\right)+ {\cal O}(\hat{\varkappa}^{-4}) \right] \\
\hat{\varkappa}^2 b_\bullet^2 r^2 &=  \left[1+\hat{\varkappa}^{-2}\frac{2\rho}{\hat{\alpha}_\bullet^2\rho_D d}+{\cal O}(\hat{\varkappa}^{-4}) \right]\\
\hat{\varkappa}^2 b_\bullet^2 r^2\left[1- \hat{\alpha}^2\, f(br) \right] &=\left(1-\frac{\rho}{\rho_D}\right)
\left[1 + 2\hat{\varkappa}^{-2}  \left( \hat{p}_\star  -\frac{d-3}{4d\hat{\alpha}_\bullet^2}\frac{\rho}{\rho_D}\right)+ {\cal O}(\hat{\varkappa}^{-4}) \right] \\
-\hat{\varkappa}^2 b_\bullet^2 r^2 \hat{\alpha}^2\, f(br)&= -\frac{\rho}{\rho_D} +   2\hat{\varkappa}^{-2}\hat{p}_\star\left(1-\frac{\rho}{\rho_D}\right)\\
&\quad +\hat{\varkappa}^{-2}\frac{\rho}{2\hat{\alpha}_\bullet^2\,\rho_D} \left(\frac{(d-3)\, \rho- (d+1) \, \rho_D}{\rho_D d} \right) + {\cal O}(\hat{\varkappa}^{-4}) \\
\frac{2\hat{\alpha}^2}{r_D\left(1+\frac{d}{2}(\hat{\alpha}^2-1)\right)} &= 4 \rho_D \hat{\varkappa}\hat{\alpha}_\bullet\left[1
+ {\cal O}(\hat{\varkappa}^{-2}) \right]\\
2\,r\,\frac{\hat{\alpha}(2\hat{\xi}-1)}{1+\frac{d}{2}\, (\hat{\alpha}^2 -1)} &= \frac{4\rho_D}{\hat{\varkappa}^2 b_\bullet^2}\left(1-\frac{\rho}{\rho_D}\right)\left[1
+ {\cal O}(\hat{\varkappa}^{-2}) \right]\\           
\end{split}\end{equation}
Using these, we get the metric as
\begin{equation}
b_\bullet^2\, \hat{\varkappa}^2\,ds^2 = ds_0^2  + \hat{\varkappa}^{-1} \, ds_1^2 +  \hat{\varkappa}^{-2} \, ds_2^2 +  \hat{\varkappa}^{-3} \, ds_3^2+ {\cal O}(\hat{\varkappa}^{-4}) 
\label{}
\end{equation}	
where 
 \begin{eqnarray}
 ds_0^2 &=& 
	2\, dt \, d\rho
	-\,\frac{\rho}{\rho_D} dt^2 + \delta_{ij}\,  dx^i\, dx^j 
\nonumber \\
ds_1^2 &= & 
	- 2\,  \hat{v}^\star_i  \,  d\rho\, dx^i - 2 \, \left(1-\frac{\rho}{\rho_D} \right) \hat{v}^\star_i \, dt\, dx^i 
\nonumber \\ 
ds_2^2 &= &  
	2\, \left( \frac{1}{2}\,\hat{v}_\star^2  + \hat{p}_\star +
	    \frac{d+1}{4\,d} \frac{1}{\hat{\alpha}_\bullet^2}\right)  dt \, d\rho
 \nonumber \\
 && \qquad  
 	+ \; \left[ \left(1-\frac{\rho}{\rho_D} \right)
	 \left( \hat{v}_\star^2 
	 + 2 \hat{p}_\star  \right) 
	+ \frac{\rho}{2\hat{\alpha}_\bullet^2\,\rho_D} \left(\frac{(d-3)\, \rho- (d+1) \, \rho_D}{\rho_D d} \right)\right]  dt^2 
\nonumber \\
&& \qquad 	 
	+\; \left[\left(1-\frac{\rho}{\rho_D} \right) \hat{v}^\star_i\,\hat{v}^\star_j
	 + \,  \frac{2 \, \rho}{\hat{\alpha}_\bullet^2\rho_D d}\, \delta_{ij} \right] dx^i \, dx^j  
\nonumber \\
ds_3^2 &= & 
	2 \left(1-\frac{\rho}{\rho_D} \right) 
	 \left[2\rho_D \left( \partial_{t}\hat{v}_{i}^\star  
	 +\hat{v}_\star^{j} \partial_j \hat{v}_{i}^\star\right)
	 -  \left(\hat{v}_\star^2 + 2 \hat{p}_\star 
	 - \frac{d-3}{2d\hat{\alpha}_\bullet^2}\, \frac{\rho}{\rho_D}\right) \hat{v}_i^\star \right] dx^{i}\, dt
\nonumber \\
&& \qquad 
	  +\; 2\left[2\rho_D \left( \partial_{t}\hat{v}_{i}^\star  
	 +\hat{v}_\star^{j} \partial_j \hat{v}_{i}^\star\right)-\;\, 
	   \left( \frac{1}{2}\,\hat{v}_\star^2   
	  + \hat{p}_\star + \frac{d+1}{4\,d} \frac{1}{\hat{\alpha}_\bullet^2}\right) \, \hat{v}_i^\star \,\right] d\rho\, dx^i \, 
\nonumber \\
&& \qquad
	-\,\left[\rho^2-\rho_D^2+4\, \rho_D\,(\rho_D-\rho)\right]\,\partial_j\partial^j \hat{v}^{\star}_i \, dx^i\,dt 
	+ \frac{1}{d}\, \partial_j\partial^j \hat{v}^{\star}_i \, d\rho\, dx^i \,.  
\label{hypnrscale}
  \end{eqnarray}
We have restricted attention to the simplest setting where $\hat{g}_{\mu\nu} = \eta_{\mu\nu}$  (hence $\hat{g}^{(0)}_{ij} = \delta_{ij}$ above) and $\hat{v}_\star^2 = \delta_{ij} \, \hat{v}_\star^i \, \hat{v}_\star^j$. In deriving the expressions as in \sec{s:nrhyp2} we have used the incompressibility condition, which appears as before as the leading order fluid equation of motion. The Navier-Stokes equations themselves can also be used to simplify the last term in \req{hypnrscale}: one can replace the convective derivative of the spatial velocity $\partial_{t}\hat{v}_{i}^\star  +\hat{v}_\star^{j} \partial_j \hat{v}_{i}^\star$ by $-\partial_i \hat{p}_\star + 2\,\hat{\nu}_\bullet \, \nabla^2 v_i^\star$.

This is in fact closely related (but not identical) to the metric derived in  \cite{Bredberg:2011jq} and generalized further in \cite{Compere:2011dx} to higher orders (compare with for instance Eq (3.2) and (6.5) of the latter paper). We recall that these works consider solutions to vacuum Einstein's equations without a cosmological constant. In particular, \cite{Bredberg:2011jq} solves the Dirichlet problem in flat space by looking for small gradient fluctuations around a Rindler geometry, while we are dealing with solutions of \req{eins} which has a non-vanishing negative cosmological constant. 

We see that to leading order we nevertheless should expect agreement to the analysis of \cite{Bredberg:2011jq,Compere:2011dx} -- this is clear from the first  two lines of \req{hypnrscale}. This is reflecting the universal Rindler nature of a non-degenerate horizon such as that of the planar 
Schwarzschild-AdS$_{d+1}$ solution we started with. Moreover, we can also physically understand how a solution of \req{eins} reduces to that of vacuum Einstein's equations by noting that in the limit we consider here, the cosmological constant gets diluted away. The factor of $\hat{\varkappa}$ on the l.h.s, is essentially indicative of $R_\text{AdS} \to \hat{\varkappa}^2 \, R_\text{AdS}$ in our near horizon limit. This is to be expected; by zooming in onto the region between $\Sigma_D$ and ${\cal H}^+$ we are effectively blowing up a small sliver of the spacetime and  are thus losing any information about the background curvature scale $R_\text{AdS}$. In some sense the near horizon limit is like the limit we take to decouple the asymptotically flat region from the AdS throat in the D3-brane geometry; what is unclear is whether some notion of decoupling exists in the present context.

But starting at ${\cal O}(\hat{\varkappa}^{-2})$ on the l.h.s. of \req{hypnrscale} we start to see deviations from the metric presented in \cite{Bredberg:2011jq,Compere:2011dx}. Even though we are zooming close to the horizon of a Schwarzschild-AdS$_{d+1}$ black hole, starting at second order we should expect to see curvature contributions.  The various terms at this and higher orders originate from how the Rindler geometry gets corrected as we step away from the strict limit. In particular, using the near-horizon scaling of variables we see that \req{eins} reduces to 
\begin{equation}
R_{AB}+\frac{d}{\hat{\varkappa}^2} \; \frac{1}{(d\, \hat{\alpha}_\bullet\, \rho_D)^2}\;  {\mathcal G}_{AB}=0\,.
\label{}
\end{equation}	
It follows that all terms which involve $\frac{1}{\hat{\alpha}_\bullet^2}$ in the denominator correspond to the corrections due to the AdS curvature. Setting such terms in \req{hypnrscale} to zero i.e., taking $\hat{\alpha}_\bullet \to \infty$ leads us to the metrics derived in \cite{Bredberg:2011jq,Compere:2011dx}.

We are here restricting attention to the metric to leading orders in the $\hat{\varkappa}$ expansion; upto ${\cal O}(\hat{\varkappa}^{-3})$. In principle since the fluid/gravity metrics are known to higher orders one can carry out the construction to higher orders and we expect that we should be able to reproduce the results of \cite{Compere:2011dx} who have derived the expressions in the near horizon limit to ${\cal O}(\hat{\varkappa}^{-6})$ in our notation. Note that this simplification of the near horizon region construction happens because of the correlated amplitude and gradient scaling, as already emphasized in \cite{Bhattacharyya:2008kq}.

\subsection{Near horizon limit and Galilean degenerations}
\label{s:nhgal}

Having seen that it is possible to take the near horizon scaling limit and retain both interesting dynamics on $\Sigma_D$ and further have a sensible geometry between $\Sigma_D$ and ${\mathcal H}^+$, our next goal is to address whether we can make sense of the metric between the hypersurface and the boundary. There are already many indications that this is going to be problematic; the vanishing of the boundary temperature $ \propto b^{-1}$ being the most prominent one, which suggests break-down of the gradient expansion.

The bulk metric turns out not to make sense, but we shall see that there is an object that does -- this is the inverse bulk metric which we shall refer to as the co-metric (see below). Similarly, the boundary co-metric also is well behaved and we will argue naturally provides a Newton-Cartan like structure on the boundary so that the boundary geometry degenerates from a Lorentzian manifold into a Galilean manifold. 

In the following, we will call the metric on the co-tangent bundle $g^{\mu\nu}$
as the co-metric. This is a more accurate terminology in the context of 
non-relativistic limit (the Galilean or Newton-Cartan limit)  than the usually preferred 
`inverse metric' since in this limit the co-metric degenerates
(becomes non-invertible) and the usual metric ceases to exist. Hence, we will prefer as
much as possible to work with the co-metric instead of the metric. 

\subsubsection{Emergence of Galilean structure on the boundary}

Let us first examine what happens to the boundary metric data when we try to push the Dirichlet hypersurface towards the horizon. This corresponds to $\hat{\alpha}$  tending to infinity. We will work with the relativistic expressions of \sec{s:dgrav} to maintain covariance; it is easy to then pass over to the near horizon scaling regime discussed in \sec{s:dhypdyn} and obtain explicit parameterization of the results. 

 From the formulae  for the boundary metric in terms of $u^\mu$  \req{bdyggu} 
we see that there are various terms which blow up. Moreover, in some cases the higher derivative terms overwhelm the lower derivatives suggesting a breakdown of derivative expansion, as already suspected from the vanishing of the boundary temperature. It is clear that this limit if it exists cannot be straightforward.

As the Dirichlet hypersurface tries to approach the horizon, it first hits the 
hypersurface $r_D=r_{D,snd}$. From the boundary viewpoint, the interaction 
between the fluid with its gravitational potential packet drives the system to have superluminal sound propagation, $\hat{c}_{snd}$ exceeds the dressed speed of
light as determined by $\hat{g}^{\mu\nu}$. By this point 
the interaction between the boundary metric and the fluid has become
so important that the bare velocities $u^\mu$ have no more physical
significance. We should rather think in terms of the dressed fluid
and see how we can make sense out of the approach to the horizon.

So, we will first rewrite the above formulae in terms of the dressed velocity $\hat{u}^\mu$.
The dictionary for the (normalized) metric and the co-metric are
\begin{equation}
\begin{split}
g_{\mu\nu} 
&= \hat{g}_{\mu\nu} + \left[1-\hat{\alpha}^2-\frac{2\,\hat{\alpha}^3\,\hat{\theta}}{r_D(\,d-1)}\right]\hat{u}_\mu \hat{u}_\nu+\frac{2\,\hat{\alpha}^3 \, \hat{u}_{(\mu} \hat{a}_{\nu)}}{r_D\left[1+\frac{d}{2}(\hat{\alpha}^2-1)\right]}  - \frac{2\,b}{\hat{\alpha}} \,F(br_D)\,\hat{\sigma}_{\mu\nu} \\
g^{\mu\nu} &= \hat{g}^{\mu\nu}+ \left[1-\frac{1}{\hat{\alpha}^2}+\frac{2\, \hat{\theta}}{\hat{\alpha}\, r_D\, (d-1)}\right]\hat{u}^\mu \hat{u}^\nu+\frac{2\,b}{\hat{\alpha}} \,F(br_D)\, \hat{\sigma}^{\mu\nu}-\frac{2\,\hat{\alpha}\, \hat{u}^{(\mu} \hat{a}^{\nu)}}{r_D\left[1+\frac{d}{2}(\hat{\alpha}^2-1)\right]}\\
\end{split}
\end{equation}

We have already remarked on the necessity to project out the dressed sound mode and take 
an incompressible limit of the dressed fluid a la BMW as the Dirichlet 
hypersurface crosses $r_D=r_{D,snd}$ in \sec{s:dgravnr}. The formulae above reinforce that 
intuition -- we see for example that unless $\hat{\theta}$ is sent to 
zero, the first derivative terms in the boundary metric overwhelm
the zeroth order answer thus leading to a breakdown of the derivative expansion.
So, we will drop the $\hat{\theta}$ terms with the understanding
that we are projecting into the incompressible sector of the dressed fluid.
Even this does not seem to help the case for the metric, since it still
diverges -- keeping only leading order terms (and assuming none of the 
subsequent terms grow with $\hat{\alpha}$ -- we will postpone a more careful
analysis for the future) we get
\begin{equation}
\begin{split}
g_{\mu\nu} 
&= -\hat{\alpha}^2\hat{u}_\mu \hat{u}_\nu+\ldots\\
g^{\mu\nu} 
&= \hat{g}^{\mu\nu}+ \hat{u}^\mu \hat{u}^\nu+\ldots\\
\end{split}
\end{equation}

We are thus led to a remarkable conclusion:  the near horizon Dirichlet constitutive
relation is that the boundary co-metric  degenerates  along $t_{\mu}\equiv\hat{u}_\mu$
i.e., $g^{\mu\nu}t_\nu=0$ and the boundary metric has one divergent time-like eigenvalue
along the same direction $g_{\mu\nu}= -\hat{\alpha}^2 t_\mu t_\nu$. This is the 
signature that the boundary metric is becoming Newtonian/non-relativistic -
crudely speaking this is analogous to the $c\to\infty$ behavior of Minkowski
metric/co-metric
\[ \eta^{\mu\nu}=\text{diag}(-\frac{1}{c^2},1,1,\ldots) \quad\text{and}\quad
   \eta_{\mu\nu}=\text{diag}(-{c^2},1,1,\ldots)
\]
The way to make this mathematically precise is to resort to what is called as
Galilean manifold.

A Galilean manifold is a manifold with a Newtonian time co-vector $t_\mu$,
and a degenerate co-metric $g^{\mu\nu}$ satisfying $g^{\mu\nu}t_\nu=0$. 
As in standard differential geometry,  we can demand a connection that is compatible with this structure. This requires that there exists a torsionless connection defining a covariant derivative $\nabla^{\text{NC}}_\mu$ which is compatible with both the co-metric and the time co-vector, i.e., $\nabla^{\text{NC}}_\lambda g^{\mu\nu}=0$ and $\nabla^{\text{NC}}_\mu t_\nu=0$. If in addition we have such a a torsionless covariant derivative $\nabla_\mu$  compatible with the Galilean structure then we  call $\nabla^{\text{NC}}$ as the Newton-Cartan connection and the corresponding
Galilean manifold is termed a Newton-Cartan manifold. See \cite{Misner:1973by,Ruede:1996sy} for further discussions and \cite{Bagchi:2009my} for another AdS/CFT perspective on the Galilean structures.

Thus we have just argued that that there is a natural emergence of  Galilean structure in the boundary theory as we take our Dirichlet surface nearer and nearer to the horizon. 
In this case, the metric that we have to put the field theory in gets enormously simplified and
we get just a CFT on a Galilean manifold with a Galilean co-metric having 
$\hat{u}_\mu$ as the degenerate direction. This is a very precise way of stating what
 the membrane paradigm is from the boundary theory viewpoint -- the membrane paradigm is a particular Galilean limit of the field theory.\footnote{Once we include the new corrections highlighted for e.g., in \eqref{hypbmwf}, we find that the boundary Dirichlet constitutive relations and the boundary gradient expansion are in tension. For instance we find terms which originate at ${\cal O}(\hat{\aleph}^{-3})$ migrate down to ${\cal O}(\hat{\aleph}^{-1})$. However, to the order we have looked at all such terms are related to enforcing Landau frame choice both on the hypersurface and the boundary, and hence to a gauge choice for the bulk metric. Our statements in this section assume that such pure gauge terms are irrelevant, but this issue deserves further investigation.}

Having shown that we have a Galilean structure on the boundary, we further ask: ``Is there a Newton-Cartan structure"?  Consider the Galilean limit of the Christoffel connection $\nabla_\mu$ which is by construction torsionless and compatible with the Galilean co-metric $g^{\mu\nu}$. This could be a sensible Newton-Cartan structure if it annihilates the time co-vector; to see what $\nabla_\mu t_\nu$ is we use
\begin{equation}
\begin{split}
\tilde{\Gamma}_{\mu\nu}{}^\rho\hat{u}_\rho &=  -(1-\frac{1}{\hat{\alpha}^2})\left[ \hat{\sigma}_{\mu\nu}+\frac{\hat{\theta}}{d-1}\ \left(\hat{P}_{\mu\nu}+\frac{d}{2}\hat{\alpha}^2\hat{u}_\mu \hat{u}_\nu\right)-\frac{d\hat{\alpha}^2}{\left[1+\frac{d}{2}(\hat{\alpha}^2-1)\right]}\hat{a}_{(\mu}\hat{u}_{\nu)}\right]\\
\end{split}
\end{equation}
so that 
\begin{equation}
\begin{split}
\nabla_\mu t_\nu &= \hat{\nabla}_\mu\hat{u}_\nu+\tilde{\Gamma}_{\mu\nu}{}^\rho\hat{u}_\rho\\
&= -\frac{d}{2}(\hat{\alpha}^2-1)\frac{\hat{\theta}}{d-1}\hat{u}_\mu \hat{u}_\nu+\hat{\omega}_{\mu\nu}+\hat{a}_\mu \hat{u}_\nu
+\frac{1}{\hat{\alpha}^2}\left[\hat{\sigma}_{\mu\nu}+\frac{\hat{\theta}}{d-1}\hat{P}_{\mu\nu}\right] -\frac{2\hat{a}_{(\mu}\hat{u}_{\nu)}}{\left[1+\frac{d}{2}(\hat{\alpha}^2-1)\right]}
\end{split}
\label{nablat}
\end{equation}
This in fact determines a sensible answer in the infinite $\hat{\alpha}$ limit , once we factor in our earlier observation that $\hat{\theta}$ is suppressed in the near horizon limit (it is ${\cal O}(\hat{\alpha}^{-4})$) as are any other gradients of the velocity field. These scalings can be read off essentially from \sec{s:dgravnr} as the near horizon limit is the BMW limit as far as the velocities are concerned.

The leading contribution to \req{nablat} comes at ${\cal O}(\hat{\alpha}^{-2})$ from the vorticity and expansion.
It is tempting to speculate that the degeneration of the co-metric along with  \req{nablat} define a 
Newton-Cartan like limit and we will postpone a structural analysis of this limit for future work. In the
next subsection, we will try to examine the bulk co-metric to see what this Galilean limit entails in the bulk.

\subsection{The bulk co-metric in the Near-Horizon limit}

In the preceding subsection, we argued for the emergence of Galilean structures in the Near-Horizon
limit thus forcing us to formulate the boundary geometry in terms of a co-metric than a 
metric. Since the boundary metric is ill-behaved in this limit, it is clear that the 
bulk metric should be ill-behaved too. This poses a conundrum since it seems as if in 
taking the near-horizon limit we have destroyed the \AdS{}-asymptopia and one might
wonder whether there is any sense in talking about the bulk geometry far away from
the horizon. We will in this subsection take the description of the boundary 
geometry via a co-metric as a clue and argue that the bulk geometry is also
well-described by a bulk co-metric. Thus while the metric description might
break down the co-metric description with its associated Galilean structures
continue to describe the bulk and the boundary geometry.

Now we present the co-metric as a function of the Dirichlet data. 
\begin{eqnarray}
\mathcal{G}^{AB}\partial_A\otimes\partial_B &=&
 \left[r^2\, f(br)-\frac{2\, r\, \theta}{d-1}\right]\partial_r\otimes\partial_r+2\left[u^\mu -r^{-1} a^{\mu}\right]\partial_\mu\otimes_s\partial_r \nonumber \\
&&\qquad \qquad +\; r^{-2}\left[P^{\mu\nu}   -2b F(br)\  \sigma^{\mu\nu}\right]\partial_\mu\otimes\partial_\nu \nonumber \\
&=& \left[r^2\, f(br)-\frac{2\,r\,\hat{\theta}}{\hat{\alpha}\,(d-1)}\right]\partial_r\otimes\partial_r \nonumber \\
&&\quad +\; 2\left[\frac{\hat{u}^\mu}{\hat{\alpha}}\left(1-\frac{\hat{\alpha}\,\hat{\theta}}{r_D\,(d-1)}\right) - \frac{\hat{a}^\mu}{r\left[1+\frac{d}{2}(\hat{\alpha}^2-1)\right]}\right]\partial_\mu\otimes_s\partial_r \nonumber \\
&&\quad +\;r^{-2}\left[\hat{P}^{\mu\nu} -\frac{\hat{\alpha}\left[\hat{u}^\mu \hat{a}^\nu + \hat{a}^{\mu}\hat{u}^\nu\right]}{r_D\left[1+\frac{d}{2}(\hat{\alpha}^2-1)\right]}   -2b\, \hat{F}(br)\   \hat{\sigma}^{\mu\nu}\right]\partial_\mu\otimes\partial_\nu
\end{eqnarray}
where
\begin{equation}\begin{split}
\hat{F}(br) &\equiv \frac{1}{\hat{\alpha}}\left(F(br)-F(br_D)\right) = \frac{1}{\hat{\alpha}} \;  \int_{br}^{br_D}\; \frac{y^{d-1}-1}{y(y^{d}-1)}dy\,.
\end{split}\end{equation}
As we anticipated, there are no problems evident in the infinite $\hat{\alpha}$ limit (provided one takes 
$\hat{\theta}$ to zero appropriately to preserve the validity of derivative expansion). 
Contrast this with the bulk metric (see \eqref{appeq:gbulk}) whose large $\hat{\alpha}$ limit seems dubious.
This supports our contention that there is a completely sensible description of the 
bulk and the boundary geometry in terms of co-metrics everywhere when we take
the near-horizon Dirichlet problem along with an incompressible limit on 
the dressed fluid.

Our exercise of rewriting the near-horizon bulk metric  in terms of the non-relativistic
variables can be repeated in the case of co-metric and we can easily convince ourselves
that this goes through without any new subtleties. In fact, most terms in the above
expression are sub-dominant in the $\hat{\varkappa}$ expansion introduced in the 
previous subsections. In fact, at the zeroth order the spatial part of the co-metric is just that of vacuum AdS; with the temporal part contributing only at a higher order. The horizon structure encoded in $(br)^{-d}$ is completely invisible until quite high orders in $\hat{\varkappa}$ expansion, which is to be expected since the boundary temperature is being scaled to zero. To be specific using the scaling \req{nhscaling} we find 
\begin{equation}
\mathcal{G}^{AB}\partial_A\otimes\partial_B = r^2 \, \partial_r \otimes \partial_r + \frac{1}{r^2} \,
\partial_i \otimes \partial_i + \hat{\varkappa}^{-1} \left(2 \, \partial_t \otimes_s \partial_r \right) + {\cal O}(\hat{\varkappa}^{-2}) \,.
\end{equation}	
Basically we seem to find that the co-metric description is mostly oblivious to the near-horizon geometry (which as we have seen is well-described in the metric description). Hence, we are naturally led to an effective description where there are two regions of the geometry:
one well-described by a metric, and another by a co-metric. 

The interesting question is to see whether there is an overlap region where 
the two descriptions are valid and the metric is just the inverse of 
the co-metric. Having detained the reader for so long, we will 
leave the detailed answer of this question to future work. However,
we would like to draw the attention of the reader  to the following
fact -- the near-horizon limit of the Dirichlet problem 
naturally seems to lead to novel geometric
structures closely associated with the Galilean limit. These
Galilean structures evidently call for more detailed studies especially
since they might hold valuable lessons for non-relativistic holography (as previously pointed out in \cite{Bagchi:2009my}).

\section{Discussion}
\label{s:discuss}

The bulk Dirichlet problem in \AdS{}, which we defined as the gravitational dynamics in a spacetime with negative cosmological constant, subject to Dirichlet boundary conditions on a preferred timelike hypersurface $\Sigma_D$, is interesting in the AdS/CFT context for several reasons. For one, it allows us to investigate questions in a cut-off AdS spacetime which  might be dual to a field theory with a rigid UV cut-off via the usual AdS/CFT dictionary. Furthermore, it allows us to touch upon the ideas involving holographic renormalisation and the implementation of the Wilsonian RG ideas in the gravitational description. The main motivation behind our work was to ask, what is it that one is doing on the dual field theory living on the boundary of the AdS spacetime that ensures these Dirichlet boundary conditions on $\Sigma_D$, and whether such a problem is always well-posed.

For linear systems in AdS it is easy to see that there is a simple map between data on the hypersurface and that on the boundary. In particular, Dirichlet boundary conditions on $\Sigma_D$ translate into non-local multi-trace deformations of the dual field theory. One can think of the boundary theory being deformed by some `state dependent' sources. Such a boundary condition while seemingly bizarre from standard field theoretic perspective, has an effective description in terms of a much simpler and local source, which is the Dirichlet data on $\Sigma_D$.  
 
In this paper we have further argued that such bulk Dirichlet problems for full non-linear gravitational dynamics in \AdS{} are also amenable to solution in a certain long wavelength regime.  Using  results from the fluid/gravity correspondence \cite{Bhattacharyya:2008jc,Bhattacharyya:2008mz} we have constructed the bulk spacetime resulting from the specification of a hypersurface metric on $\Sigma_D$. The solution is contingent upon the hypersurface dynamics, as determined by the conservation of the stress tensor induced on $\Sigma_D$, be conserved. In the long wavelength regime this stress tensor can be written as that of a non-conformal fluid propagating on the fixed geometry of $\Sigma_D$. Armed with such a solution we can examine the dual CFT to determine how one is achieving the Dirichlet dynamics.

In the long wavelength regime the boundary theory is described by fluid dynamics with what we term as Dirichlet constitutive relations. In particular, while the stress tensor of the boundary fluid is of the conventional conformal fluid form, it lives on a background geometry whose metric is `dynamical' in the following sense: the boundary metric depends on the dynamical fluid degrees of freedom. One should think of this in terms of a fluid being subject to a gravitational potential well which is furthermore carried along with the fluid. A useful analogy is to think of a charge carrier in a polarizable medium. The medium or more precisely in our case the `dynamical' background exerts a force on the fluid. However, one can subsume the `dynamical' background's effect on the fluid, and rewrite the dynamics as that of a `dressed fluid' on a fixed background. This `dressed fluid' is a collective effect and moreover from the geometry we know what its description should be -- it is simply the non-conformal fluid on the Dirichlet surface. This can be independently verified by starting with the boundary fluid and the `dynamical' boundary metric and showing that the boundary conservation equation can be rewritten as the hypersurface conservation equation, thus deriving the non-conformal fluid stress tensor on $\Sigma_D$. 

We also note that in the long wavelength regime because one is working  order by order in boundary gradients, the boundary fluid is deformed locally -- at any spacetime point the source $g_{\mu\nu}$ depends only on the fluid degrees of freedom (velocity and temperature) at that point. This is what allows us to think of the fluid as carrying around with it a local gravitational cloud. From a formal point of view, the gravitational Dirichlet problem involves turning on local multi-trace deformations for the field theory.

From the above discussion one is then tempted to say that the bulk geometry provides a way to repackage non-local deformations into a local perturbation at a lower radius or scale. This is highly suggestive of some sort of renormalisation of sources as one propagates boundary conditions into the bulk. Let us compare this picture with the recent discussion of the holographic Wilsonian RG flow idea of \cite{Heemskerk:2010hk,Faulkner:2010jy}. In these works starting with a field theory with given sources for say just single trace operators in the planar theory, one derives an effective action containing not just renormalized single trace sources, but also multi-traces on some chosen cut-off surface. The flow equation governing the radial evolution of such sources was argued to arise  by effectively integrating out a part of the bulk geometry between the boundary and the  cut-off surface. It was also important for that discussion given the particularization to the Wilsonian perspective that one does not prejudice oneself with the boundary conditions in the interior of the spacetime below the cut-off (the infra-red of the field theory). 

In the current context we see that by transferring the boundary conditions onto the Dirichlet surface we have shifted the burden of multi-traces onto the boundary -- as described above there is a suitable set of non-local multi-trace deformations on the boundary which `renormalizes' into a single trace source on the cut-off. Whilst this is not totally surprising for linear systems such as the scalar problem discussed in the text, it is satisfying that a similar statement can be made for non-linear gravitational dynamics by invoking the long wavelength gradient expansion. We should however note that our discussion explicitly assumes knowledge of the interior (infra-red) boundary conditions; in the fluid/gravity solutions constructed herein we have demanded that there be a regular future event horizon to single out a sensible solution.

Having obtained the effective dynamics on the hypersurface, we learnt that the conservation equations result in a sound mode which travels outside the light-cone of the fixed hypersurface metric once one pushes the surface too far into the interior. Past this sonic threshold, it is probably not sensible to maintain a relativistic fluid description on $\Sigma_D$. We argued that we should pass over into a non-relativistic regime, by a suitable scaling of fluid variables (a la BMW \cite{Bhattacharyya:2008kq}). It is interesting to ask whether the acausality manifesting itself in the sound mode of the hypersurface dynamics can be discerned (without invoking the `dressed' picture) from the boundary. More importantly, one wonders whether the gravitational Dirichlet problem suffers from such pathologies in general.\footnote{We believe the answer to this question is most likely to be in the affirmative. If we consider deforming the boundary conditions at infinity by making the boundary metric an arbitrary local function of multi-traces of the stress tensor, we can think of the problem effectively as a mixed boundary condition a la \cite{Compere:2008us}, which has been argued to suffer from ghosts generically (see also \cite{Andrade:2011dg}).}
Does one encounter any acausal behavior the moment the hypersurface is inside the bulk outside the long wavelength regime? What about their validity where say stringy effects are  taken into account? Can we engineer such a boundary condition consistently using objects like orientifolds in string theory? What happens once we go beyond large N? Does the field  theory with these non-local deformations make sense beyond large N for at least some subclass of these deformations?
What is the field theory interpretation of other kinds of boundary conditions other than Dirichlet
boundary conditions (say if we fix the mean extrinsic curvature as proposed in \cite{Lysov:2011xx}) ? 
These are fascinating questions which deserve to be explored further. 

Finally, by examining the Dirichlet problem in the vicinity of the event horizon of the fluid/gravity solutions we made contact with  the recent constructions of flat space gravitational duals to incompressible Navier-Stokes flows on a cut-off hypersurface \cite{Bredberg:2011jq,Compere:2011dx}, deriving in effect the membrane paradigm from the boundary field theory. Focussing on a tiny sliver of the geometry between the horizon and $\Sigma_D$ leads to the long wavelength solution around the Rindler horizon found in these works. This is reminiscent of the Penrose limit; while in the latter we focus on the geometry close to a null geodesic and blow it up, here we zoom close to a null surface and blow up the spacetime in its vicinity (perhaps a better analogy is the near horizon limit of black D3-branes without any statement about decoupling). The process of blowing up the near horizon region dilutes away the cosmological constant; hence rather than obtaining a solution to Einstein's equations with negative cosmological constant we  end up with a geometry that solves vacuum Einstein's equations at leading and next-to-leading order. However, starting at higher orders we start to see the effect of the cosmological constant, revealing the throat region between  the near-horizon Rindler geometry and the asymptotic AdS spacetime.\footnote{It has been pointed out in \cite{Bredberg:2011xw} that there is an obstruction to finding solutions with Dirichlet boundary conditions in the near horizon Rindler like region, when the fluids live on compact curved spatial manifolds, such a a sphere (as would be relevant for the near horizon of asymptotically flat Schwarzschild black hole). In our construction the near-horizon limit described in \sec{s:nh} requires that we simultaneously scale the curvatures to zero, and so we are unable to see the origins of such obstructions.}

From our embedding of the construction of \cite{Bredberg:2011jq,Compere:2011dx} into the fluid/gravity correspondence \cite{Bhattacharyya:2008jc,Bhattacharyya:2008mz} we learn that the boundary fluid lives on a manifold with Galilean/Newton-Cartan like structure in the near horizon limit.  This degeneration of the Lorentzian structure into an effective Galilean one is what enforces the incompressible Navier-Stokes scaling from the viewpoint of the CFT. One can simply say that the membrane paradigm is a particular Galilean limit of the field theory.  This perspective also clarifies the universality of the membrane paradigm -- since we are zooming down to the Rindler region in the vicinity of a non-extremal black hole horizon, we should expect the same geometry for any non-degenerate horizon.  Hence the dual description should always be in terms of an incompressible Navier-Stokes fluid as long as we deal with systems carrying only conserved charges \cite{Bhattacharyya:2008kq}. In the presence of other light degrees of freedom, as happens in systems with spontaneously broken symmetries, say the holographic superfluids discussed in \cite{Sonner:2010yx,Bhattacharya:2011ee,Herzog:2011ec, Bhattacharya:2011tr} we would have to project out all linearly dispersing modes to achieve the same. 

In the near horizon limit we find that both the bulk metric and the boundary metric degenerate into a Newton-Cartan time-metric, but the co-metric is spacelike and well behaved. We speculate that the bulk spacetime in this limit should be described in two patches: the region between the horizon and the Dirichlet hypersurface enjoys a description in terms of a conventional metric while the region between $\Sigma_D$ and the boundary requires use of the   co-metrics and Newton-Cartan structures. We conclude that the language of Galilean geometries is what is necessary to implement the membrane paradigm in the AdS/CFT correspondence.

The idea of implementing a Galilean limit of AdS/CFT correspondence was proposed recently in \cite{Bagchi:2009my} who were also motivated by considerations involving the incompressible Navier-Stokes equations, which turn out to enjoy an enhanced symmetry algebra \cite{Gusyatnikova:1989nx}. It was  argued there based on this algebra that  appropriate Newton-Cartan manifold should correspond to an \AdS{2} slice of the bulk \AdS{d+1} spacetime. Here in contrast we are retaining the entire manifold, not just the two dimensional slice.  However, we achieve this at the expense of discarding the Lorentzian metric structure, and work with a co-metric and a time co-vector.  

Clearly the role of Galilean/Newton-Cartan structures in the context of AdS/CFT requires further investigation. The bulk geometries we construct here seem to provide an interesting interpolation between Lorentzian and Galilean structures somewhat reminiscent of the Galilean limit procedure described in \cite{Kunzle:1976fk,Gonzalez:2003fk}.  It is also tempting to contemplate the possibility that apart from the context discussed herein, these structures  could also provide useful clues on how to work with the gravity duals of non-relativistic field theories. In the context of Schr\"odinger spacetimes which were proposed as duals to non-relativistic conformal field theories \cite{Son:2008ye,Balasubramanian:2008dm}
the Galilean limit is usually implemented as a DLCQ limit which can naturally be studied in  the Newton-Cartan language.

There are many directions in which the constructions we have described here can be generalized. For one it would be very useful to understand the gravitational Dirichlet problem outside of the long wavelength regime and to investigate its well-posedness. Further it would be useful to flesh out in precise detail the connections (if any) between the ideas around holographic RG flows and the Dirichlet problem and ask whether one can use the ideas developed here to implement the Wilsonian holographic flow of \cite{Heemskerk:2010hk,Faulkner:2010jy} in the non-linear gravity context. 

Even within the long-wavelength regime there are tantalizing similarities with the blackfold approach \cite{Emparan:2009at,Emparan:2011br}, where one considers hypersurfaces which have intrinsic as well as extrinsic dynamics. Freezing out the extrinsic dynamics should lead one to the gravitational Dirichlet problem (for instance in the analysis of the black string and membrane instabilities \cite{Camps:2010br} the extrinsic dynamics was frozen and the blackfold equations are precisely those of fluid dynamics). While the blackfold analysis is generically well suited for co-dimension three of higher hypersurfaces, it appears that one could recover some of the results discussed herein by extrapolating the equations to a co-dimension one hypersurface.\footnote{We thank Roberto Emparan for useful discussions on this issue.}

Of more immediate interest is to complete the derivation of the hypersurface dynamics from that on the boundary to higher orders in the gradient expansion. In particular, while we have shown that the ideal fluid equations on $\Sigma_D$ arise from those on the boundary, moving to second order in gradients in the conservation equation will allow us to show at the non-linear level that the shear viscosity of the hypersurface fluid is the same as that on the boundary. This statement would establish the non-linear version of the non-renormalisation of $\eta$ (or equivalently $\eta/s$ since we know that the entropy density being associated with the horizon remains unchanged) complementing the earlier  analysis of \cite{Iqbal:2008by} who showed this in the regime of linearized hydrodynamics using a flow equation. One would hope to also understand by this study why the bulk viscosity term on the hypersurface vanishes despite the fluid being non-conformal and having a non-vanishing trace for the stress tensor. At the same time it would be interesting to understand how the higher order transport coefficients evolve from the boundary to $\Sigma_D$ which should be possible to examine within our framework with a little bit of work.

Another interesting avenue for exploration is to ask about the gravitational Dirichlet problem in the presence of degenerate horizons. All of the discussion in the present paper and indeed in the fluid/gravity correspondence has focussed on situations of thermal equilibrium at non-zero temperature and hence one naturally studies spacetimes with non-degenerate horizons. Degenerate horizons pose an intriguing challenge and it would be useful to understand how the boundary theory (which should be more than just a fluid dynamical system) reorganizes itself as we pass over to the hypersurface description. It would also be interesting to extend our considerations to stationary black holes where it would be useful to understand the Dirichlet dynamics when the hypersurface is inside the ergorsurface (as defined by the asymptotic observer). 

Finally, another interesting avenue for contemplation is whether the ideas discussed herein can be ported to the context of brane-worlds, where we have gravity induced on a cut-off surface in AdS. Recently \cite{Figueras:2011gd}  attempts to derive the brane gravitational equations by working on a hypersurface near the boundary of AdS and invoking the Fefferman-Graham expansion.  It is clear from our discussion that no matter what boundary condition we choose on the hypersurface, it is most likely to involve non-local deformations on the dual boundary field theory. What this implies for induced gravity scenarios is an issue that deserves further study.

\subsection*{Acknowledgements}
\label{s:acks}
We would especially like to thank Sayantani Bhattacharyya
for collaborating with us at the various steps of this
paper. It is a pleasure to thank Jyotirmoy Bhattacharya,  
Amar V Chandra, Atish Dabholkar, Roberto Emparan, Sean Hartnoll, 
Veronika Hubeny, Nabil Iqbal, Cynthia Keeler, Vijay Kumar, Hong Liu,
Donald Marolf, Shiraz Minwalla, Ricardo Monteiro, Niels Obers, David Poland,
Suvrat Raju, Andrew Strominger and Piotr Surowka for extremely useful 
discussions on ideas presented in this paper.  We further would like to thank 
Roberto Emparan, Rajesh Gopakumar, Cynthia Keeler, Hong Liu, Donald Marolf and Andrew Strominger
for comments on a draft version of the manuscript. RL and MR would 
like to thank ICTS, TIFR for wonderful hospitality during 
the Applied AdS/CFT workshop. MR in addition would also like
to thank HRI and University of Amsterdam for their kind 
hospitality. DB is supported by a STFC studentship, while
JC, MR are supported by a STFC Rolling grant. RL is 
supported by the Harvard Society of Fellows through a junior fellowship.
Finally, RL would  like to thank various colleagues at the 
society for interesting discussions.

\appendix

\section{Notation}
\label{app:notation}

We work in the $(-++\ldots)$ signature. The dimensions of the spacetime in which the conformal fluid lives is denoted by $d$ . We usually assume $d>2$ unless otherwise specified. In the context of AdS/CFT, the dual AdS$_{d+1}$ space has $d+1$ spacetime dimensions.The lower-case Greek indices $\mu,\nu= 0,1,\ldots,d-1$ are used as boundary space-time indices, whereas the upper-case Latin indices $A,B=0,1,\ldots, d$ are used as the bulk indices. The lower-case Latin
indices $i=1,\ldots,d-1$ index the different spatial directons at  the boundary.

Throughout this paper, we take the extra holographic co-ordinate to be $r$ 
with the boundary of the bulk spacetime at $r\rightarrow\infty$.

Among the objects carrying Greek indices the hatted variables belong to a 
hypersurface $r=r_D$ whereas the unhatted objects naturally belong to the
boundary $r=\infty$. Further we use $*$ to mark non-relativistic objects.

We use round brackets to denote symmetrisation and square brackets to denote antisymmetrisation. For example, $B_{(\mu\nu)}\equiv \frac{1}{2}\, \left(B_{\mu\nu}+B_{\nu\mu}\right)$ and $B_{[\mu\nu]}\equiv \frac{1}{2}\left( B_{\mu\nu}-B_{\nu\mu}\right)$. For tensor products we denote symmetrisation with an explicit subscript as $X \otimes_s Y = \frac{1}{2} (X \otimes Y + Y \otimes X)$.

\subsection{Quick reference table}
\label{s:tabsum}
We have included a table with other useful parameters used in the text. In the table~\ref{notation:tab}, the relevant equations are denoted by their respective equation numbers appearing inside parentheses.

\begin{table}[htp]
\label{notation:tab}
 \centering
 \begin{tabular}{||r|l||r|l||}
   \hline
   \multicolumn{4}{||c||}{\textbf{Table of Notation I}} \\
   \hline 
   Symbol & Definition & Symbol & Definition \\
   \hline
   \multicolumn{4}{||c||}{Bulk Spacetime} \\
   \hline
   $\mathcal{G}_{AB}$ & Bulk metric with  & $\mathcal{G}^{AB}$ & Bulk co-metric with \\
    & components $\mathfrak{u}_\mu,\mathfrak{V}_\mu,\mathfrak{G}_{\mu\nu}$ &  & components $\mathfrak{u}^\mu,\mathfrak{P}^{\mu\nu},\mathfrak{P}^{\mu\nu}\mathfrak{V}_\nu$ \\ 
   $G_{d+1}$ & Bulk Newton constant  &  $\Lambda_{d+1}$ & $-\frac{1}{2}d(d-1)$ Bulk C.C.\\
   $b$ & Inverse Horizon radius & $f(br)$ & $1-(br)^{-d}$ \\
   $F(br)$ & $\int_{br}^{\infty}\; \frac{y^{d-1}-1}{y(y^{d}-1)}dy$ &
    $\hat{F}(br)$ & $\frac{\left(F(br)-F(br_D)\right)}{\sqrt{f(br_D)}} $  \\
   $\hat{\xi}$ &$1-\frac{rf(br)}{2r_Df(br_D)}$ & & \\
   \hline
   \multicolumn{4}{||c||}{Boundary at $r=\infty$} \\
   \hline
   $g_{\mu\nu}$ & Boundary metric $[r^{-2}ds_{d+1}^2]_{r=\infty}$ & $g^{\mu\nu}$ & Boundary co-metric\\ 
   $u^\mu$ & Fluid velocity  & $P_{\mu\nu}$ & $g_{\mu\nu}+u_\mu u_\nu$ \\
   $p$ & Fluid pressure \eqref{enmom:eq} & $\varepsilon$ & Fluid energy density \eqref{enmom:eq} \\
   $b$ & $(4 G_{d+1} s)^{-\frac{1}{d-1}}$ & $s$ & Fluid entropy density \\
   $\zeta$& Bulk viscosity \eqref{enmom:eq} & $\eta$ & $\frac{s}{4\pi}$ Shear Viscosity \eqref{enmom:eq}  \\
   $\nabla$ & Christoffel covariant derivative & ${\Gamma}_{\mu\nu}{}^\lambda$ & Christoffel symbols \\
   $\tilde{\Gamma}_{\mu\nu}{}^\lambda$ & $\equiv\hat{\Gamma}_{\mu\nu}{}^\lambda-{\Gamma}_{\mu\nu}{}^\lambda$ 
      & $\mathcal{A}_\mu$ & Weyl-Connection \eqref{eqn:Adef} \\
   $\sigma_{\mu\nu}$ & Shear strain rate \eqref{eqn:sigdef} & $\omega_{\mu\nu}$ & Fluid vorticity \eqref{eqn:omdef} \\
   $a_{\mu}$ & Acceleration field \eqref{eqn:adef} & $\theta$ & expansion rate \eqref{eqn:thdef} \\
   \hline
   $c^2_{snd}$ &  \multicolumn{3}{c||}{Speed of sound-squared $\frac{\partial p}{\partial\varepsilon}=\frac{1}{d-1}$}\\
   \hline
   \multicolumn{4}{||c||}{Dirichlet Hypersurface at $r=r_D$ (Dressed dynamics)} \\
   \hline
   $\hat{g}_{\mu\nu}$ & Dirichlet metric $[r^{-2}ds_{d+1}^2]_{r=r_D}$ & $\hat{g}^{\mu\nu}$ & Dirichlet co-metric\\ 
   $\hat{u}^\mu$ & Fluid velocity  & $\hat{P}_{\mu\nu}$ & $\hat{g}_{\mu\nu}+\hat{u}_\mu \hat{u}_\nu$ \\
   $\hat{p}$ & Fluid pressure \eqref{enmomh:eq} & $\hat{\varepsilon}$ & Fluid energy density \eqref{enmomh:eq} \\
   $\hat{\alpha}$ & $(1-(br_D)^{-d})^{-1/2}=f^{-1/2}(br_D)$ & $\hat{s}$ & Fluid entropy density \\
   $\hat{\zeta}$& Bulk viscosity \eqref{enmom:eq} & $\hat{\eta}$ & $\frac{\hat{s}}{4\pi}$ Shear Viscosity \eqref{enmom:eq}  \\
   $\hat{\nabla}$ & Christoffel covariant derivative & $\hat{\Gamma}_{\mu\nu}{}^\lambda$ & Christoffel symbols \\
   $\tilde{\Gamma}_{\mu\nu}{}^\lambda$ & $\equiv\hat{\Gamma}_{\mu\nu}{}^\lambda-{\Gamma}_{\mu\nu}{}^\lambda$ &
     $\hat{\mathcal{A}}_\mu$ & Weyl-Connection    \\
   $\hat{\sigma}_{\mu\nu}$ & Shear strain rate & $\hat{\omega}_{\mu\nu}$ & Fluid vorticity  \\
   $\hat{a}_{\mu}$ & Acceleration field  & $\hat{\theta}$ & expansion rate  \\
   \hline 
   $r_{D,snd}$ & \multicolumn{3}{c||}{radius where dressed sound becomes superluminal
  w.r.t $\hat{g}$ i.e., where $\hat{c}_{snd}=1$ }\\
   \hline
   $\hat{c}^2_{snd}$ & \multicolumn{3}{c||}{Speed of sound-squared $\frac
  {\partial \hat{p}}{\partial\hat{\varepsilon}}=\frac{1}{d-1}\left[1+\frac{d}{2}(\hat{\alpha}^2-1)\right]$ } \\
   \hline
\end{tabular}
\end{table}

\begin{table}[h]
\label{notation2:tab}
 \centering
 \begin{tabular}{||r|l||r|l||}
   \hline
   \multicolumn{4}{||c||}{\textbf{Table of Notation II}} \\
   \hline 
   Symbol & Definition & Symbol & Definition \\
   \hline
   \multicolumn{4}{||c||}{Incompressible Limit ala BMW(Reference \cite{Bhattacharyya:2008kq}) (Add hats for fluid at $r=r_D$)} \\
   \hline
   $\aleph^{-1}$ & \multicolumn{3}{c||}{BMW expn. parameter with $p_*,b_*,\partial_t\sim\aleph^{-2}$ 
   and $v_*^i,\nabla_i\sim\aleph^{-1}$ }\\
   \hline
   $g^{(0)}_{ij}$ & Spatial metric backgnd. & $g^{ij}_{(0)}$ & Spatial co-metric backgnd.\\ 
   $h_{tt}^*,h_{ij}^*$ &  $\delta g_{tt},\delta g_{ij}\sim \aleph^{-2}$ & $k_i^*$ & $\delta g_{ti}\sim \aleph^{-1}$\\
   $\nabla^{(0)}_i$ & Covariant derivative using $g^{(0)}_{ij}$ &  $q^*_{ij}$ & $\nabla^{(0)}_i  k_j^*-\nabla^{(0)}_j  k_i^*$ \\
   $v_*^i$ & Fluid velocity $\frac{u^i}{u^t}\sim  \aleph^{-1}$ & $\rho_0$ & Mass density $\varepsilon_0+p_0$ \\
   $b_0,\hat{\alpha}_0,\ldots$ & Backgnd. values & $\delta b_*,\hat{\alpha}_*,\ldots$ & Variation-typically $\sim \aleph^{-2} $ \\  
   $p_0$ & Backgnd. pressure & $\varepsilon_0$ & Backgnd. energy density \\
   $p_*$ & Pressure/mass density   
    & $\nu_0$ & Kinematic viscosity $\frac{\eta_0}{\rho_0}$  \\
   &$\frac{p-p_0}{\rho_0}\sim c_{snd}^2\frac{T_*}{T_0}\sim \aleph^{-2} $ & $\hat{\xi}_0$ &$1-\frac{rf(b_0r)}{2r_Df(b_0r_D)}$\\
   \hline
   \multicolumn{4}{||c||}{Incompressible Limit ala BKLS (References \cite{Bredberg:2010ky,Bredberg:2011jq,Compere:2011dx}) at $r=r_D$ (with $r_D$ near Horizon)} \\
   \hline
   $\hat{\varkappa}^{-1}$ & \multicolumn{3}{c||}{BKLS expn. parameter with $\hat{p}_\star,\partial_t\sim\hat{\varkappa}^{-2}$ 
   and $\hat{v}_\star^i,\nabla_i\sim\hat{\varkappa}^{-1}$.} \\
   & \multicolumn{3}{c||}{ Further $b_\bullet,\hat{\alpha}_\bullet,\hat{c}_{snd}\sim \hat{\varkappa}$ and
   $b_\star \sim \hat{\varkappa}^{-3}$ }\\
   \hline
   $\hat{g}^{(\bullet)}_{ij}$ & Spatial metric backgnd. & $\hat{g}^{ij}_{(\bullet)}$ & Spatial co-metric backgnd.\\ 
   $\hat{\nabla}^{(\bullet)}_i$ & Covariant derivative using $\hat{g}^{(\bullet)}_{ij}$ & $\rho_\bullet$ & Mass density\\
   $\hat{v}_\star^i$ & Fluid velocity $\frac{\hat{u}^i}{\hat{u}^t}\sim  \hat{\varkappa}^{-1}$ & & $\varepsilon_\bullet+p_\bullet$ $\sim \hat{\varkappa}^{-(d-1)} $ \\
   $b_\bullet,\hat{\alpha}$ & Backgnd. values $\sim \hat{\varkappa}$ & $\delta b_\star$ & Variation- $\sim \hat{\varkappa}^{-3} $ \\  
   $\hat{p}_\bullet$ & Backgnd. pressure $\sim \hat{\varkappa}^{-(d-1)} $ & $\hat{\varepsilon}_\bullet$ & Backgnd. energy density $\sim \hat{\varkappa}^{-d} $ \\
   $\hat{p}_\star$ & Pressure/mass density   
    & $\hat{\nu}_\bullet$ & Kinematic viscosity $\frac{\hat{\eta}_\bullet}{\hat{\rho}_\bullet}$  \\
    & $\frac{\hat{p}-\hat{p}_\bullet}{\hat{\rho}_\bullet}\sim \hat{c}_{snd}^2\frac{\hat{T}_*}{\hat{T}_0}\sim \hat{\varkappa}^{-2} $ & $\rho_D$ & $\rho$ co-ordinate of Dirichlet surface \\ \hline
    $\rho$ & \multicolumn{3}{|c||}{Near-horizon radial co-ordinate defined via $r =  \frac{1}{\hat{\varkappa}\, b_\bullet} \left(1 + \frac{\rho}{\hat{\varkappa}^2 \hat{\alpha}_\bullet b_\bullet}\right)$ } \\
   \hline
\end{tabular}
\end{table}

\subsection{Notation in the Bulk}

We denote the bulk metric by $\mathcal{G}_{AB}$. The inverse metric in the bulk 
(we will call this the co-metric - since it is the metric on the cotangent bundle)
is denoted by $\mathcal{G}^{AB}$.

We take the bulk AdS radius to be unity which is equivalent to setting the bulk cosmological constant to be $\Lambda_{d+1}=-\frac{d(d-1)}{2}$. We denote the bulk Newton constant by $G_{d+1}$. For the ease of reference, we now give the value of $G_{d+1}$ for some of the well-known CFTs with gravity duals (see \cite{Maldacena:1997re,Aharony:1999ti,Aharony:2008ug} for further details):
\begin{enumerate}
\item The d=4, $\mathcal{N}$=4 Super Yang-Mills theory on $N_c$ D3-Branes with a gauge group SU(N$_c$) and a `t Hooft coupling $\lambda\equiv g_{YM}^2 N_c$ is believed to be dual to IIB string theory on  AdS$_5\times$S$^5_{R=1}$ with  $G_{5}=\pi /(2 N_c^2)$ and $\alpha'=(4\pi\lambda)^{-1/2} $.
\item A d=3, $\mathcal{N}$=6 Superconformal \footnote{In the case of $k=1,2$, the supersymmetry should get enhanced to d=3, $\mathcal{N}$=8.} Chern-Simons theory on $N_c$ M2-Branes with a gauge group U(N$_c$)$_k\times$ U(N$_c$)$_{-k}$ (where the subscripts denote the Chern-Simons couplings) and a `t Hooft coupling $\lambda\equiv N_c/k$ is conjectured  to be dual to M-theory on AdS$_4\times$S$^7_{R=2}$/Z$_k$ with $G_{4}=N_c^{-2}\sqrt{9\lambda/8}=3k^{-1/2}(2N_c)^{-3/2}$.
\item A d=6, $\mathcal{N}$=(2,0) superconformal theory on  $N_c$ M5-Branes is conjectured  to be dual to M-theory on AdS$_7\times$S$^4_{R=1/2}$ with $G_{7}=3\pi^2/(16 N_c^{3})$.
\end{enumerate}

The general bulk metric which is Weyl-covariant is given by 
\begin{equation} ds^2 =\mathcal{G}_{AB}dx^Adx^B= -2\,{\mathfrak u}_\mu dx^\mu\left(dr+r \,{\mathfrak V} _\nu dx^\nu\right) + r^2\, {\mathfrak G}_{\mu\nu} dx^\mu dx^\nu \end{equation}
which is invariant under the boundary Weyl transformations
\begin{equation}\left\{r,{\mathfrak u}_\mu, {\mathfrak V}_\nu,{\mathfrak G}_{\mu\nu} \right\} \mapsto \left\{e^{-\phi}r,e^{\phi}{\mathfrak u}_\mu, {\mathfrak V}_\nu+\partial_\nu\phi,e^{2\phi}{\mathfrak G}_{\mu\nu} \right\} \end{equation} 
where $\phi=\phi(x)$ is an arbitrary function at the boundary. Without loss of generality assume that ${\mathfrak G}_{\mu\nu}$ is transverse to ${\mathfrak u}_\mu$, i.e. , ${\mathfrak G}_{\mu\nu}\mathfrak{u}^\nu=0$. Further we have 
\begin{equation}
{\mathfrak u}_\mu = \mathfrak{u}_\mu(x), \quad {\mathfrak V}_\nu={\mathfrak V}_\nu(r,x),\quad {\mathfrak G}_{\mu\nu}={\mathfrak G}_{\mu\nu}(r,x)\end{equation}
We will raise/lower/contract the unhatted greek indices using the boundary metric
\begin{equation} g_{\mu\nu} dx^\mu dx^\nu \equiv [r^{-2}ds^2]_{r\to \infty} = \left[{\mathfrak G}_{\mu\nu}-\frac{2}{r}{\mathfrak u}_{(\mu} {\mathfrak V}_{\nu)} \right]_{r\to \infty}  dx^\mu dx^\nu \end{equation}
and the velocity field $\mathfrak{u}_\mu$ is a unit time-like vector of this metric
\begin{equation} \mathfrak{u}^\mu \mathfrak{u}_\mu \equiv g^{\mu\nu}\mathfrak{u}_\mu \mathfrak{u}_\nu = -1 \end{equation}

A Weyl-covariant basis for the bulk cotangent bundle is 
given by $\{dr+r\,{\mathfrak V}_\nu dx^\nu\ ,dx^\mu\}$ with a corresponding dual 
basis $\left\{\partial_r\ , \partial_\mu - r\,{\mathfrak V}_\mu\partial_r\right\}$. In this dual 
basis, the co-metric (or the inverse metric) is given by 
\begin{equation}
\begin{split}
\mathcal{G}^{AB}&\partial_A\otimes\partial_B \\
&=2\, {\mathfrak u}^\mu\left[\partial_\mu - r\,{\mathfrak V}_\mu\partial_r\right]\otimes_s\partial_r + r^{-2}\,\mathfrak{P}^{\mu\nu}\left[\partial_\mu - r\,{\mathfrak V}_\mu\partial_r\right]\otimes\left[\partial_\nu - r\,{\mathfrak V}_\nu\partial_r\right] \\
\end{split} 
\end{equation}
where $\mathfrak{P}^{\mu\nu}$ is the unique transverse tensor that satisfies
\begin{equation} \mathfrak{P}^{\mu\nu} {\mathfrak u}_\nu=0 \quad\text{and}\quad \mathfrak{P}^{\mu\lambda}{\mathfrak G}_{\lambda\nu}= \delta^\mu_\nu + {\mathfrak u}^\mu \, {\mathfrak u}_\nu \end{equation}

The unit normal vector of a hypersurface $r=r_D$ is given by\footnote{The corresponding problem
 for general function $r=r_D(x)$ can be reduced to this
problem by a suitable choice of Weyl frame.}
\[ n_A dx^A = \frac{dr}{\sqrt{\mathcal{G}^{rr}}} = \frac{dr}{\sqrt{\mathfrak{P}^{\alpha\beta}{\mathfrak V}_\alpha{\mathfrak V}_\beta-2r u^\alpha{\mathfrak V}_\alpha}} \]
\begin{equation*}
\begin{split}
n^A \partial_A &= \frac{\mathcal{G}^{rr}\partial_r+\mathcal{G}^{r\mu}\partial_\mu}{\sqrt{\mathcal{G}^{rr}}}\\
&= \frac{\left(\mathfrak{P}^{\alpha\beta}\,{\mathfrak V}_\alpha\,{\mathfrak V}_\beta-2\,r \,{\mathfrak u}^\alpha\,{\mathfrak V}_\alpha\right)\partial_r+\left({\mathfrak u}^\mu-r^{-1}\,\mathfrak{P}^{\mu\alpha}\, {\mathfrak V}_\alpha\right)\partial_\mu}{\sqrt{\mathfrak{P}^{\alpha\beta}\,{\mathfrak V}_\alpha{\mathfrak V}_\beta-2\,r\, {\mathfrak u}^\alpha\,{\mathfrak V}_\alpha}}\\
&= \frac{-r\,{\mathfrak u}^\alpha\,{\mathfrak V}_\alpha\partial_r+\left({\mathfrak u}^\mu-r^{-1}\,\mathfrak{P}^{\mu\alpha}\,{\mathfrak V}_\alpha\right)\left[\partial_\mu - r\,{\mathfrak V}_\mu\partial_r\right]}{\sqrt{\mathfrak{P}^{\alpha\beta}\,{\mathfrak V}_\alpha\,{\mathfrak V}_\beta-2\,r\, {\mathfrak u}^\alpha{\mathfrak V}_\alpha}}\\
\end{split} 
\end{equation*} 

From these expressions, it follows that the normalized induced metric
and co-metric on
a hypersurface $r=r_D$  is given by
\begin{equation}
\begin{split}
\hat{g}_{\mu\nu} &\equiv \left\{ r^{-2}\mathcal{G}_{\mu\nu}
\right\}_{r\to r_D}\\
&=  \left\{ \mathfrak{G}_{\mu\nu} -
\frac{2}{r}\mathfrak{u}_{(\mu}\mathfrak{V}_{\nu)} \right\}_{r\to r_D} \\
\hat{g}^{\mu\nu} &\equiv \left\{
r^2\left(\mathcal{G}^{\mu\nu} - \frac{\mathcal{G}^{\mu
r}\mathcal{G}^{r\nu}}{\mathcal{G}^{rr}}\right)\right\}_{r\to r_D}\\
&= \left\{ \mathfrak{P}^{\mu\nu} -
\frac{\left[r \, {\mathfrak u}^\mu-\mathfrak{P}^{\mu\alpha}\mathfrak{V}_\alpha\right]\left[r\, 
{\mathfrak u}^\nu-\mathfrak{P}^{\nu\beta}\mathfrak{V}_\beta\right]}{\mathfrak{P}^{\alpha\beta}\mathfrak{V}_\alpha\mathfrak{V}_\beta-2\, r\, 
{\mathfrak u}^\alpha\mathfrak{V}_\alpha}\right\}_{r\to r_D}
\end{split}
\end{equation}
The hatted greek indices are raised/lowered/contracted using this hatted metric/co-metric.

\subsection{Notation at the Boundary}

The metric on the $d$ dimensional boundary is denoted by $g_{\mu\nu}$ which is a representative
the class of metrics on the conformal boundary of the bulk spacetime. 
\begin{equation} g_{\mu\nu}\equiv \left\{r^{-2}ds^2\right\}_{r\to \infty} \end{equation}
The inverse of this metric (we will call this the co-metric -- since it is the metric
on the cotangent bundle) is denoted by $g^{\mu\nu}$. We denote with $\nabla$ the 
corresponding Christoffel connection/covariant derivative. Our conventions for 
Christoffel symbols and the curvature tensors are fixed by the relations
\begin{equation}
\begin{split}
\nabla_{\mu}V^{\nu}&=\partial_{\mu}V^{\nu}+\Gamma_{\mu\lambda}{}^{\nu}V^{\lambda} \qquad \text{and}\qquad [\nabla_\mu,\nabla_\nu]V^\lambda=-R_{\mu\nu\sigma}{}^{\lambda}V^\sigma .
\end{split}
\end{equation}

On this spacetime lives a conformal fluid with velocity field $u^\mu$ (with $u^\mu u_\mu =-1$) ,
pressure $p$, energy density $\varepsilon=(d-1)p$, and shear viscosity $\eta$.
We introduce the projector $P_{\mu\nu}\equiv g_{\mu\nu}+u_\mu u_\nu$
which projects onto the space transverse to $u^\mu$. 

The gradients of the velocity field are decomposed as:
\begin{equation}
\begin{split}
\nabla_\mu u_\nu &= \sigma_{\mu\nu}+\omega_{\mu\nu}-u_\mu a_\nu + \frac{\theta}{d-1} P_{\mu\nu}\\
&= \sigma_{\mu\nu}+\omega_{\mu\nu}-u_\mu \mathcal{A}_\nu + \frac{\theta}{d-1} g_{\mu\nu}\\
\end{split}
\end{equation}
where we have introduced 
\begin{eqnarray}
&&
\bullet \; \text{the shear strain rate:} \qquad  \sigma_{\mu\nu}\equiv \left[P_{(\mu}^\alpha P_{\nu)}^\beta- \frac{P_{\mu\nu}}{d-1}P^{\alpha\beta} \right] \nabla_\alpha u_\beta \label{eqn:sigdef}  \\
&&
\bullet \; \text{the vorticity:} \qquad \qquad \quad \ \;
\omega_{\mu\nu}\equiv  P_{[\mu}^\alpha P_{\nu]}^\beta \nabla_\alpha u_\beta 
\label{eqn:omdef} \\ 
&& 
\bullet \; \text{the acceleration field:} \qquad \  \,a_\mu \equiv  u_\alpha\nabla^\alpha u_\mu
\label{eqn:adef}  \\
&&
 \bullet \; \text{the expansion rate:} \qquad  \qquad \theta \equiv \nabla_\alpha u^\alpha 
\label{eqn:thdef} 
 \end{eqnarray}
The hydrodynamic Weyl-connection (see \cite{Loganayagam:2008is} where it was introduced for more details) is defined to be
\begin{equation}
\mathcal{A}_\mu \equiv  u_\alpha\nabla^\alpha u_\mu - \frac{\nabla_\alpha u^\alpha}{d-1} u_\mu
\label{eqn:Adef}
\end{equation}	
The bulk metric-dual for hydrodynamics is given by (see \cite{Bhattacharyya:2008mz}) 
\begin{equation}
\begin{split}
\mathfrak{u}_\mu &= u_\mu\ ,\quad\ \mathfrak{G}_{\mu\nu} =  P_{\mu\nu} +2b\, F(br)\,  \sigma_{\mu\nu}+\ldots \\
\mathfrak{V}_\mu &=  \mathcal{A}_\mu+\frac{r}{2}\, (1-(br)^{-d})\, u_\nu +\ldots \\
\mathfrak{u}^\mu &= u^\mu\ ,\quad\mathfrak{P}^{\mu\nu} = P^{\mu\nu}   -2b \,F(br)\,  \sigma^{\mu\nu}+\ldots  \\
\end{split}
\end{equation}
where
\begin{equation}
\hat{F}(br) \equiv \frac{1}{\hat{\alpha}}\left(F(br)-F(br_D)\right) = \frac{1}{\hat{\alpha}} \;  \int_{br}^{br_D}\; \frac{y^{d-1}-1}{y(y^{d}-1)}dy\,.
\end{equation}

The energy-momentum  tensor of a general relativistic fluid (till first order in the 
gradient expansion) is given as
\begin{equation}\label{enmom:eq}
T^{\mu\nu} \equiv (\varepsilon+p)\, u^\mu\, u^\nu + p\, g^{\mu\nu}- 2\,\eta\, \sigma^{\mu\nu}-\zeta\, \theta\, P^{\mu\nu}+\ldots 
\end{equation}
which can be computed from the bulk data (of the metric dual to hydrodynamics) via
\begin{equation}\label{eq:BYT}
T_{\mu\nu} \equiv \left\{\frac{r^d}{16\pi \,G_{d+1}}\left[-2\,{K}_{\mu\nu}+2\,K\,{g}_{\mu\nu}-2\,(d-1)\,{g}_{\mu\nu}+\ldots \right]\right\}_{r\to\infty}
\end{equation}
where $r^2 \,K_{\mu\nu}$ is the extrinsic curvature of the constant $r$ hypersurface
and $K\equiv g^{\mu\nu}\, K_{\mu\nu}$ is its trace. This computation gives
\begin{equation}
p=\frac{\varepsilon}{d-1} = \frac{1}{16\pi G_{d+1}} \; \frac{1}{b^d} ,\quad\ \eta=\frac{1}{16\pi G_{d+1}}\; \frac{1}{b^{d-1}} \quad\text{and}\quad \zeta =0 
\end{equation}
and the Bekenstein-Hawking argument in the bulk\footnotemark\ gives the entropy density and the
temperature of this fluid as
\begin{equation}
s= \frac{1}{4\, G_{d+1}}\; \frac{1}{b^{d-1}} ,\quad\text{and}\quad T=\frac{d}{4\pi\, b}  
\end{equation}
where we have found it convenient to introduce a variable $b\equiv (4 G_{d+1} s)^{-\frac{1}{d-1}}$.

\footnotetext{Note that the above forms can be deduced upto a normalization constant from
just the thermodynamic relations
\begin{equation}
\frac{ds}{s}= \frac{d\varepsilon}{\varepsilon+p} ,\quad \frac{dT}{T}= \frac{dp}{\varepsilon+p} \quad\text{and}\quad \varepsilon+p = T s 
\end{equation}
So the Bekenstein-Hawking argument  is needed just to fix the normalization  constant.}

\subsection{Notation on the Dirichlet Hypersurface}

We choose a hypersurface $r=r_D$ in the bulk to impose Dirichlet boundary condition. We will work
in a boundary Weyl-frame where $r_D$ is independent of $x$. This Weyl-frame change is
consistent with  the gradient expansion in the boundary hydrodynamics provided the initial
surface $r=\rho(x)$ had a slowly varying $\rho(x)$. We will introduce a parameter 
\begin{equation} \hat{\alpha} \equiv \frac{1}{\sqrt{1-(b r_D)^{-d}}} \end{equation}
which parameterizes how far the Dirichlet surface is to the boundary/ how close it is to 
the horizon. We have  $\hat{\alpha}(r_D=\infty) =1$ and $\hat{\alpha}(r_D=1/b) =\infty$.

After this we proceed as we did in the $r=\infty$ case in the previous subsection. All the 
same definitions can be repeated - we will just distinguish the objects in the Dirichlet 
hypersurface by a hat - so we have $\hat{u}^\mu ,\ \hat{g}^{\mu\nu} ,\ \hat{p}$ and so on.
Unless specified, all the hatted tensors are raised and lowered by the hatted metric/co-metric.

The energy momentum tensor is calculated using the same expression as in equation \eqref{eq:BYT}
except that we evaluate  it at $r=r_D$ now. We get 
\begin{equation}\label{enmomh:eq}
\hat{T}^{\mu\nu} \equiv (\hat{\varepsilon}+\hat{p})\,\hat{u}^\mu \,\hat{u}^\nu + \hat{p}\, \hat{g}^{\mu\nu}- 2\,\hat{\eta}\, \hat{\sigma}^{\mu\nu}-\hat{\zeta} \,\hat{\theta}\, \hat{P}^{\mu\nu}+\ldots 
\end{equation}
where 
\begin{equation}\label{enmomh2:eq}
\begin{split}
\hat{\varepsilon} &\equiv \frac{(d-1)}{8\pi \,G_{d+1}}\; \frac{\hat{\alpha}}{\hat{\alpha}+1} ;\; \frac{1}{b^d}=  \frac{2\hat{\alpha}}{\hat{\alpha}+1}\ \varepsilon\\
\hat{p} &\equiv \frac{\left[1+\frac{d}{2}(\hat{\alpha}-1)\right]}{8\pi\, G_{d+1}}\; \frac{\hat{\alpha}}{\hat{\alpha}+1} \; \frac{1}{b^d}=  \frac{2\hat{\alpha}}{\hat{\alpha}+1}\left[1+\frac{d}{2}(\hat{\alpha}-1)\right]p\\
\hat{\eta} &\equiv \frac{1}{16\pi \,G_{d+1}}\; \frac{1}{b^{d-1}}=\eta \quad\text{and}\quad \hat{\zeta} \equiv 0 =\zeta\\  
\end{split}
\end{equation}
%

\section{Dictionary for the Dirichlet Problem}
\label{A:dirdict}

In this subsection, we collect the formulae which translate between the boundary data and the Dirichlet data
to serve as a ready reference for the reader. The (normalized) metric and the co-metric on 
the Dirichlet hypersurface are given by
\begin{equation}
\begin{split}
\hat{g}_{\mu\nu}&= g_{\mu\nu}   +  \left(1-\frac{1}{\hat{\alpha}^2}\right)u_\mu u_\nu+2b  F(br_D)\  \sigma_{\mu\nu} -\frac{1}{r_D}\left[u_\mu \mathcal{A}_\nu+\mathcal{A}_\mu u_\nu \right]+\ldots\\
&= g_{\mu\nu}   + \left(1-\frac{1}{\hat{\alpha}^2}+\frac{2\theta}{(d-1)r_D}\right)u_\mu u_\nu-\frac{1}{r_D}\left[u_\mu a_\nu+a_\mu u_\nu \right] +2b  F(br_D)\  \sigma_{\mu\nu} +\ldots\\
\hat{g}^{\mu\nu} &= P^{\mu\nu}-2b F(br_D)\  \sigma^{\mu\nu} -\hat{\alpha}^2\left[u^\mu - r_D^{-1}a^{\mu}\right]\left[u^\nu - r_D^{-1}a^{\nu}\right]\left[1+\frac{2\hat{\alpha}^2\theta}{(d-1)r_D}\right] \\
&= g^{\mu\nu}+\left(1-\hat{\alpha}^2-\frac{2\hat{\alpha}^4\theta}{(d-1)r_D}\right)u^\mu u^\nu \ -2b F(br_D)\  \sigma^{\mu\nu} +\frac{\hat{\alpha}^2}{r_D}\left[u^\mu a^\nu + a^{\mu}u^\nu\right]  \\
\end{split}
\end{equation}
and the correctly normalized velocities at the hypersurface are
\begin{equation}
\begin{split}
\hat{u}_\mu &= \frac{u_\mu}{\hat{\alpha}} + \frac{\hat{\alpha}}{r_D} \mathcal{A}_\mu = \left(1-\frac{\hat{\alpha}^2\theta}{r_D(d-1)}\right)\frac{u_\mu}{\hat{\alpha}} + \frac{\hat{\alpha}}{r_D} a_\mu \\
\hat{u}^\mu &=  \hat{\alpha}u^\mu\left(1+\frac{\hat{\alpha}^2\theta}{r_D(d-1)}\right) +\ldots \\
\end{split}
\end{equation}

From these it follows that the transverse projectors are related by
\begin{equation}
\begin{split}
\hat{P}_{\mu\nu} &= P_{\mu\nu} + 2b F(br_D)\sigma_{\mu\nu}\\
\hat{P}^\mu{}_\nu &= {P}^\mu{}_\nu+\frac{\hat{\alpha}^2}{r_D}u^\mu a_\nu \\
\hat{P}^{\mu\nu} &= {P}^{\mu\nu} -2b F(br_D) \sigma^{\mu\nu}+\frac{\hat{\alpha}^2}{r_D}\left[u^\mu a^\nu + a^{\mu}u^\nu\right]\\
\end{split}
\end{equation}

Given  a relation of the form $\hat{U}_\mu=V_\mu$, we can always write
\begin{equation}
\begin{split}
\hat{\nabla}_\mu \hat{U}_\nu = \nabla_\mu V_\nu -\tilde{\Gamma}_{\mu\nu}{}^\rho V_\rho
\end{split}
\end{equation}
which defines the tensor $\tilde{\Gamma}_{\mu\nu}{}^\rho=\hat{\Gamma}_{\mu\nu}{}^\rho-{\Gamma}_{\mu\nu}{}^\rho$. We evaluate this tensor in the 
\App{app:CovDeriv} to get
\begin{equation}
\begin{split}
\tilde{\Gamma}_{\mu\nu}{}^\rho &= (\hat{\alpha}^2-1)\left[\sigma_{\mu\nu}+\frac{\theta}{d-1}\ \left(P_{\mu\nu}+\frac{d}{2}u_\mu u_\nu\right)-da_{(\mu}u_{\nu)}\right]u^\rho\\
&\qquad + \frac{\hat{\alpha}^2-1}{\hat{\alpha}^2}\left[-2\omega^\rho{}_{(\mu}u_{\nu)}+(\frac{d}{2}-1)u_{\mu}u_{\nu}a^\rho\right] \\
\end{split}
\end{equation}
in particular
\begin{equation}
\begin{split}
\tilde{\Gamma}_{\mu\nu}{}^\rho u_\rho 
&=-(\hat{\alpha}^2-1)\left[\sigma_{\mu\nu}+\frac{\theta}{d-1}\ \left(P_{\mu\nu}+\frac{d}{2}u_\mu u_\nu\right)-da_{(\mu}u_{\nu)}\right] \\
\end{split}
\end{equation}
From this it follows that 
\begin{equation}
\begin{split}
\hat{\sigma}_{\mu\nu} &= \hat{\alpha} \,\sigma_{\mu\nu}\ ,\quad \hat{\omega}_{\mu\nu} = \frac{1}{\hat{\alpha}} \,\omega_{\mu\nu} ,\quad \hat{\theta}=\hat{\alpha}\,\theta,\\
\hat{a}_\nu &=\left[1+\frac{d}{2}(\hat{\alpha}^2-1)\right]a_\nu ,\quad \hat{\mathcal{A}}_\nu =\mathcal{A}_\nu+\frac{d}{2}\,(\hat{\alpha}^2-1)\,a_{\nu} \\
\end{split}
\end{equation}
We refer the reader to the previous subsection for the dictionary involving the energy-momentum tensor.

Now we are ready to present the inverse relations. We will first invert the equations above to get
\begin{equation}
\begin{split}
{\sigma}_{\mu\nu} &= \frac{1}{\hat{\alpha}}\, \hat{\sigma}_{\mu\nu}\ ,\quad {\omega}_{\mu\nu} = {\hat{\alpha}} \,\hat{\omega}_{\mu\nu} ,\quad {\theta}=\frac{1}{\hat{\alpha}}\,\hat{\theta},\\
{a}_\nu &=\frac{\hat{a}_\nu}{\left[1+\frac{d}{2}(\hat{\alpha}^2-1)\right]} ,\quad {\mathcal{A}}_\nu =\hat{\mathcal{A}}_\nu-\frac{(\hat{\alpha}^2-1)\frac{d}{2}}{\left[1+\frac{d}{2}(\hat{\alpha}^2-1)\right]}\hat{a}_{\nu} =\frac{\hat{a}_\nu}{\left[1+\frac{d}{2}(\hat{\alpha}^2-1)\right]}-\frac{\hat{\theta}}{d-1}\hat{u}_\nu \\
\end{split}
\end{equation}
and then the velocities
\begin{equation}
\begin{split}
{u}_\mu &= \left(1+\frac{\hat{\alpha}\hat{\theta}}{r_D(d-1)}\right)\hat{\alpha}\,\hat{u}_\mu - \frac{\hat{\alpha}^2}{r_D} \frac{\hat{a}_\mu}{\left[1+\frac{d}{2}(\hat{\alpha}^2-1)\right]} \\
{u}^\mu &=  \frac{\hat{u}^\mu}{\hat{\alpha}}\left(1-\frac{\hat{\alpha}\,\hat{\theta}}{r_D\,(d-1)}\right) +\ldots \\
\end{split}
\end{equation}
followed by the projectors
\begin{equation}
\begin{split}
{P}_{\mu\nu} 
&= \hat{P}_{\mu\nu} - \frac{2b}{\hat{\alpha}} \,F(br_D)\,\hat{\sigma}_{\mu\nu}\\
{P}^\mu{}_\nu 
&= \hat{P}^\mu{}_\nu-\frac{\hat{\alpha}}{r_D}\frac{\hat{u}^\mu\hat{a}_\nu}{\left[1+\frac{d}{2}(\hat{\alpha}^2-1)\right]} \\
{P}^{\mu\nu} &= \hat{P}^{\mu\nu} +\frac{2b}{\hat{\alpha}} \,F(br_D)\hat{\sigma}^{\mu\nu}-\frac{\hat{\alpha}\left[\hat{u}^\mu \hat{a}^\nu + \hat{a}^{\mu}\hat{u}^\nu\right]}{r_D\left[1+\frac{d}{2}(\hat{\alpha}^2-1)\right]}\\
\end{split}
\end{equation}

The dictionary for the (normalized) metric and the co-metric are
\begin{equation}
\begin{split}
g_{\mu\nu} 
&= \hat{g}_{\mu\nu} + \left[1-\hat{\alpha}^2-\frac{2\hat{\alpha}^3\hat{\theta}}{r_D(d-1)}\right]\hat{u}_\mu \hat{u}_\nu+\frac{\hat{\alpha}^3\left[\hat{u}_\mu \hat{a}_\nu + \hat{a}_{\mu}\hat{u}_\nu\right]}{r_D\left[1+\frac{d}{2}(\hat{\alpha}^2-1)\right]}  - \frac{2b}{\hat{\alpha}} F(br_D)\hat{\sigma}_{\mu\nu} \\
g^{\mu\nu} 
&= \hat{g}^{\mu\nu}+ \left[1-\frac{1}{\hat{\alpha}^2}+\frac{2\hat{\theta}}{\hat{\alpha}r_D(d-1)}\right]\hat{u}^\mu \hat{u}^\nu+\frac{2b}{\hat{\alpha}} F(br_D)\hat{\sigma}^{\mu\nu}-\frac{\hat{\alpha}\left[\hat{u}^\mu \hat{a}^\nu + \hat{a}^{\mu}\hat{u}^\nu\right]}{r_D\left[1+\frac{d}{2}(\hat{\alpha}^2-1)\right]}\\
\end{split}
\end{equation}

Finally, we can write the tensor $\tilde{\Gamma}_{\mu\nu}{}^\rho $  in terms of hatted variables as
\begin{equation}
\begin{split}
\tilde{\Gamma}_{\mu\nu}{}^\rho 
&=  (1-\frac{1}{\hat{\alpha}^2})\left[ \hat{\sigma}_{\mu\nu}+\frac{\hat{\theta}}{d-1}\ \left(\hat{P}_{\mu\nu}+\frac{d}{2}\hat{\alpha}^2\hat{u}_\mu \hat{u}_\nu\right)-\frac{d\hat{\alpha}^2}{\left[1+\frac{d}{2}(\hat{\alpha}^2-1)\right]}\hat{a}_{(\mu}\hat{u}_{\nu)}\right]\hat{u}^\rho\\
&\qquad + (\hat{\alpha}^2-1)\left[-2\hat{\omega}^\rho{}_{(\mu}\hat{u}_{\nu)}+\frac{\frac{d}{2}-1}{\left[1+\frac{d}{2}(\hat{\alpha}^2-1)\right]}\hat{u}_{\mu}\hat{u}_{\nu}\hat{a}^\rho\right] \\
\end{split}
\end{equation}
In particular
\begin{equation}
\begin{split}
\tilde{\Gamma}_{\mu\nu}{}^\rho \hat{u}_\rho
&= - (1-\frac{1}{\hat{\alpha}^2})\left[ \hat{\sigma}_{\mu\nu}+\frac{\hat{\theta}}{d-1}\ \left(\hat{P}_{\mu\nu}+\frac{d}{2}\hat{\alpha}^2\hat{u}_\mu \hat{u}_\nu\right)-\frac{d\hat{\alpha}^2}{\left[1+\frac{d}{2}(\hat{\alpha}^2-1)\right]}\hat{a}_{(\mu}\hat{u}_{\nu)}\right]
\end{split}
\end{equation}

Now we present the Bulk metric/co-metric as a function of the Dirichlet data. The Bulk metric is given by
\begin{equation}\label{appeq:gbulk}
\begin{split}
\mathcal{G}_{AB}dx^A dx^B 
&=-2 u_\mu dx^\mu \left( dr + r\ \left[\mathcal{A}_\nu+\frac{r}{2}(1-(br)^{-d})u_\nu\right] dx^\nu \right) \\
& \qquad +  r^2 \left[ P_{\mu\nu} +2b F(br)\  \sigma_{\mu\nu}\right] dx^\mu dx^\nu +\ldots\\
&= -2 \left[\hat{\alpha}\hat{u}_\mu - \frac{\hat{\alpha}^2}{r_D}{\mathcal{A}}_\mu \right]dx^\mu\left( dr + r\ \left[
\hat{\xi}\mathcal{A}_\nu +\frac{r}{2}(1-(br)^{-d})\hat{\alpha}\hat{u}_\nu \right] dx^\nu \right)\\
& \qquad +  r^2 \left[ \hat{P}_{\mu\nu} +2b \hat{F}(br)\  \hat{\sigma}_{\mu\nu}\right] dx^\mu dx^\nu +\ldots\\
\end{split}
\end{equation}
where 
\begin{equation}\begin{split}
\mathcal{A}_\mu  &\equiv \frac{\hat{a}_\mu}{\left[1+\frac{d}{2}(\hat{\alpha}^2-1)\right]}-\frac{\hat{\theta}}{d-1}\hat{u}_\mu=\hat{\mathcal{A}}_\mu-\frac{(\hat{\alpha}^2-1)\frac{d}{2}}{\left[1+\frac{d}{2}(\hat{\alpha}^2-1)\right]}\hat{a}_{\mu}\\
\hat{\xi} &\equiv 1-\frac{1}{2}\frac{r}{r_D}\frac{(1-(br)^{-d})}{(1-(br_D)^{-d})}
= 1-\frac{\hat{\alpha}^2}{r_D}\frac{r}{2}(1-(br)^{-d})\\
\hat{F}(br) &\equiv \frac{1}{\hat{\alpha}}\left(F(br)-F(br_D)\right) = \frac{1}{\hat{\alpha}} \;  \int_{br}^{br_D}\; \frac{y^{d-1}-1}{y(y^{d}-1)}dy\,.
\end{split}\end{equation}
The Bulk co-metric is given by 
\begin{equation}
\begin{split}
\mathcal{G}^{AB}&\partial_A\otimes\partial_B \\
&= \left[r^2(1-(br)^{-d})-\frac{2r\theta}{d-1}\right]\partial_r\otimes\partial_r+2\left[u^\mu -r^{-1} a^{\mu}\right]\partial_\mu\otimes_s\partial_r\\
&\qquad +r^{-2}\left[P^{\mu\nu}   -2b F(br)\  \sigma^{\mu\nu}\right]\partial_\mu\otimes\partial_\nu\\
&= \left[r^2(1-(br)^{-d})-\frac{2r\hat{\theta}}{\hat{\alpha}(d-1)}\right]\partial_r\otimes\partial_r\\
&+2\left[\frac{\hat{u}^\mu}{\hat{\alpha}}\left(1-\frac{\hat{\alpha}\hat{\theta}}{r_D(d-1)}\right) - \frac{\hat{a}^\mu}{r\left[1+\frac{d}{2}(\hat{\alpha}^2-1)\right]}\right]\partial_\mu\otimes_s\partial_r\\
&\qquad +r^{-2}\left[\hat{P}^{\mu\nu} -\frac{\hat{\alpha}\left[\hat{u}^\mu \hat{a}^\nu + \hat{a}^{\mu}\hat{u}^\nu\right]}{r_D\left[1+\frac{d}{2}(\hat{\alpha}^2-1)\right]}   -2b \hat{F}(br)\   \hat{\sigma}^{\mu\nu}\right]\partial_\mu\otimes\partial_\nu
\end{split}
\end{equation}
In terms of the components in the Weyl-covariant basis, we have
\begin{equation}
\begin{split}
\mathfrak{u}_\mu &= u_\mu = \hat{\alpha}\, \hat{u}_\mu - \frac{\hat{\alpha}^2}{r_D}\left[\frac{\hat{a}_\mu}{\left[1+\frac{d}{2}(\hat{\alpha}^2-1)\right]}-\frac{\hat{\theta}}{d-1}\hat{u}_\mu\right] \\
\mathfrak{V}_\mu &=  \mathcal{A}_\mu+\frac{r}{2}\, (1-(br)^{-d})\, u_\nu \\
&= \hat{\xi}\left[\frac{\hat{a}_\mu}{\left[1+\frac{d}{2}(\hat{\alpha}^2-1)\right]}-\frac{\hat{\theta}}{d-1}\, \hat{u}_\mu\right] +\frac{r}{2}\, (1-(br)^{-d})\, \hat{\alpha}\,\hat{u}_\mu\\
\mathfrak{G}_{\mu\nu} &=  P_{\mu\nu} +2b\, F(br)\,  \sigma_{\mu\nu} = \hat{P}_{\mu\nu} +2b \,\hat{F}(br)\,  \hat{\sigma}_{\mu\nu}\\
\mathfrak{u}^\mu &= u^\mu = \frac{\hat{u}^\mu}{\hat{\alpha}}\left(1-\frac{\hat{\alpha}\,\hat{\theta}}{r_D(d-1)}\right)\\
\mathfrak{P}^{\mu\nu} &= P^{\mu\nu}   -2b \,F(br)\,  \sigma^{\mu\nu} = \hat{P}^{\mu\nu} -\frac{\hat{\alpha}\left[\hat{u}^\mu \hat{a}^\nu + \hat{a}^{\mu}\hat{u}^\nu\right]}{r_D\left[1+\frac{d}{2}(\hat{\alpha}^2-1)\right]}   -2b \,\hat{F}(br)\,   \hat{\sigma}^{\mu\nu} \\
\end{split}
\end{equation}

\section{Lorentz-Covariant derivative of the induced metric}
\label{app:CovDeriv}

We wish to find the tensor $\tilde{\Gamma}_{\mu\nu}{}^\rho$ which describes the difference
between the covariant derivatives at the boundary and the Dirichlet surface, i.e.,
Given  a relation of the form $\hat{U}_\mu=V_\mu$, we can always write
\begin{equation}
\begin{split}
\hat{\nabla}_\mu \hat{U}_\nu = \nabla_\mu V_\nu -\tilde{\Gamma}_{\mu\nu}{}^\rho V_\rho
\end{split}
\end{equation}
which defines the tensor $\tilde{\Gamma}_{\mu\nu}{}^\rho$. 

For definiteness, in this subsection we will continue to 
raise/lower/contract using the boundary metric $g_{\mu\nu}$ -
this means in particular raising/lowering/contracting do not commute with the
hatted covariant derivative $\hat{\nabla}$ so we need to be a bit careful. 

As usual, zero-torsion condition implies 
$\tilde{\Gamma}_{\mu\nu}{}^\rho=\tilde{\Gamma}_{\nu\mu}{}^\rho$ and metric compatibility
with $\hat{g}_{\mu\nu}\equiv g_{\mu\nu}+h_{\mu\nu}$ gives
\begin{equation} \tilde{\Gamma}_{\mu\nu}{}^\rho\hat{g}_{\rho\lambda}=\frac{1}{2} \left[\nabla_\mu h_{\lambda\nu}+\nabla_\nu h_{\lambda\mu}-\nabla_\lambda h_{\mu\nu} \right] \end{equation}
where
\[ h_{\mu\nu} = \left\{  \frac{u_\mu u_\nu}{(br)^d}+2b  F(br)\  \sigma_{\mu\nu} -\frac{1}{r}\left[u_\mu \mathcal{A}_\nu+\mathcal{A}_\mu u_\nu \right]+\ldots \right\}_{r\to r_D}\]
Since all our expressions are exact upto second derivatives, it is enough to work with just the
zero derivative piece in $h_{\mu\nu}$. 
\begin{equation}
\begin{split}
\frac{\hat{\alpha}^2}{\hat{\alpha}^2-1}\tilde{\Gamma}_{\mu\nu}{}^\rho &=\hat{\alpha}^2\left[\sigma_{\mu\nu}+\frac{\theta}{d-1}\ \left(P_{\mu\nu}+\frac{d}{2}u_\mu u_\nu\right)-da_{(\mu}u_{\nu)}\right]u^\rho-2\omega^\rho{}_{(\mu}u_{\nu)}+(\frac{d}{2}-1)u_{\mu}u_{\nu}a^\rho \\
\end{split}
\end{equation}

It follows that
\begin{equation}
\begin{split}
\tilde{\Gamma}_{\mu\nu}{}^\rho u_\rho 
&=-(\hat{\alpha}^2-1)\left[\sigma_{\mu\nu}+\frac{\theta}{d-1}\ \left(P_{\mu\nu}+\frac{d}{2}u_\mu u_\nu\right)-da_{(\mu}u_{\nu)}\right] \\
\end{split}
\end{equation}

We can now evaluate
\begin{equation}
\begin{split}
\hat{\nabla}_\mu u_\nu &\equiv\nabla_\mu u_\nu-\tilde{\Gamma}_{\mu\nu}{}^\rho u_\rho\\
&= \hat{\alpha}^2 \sigma_{\mu\nu} + \omega_{\mu\nu}+ \hat{\alpha}^2 \frac{\theta}{d-1}\ P_{\mu\nu}-u_\mu a_\nu\left(1+\frac{d}{2}(\hat{\alpha}^2-1)\right)-\frac{d}{2}(\hat{\alpha}^2-1)\mathcal{A}_{\mu}u_{\nu}\\
\end{split}
\end{equation}

Finally, we obtain\footnote{We have used 
\[\partial_\mu \left\{\frac{1}{\hat{\alpha}}\right\}=\frac{1}{\hat{\alpha}}\frac{d}{2}(\hat{\alpha}^2-1)\mathcal{A}_\mu  \]
 }
\begin{equation}
\begin{split}
\hat{\nabla}_\mu\hat{u}_\nu&=\hat{\nabla}_\mu \left\{\frac{u_\nu}{\hat{\alpha}}\right\}+\ldots\\
&= \hat{\alpha} \sigma_{\mu\nu} +\frac{\omega_{\mu\nu}}{\hat{\alpha}} +  \frac{\hat{\alpha}\theta}{d-1}\ \hat{P}_{\mu\nu}-\hat{u}_\mu a_\nu\left(1+\frac{d}{2}(\hat{\alpha}^2-1)\right)\\
\end{split}
\end{equation}
from which it follows that
\begin{equation}
\begin{split}
\hat{\sigma}_{\mu\nu} &\equiv \hat{\alpha} \sigma_{\mu\nu}\ ,\quad \hat{\omega}_{\mu\nu} \equiv \frac{1}{\hat{\alpha}} \omega_{\mu\nu} ,\quad \hat{\theta}\equiv\hat{\alpha}\theta\\
\hat{a}_\nu &\equiv \left[1+\frac{d}{2}(\hat{\alpha}^2-1)\right]a_\nu \\
\hat{\mathcal{A}}_\nu&\equiv \mathcal{A}_\nu-(1-\hat{\alpha}^2)\frac{d}{2}a_{\nu} \\
\end{split}
\end{equation}
We invert the last relation to get
\begin{equation}
\begin{split}
\mathcal{A}_\nu &=\frac{\hat{a}_\nu}{1-\frac{d}{2}(1-\hat{\alpha}^2)}-\frac{\hat{\theta}}{d-1}\hat{u}_\nu 
\end{split}
\end{equation}
This can be used to write $u_\mu$ is terms of hatted variables
\begin{equation}
\begin{split}
u_\mu = \hat{\alpha}\hat{u}_\mu-\frac{\hat{\alpha}^2}{r_D} \mathcal{A}_\mu = \left(1+\frac{\hat{\alpha}\hat{\theta}}{r_D(d-1)}\right)\hat{\alpha}\hat{u}_\mu - \frac{\hat{\alpha}^2}{r_D} \frac{\hat{a}_\mu}{\left[1+\frac{d}{2}(\hat{\alpha}^2-1)\right]} 
\end{split}
\end{equation}

Now, we want to write $\tilde{\Gamma}_{\mu\nu}{}^\rho$ in hatted variables. We start with
\begin{equation}
\begin{split}
\tilde{\Gamma}_{\mu\nu}{}^\rho &=  (\hat{\alpha}^2-1)\left[\sigma_{\mu\nu}+\frac{\theta}{d-1}\ \left(P_{\mu\nu}+\frac{d}{2}u_\mu u_\nu\right)-da_{(\mu}u_{\nu)}\right]u^\rho\\
&\qquad + \frac{\hat{\alpha}^2-1}{\hat{\alpha}^2}\left[-2\omega^\rho{}_{(\mu}u_{\nu)}+(\frac{d}{2}-1)u_{\mu}u_{\nu}a^\rho\right] \\
\end{split}
\end{equation}
and use the Dirichlet dictionary to get
\begin{equation}
\begin{split}
\tilde{\Gamma}_{\mu\nu}{}^\rho 
&=  (1-\frac{1}{\hat{\alpha}^2})\left[ \hat{\sigma}_{\mu\nu}+\frac{\hat{\theta}}{d-1}\ \left(\hat{P}_{\mu\nu}+\frac{d}{2}\hat{\alpha}^2\hat{u}_\mu \hat{u}_\nu\right)-\frac{d\hat{\alpha}^2}{\left[1+\frac{d}{2}(\hat{\alpha}^2-1)\right]}\hat{a}_{(\mu}\hat{u}_{\nu)}\right]\hat{u}^\rho\\
&\qquad + (\hat{\alpha}^2-1)\left[-2\hat{\omega}^\rho{}_{(\mu}\hat{u}_{\nu)}+\frac{\frac{d}{2}-1}{\left[1+\frac{d}{2}(\hat{\alpha}^2-1)\right]}\hat{u}_{\mu}\hat{u}_{\nu}\hat{a}^\rho\right] \\
\end{split}
\end{equation}

\section{Non-relativistic scaling a la BMW}
\label{A:bmw}

In this appendix we review and correct the scaling limit of \cite{Bhattacharyya:2008kq}.The major change in notation we make is to replace $\epsilon_\text{BMW}$ by a parameter $\aleph^{-1}$ so that the BMW limit is a large $\aleph$ asymptotics of the expressions below.
 
For simplicity we consider a fluid without bulk viscosity (which includes conformal fluids) with an
energy-momentum tensor $T^{\mu\nu}$ given by
\begin{equation}
T_{\mu\nu} = p\, g_{\mu\nu} + (\varepsilon  + p) \, u_\mu\,u_\nu - 2 \, \eta\, \sigma_{\mu\nu} + \ldots
\label{Tbdy2}
\end{equation} 
where we assume that this fluid lives on a background spacetime with a metric $g_{\mu\nu}$ and for the moment have just written out the stress tensor to first order in gradients.

\subsection{Spacetime split for the non-relativistic scaling }
\label{s:bmwA}

We begin by  decomposing this metric into an ambient part $g^{(0)}_{\mu\nu}$ and a forcing 
part $h_{\mu\nu}$, the split being done so as to recover explicit forcing terms 
in the Navier-Stokes (in addition to the pressure gradient term). One 
picks a suitable frame for the ambient metric, and writes the geometry as
\begin{equation}
g_{\mu\nu} = g^{(0)}_{\mu\nu} + h_{\mu\nu} \ , \qquad g^{(0)}_{\mu\nu}\, dx^\mu\, dx^\nu = -dt^2 + g^{(0)}_{ij} \, dx^i\, dx^j \ ,
\end{equation}	
where ${g}^{(0)}_{ij}$ are slowly varying functions of $x^i$ with ${h}_{\mu\nu}$ being 
treated as a perturbation. The metric perturbations which force the fluid are taken to be 
\begin{equation}
{h}_{\mu\nu} \,dx^\mu\,dx^\nu= 2\, {\aleph}^{-1}\, {k}^*_{i}\,   dt\, dx^i +  {\aleph}^{-2}\, \left({h}^*_{tt}\, dt^2 + {h}^*_{ij} \, dx^i \, dx^j \right)
\end{equation} 
where $\aleph$ is the book-keeping parameter that implements the BMW scaling  (note $\aleph = \epsilon_\text{BMW}^{-1}$).

We employ the notation that all the functions which have a $*$ subscript or superscript 
(which we freely interchange to keep formulae clear) are of a specific functional form 
with anisotropic scaling of their spatial and temporal gradients.
\begin{equation}
{\cal Y}_*(t,x^i) : {\mathbb R}^{d-1,1} \mapsto {\mathbb R}\ ,  \;\; \text{such that}
\;\; \{ \partial_t {\cal Y}_*(t,x^i), \nabla^{(0)}_i {\cal Y}_*(t,x^i)\}  \sim \{{\cal O}(\aleph^{-2}) ,{\cal O}(\aleph^{-1})\} 
\end{equation}	
The co-metric corresponding to the metric above is given as
\begin{equation}
\begin{split}
g^{\mu\nu} \partial_\mu \otimes \partial_\nu &= - \partial_t\otimes \partial_t + g^{ij}_{(0)}\partial_i\otimes\partial_j + 2\, {\aleph}^{-1}\ {k}_*^{i}\  \partial_t \otimes_s \partial_i\\
&\qquad  -  {\aleph}^{-2}\left[ \left({h}^*_{tt}-k_j^*k^j_*\right)\partial_t\otimes \partial_t + \left({h}_*^{ij}+k^i_*k^j_*\right)\partial_i\otimes \partial_j \right] \\
& \qquad + 2\, {\aleph}^{-3}\left[\left({h}^*_{tt}-k_j^*k^j_*\right) {k}_*^{i} -  {h}_*^{ij} k^*_j\right]\partial_t \otimes_s \partial_i + {\cal O}(\aleph^{-4})
\end{split}
\end{equation}
where we have freely raised and lowered the spatial indices with ${g}^{(0)}_{ij}$.

The velocity field of the fluid is parameterized as
\begin{equation}
{u}^{\mu} = u^t\, \left(1, {\aleph}^{-1}\, {v}_*^i \right) 
\label{bmwd1}
\end{equation}	
where the function $u^t$ is determined via the constraint $g_{\mu\nu}\, u^\mu\, u^\nu =-1$. This
gives the full velocity field in a large $\aleph$ expansion as
\begin{equation}
\begin{split}
u^t &=1 + \frac{\aleph^{-2}}{2} \left( h^*_{tt} + 2 \,k^*_{j}\, v^{j}_{*} +{g}^{(0)}_{jk} \, v^{j}_{*}\,  v^{k}_{*} \right)+ {\cal O}(\aleph^{-4})\\
u^i  &= \aleph^{-1} \, v_*^i + \frac{\aleph^{-3}}{2} \left( h^*_{tt} + 2 \,k^*_{j}\, v^{j}_{*} + {g}^{(0)}_{jk} \, v^{j}_{*}\,  v^{k}_{*} \right)\, v_*^i + {\cal O}(\aleph^{-4})\\
{u}_t &= -1 - \frac{1}{2}\,  {\aleph}^{-2} \, \left(- {h}^*_{tt} +  {g}^{(0)}_{jk} \, {v}^j_* \,  {v}^k_* \right)+ {\cal O}(\aleph^{-4})\\
{u}_i &=  {\aleph}^{-1}\, \left( {v}^*_i +  {k}^*_i \right)+ \aleph^{-3}\left[{h}^*_{ij}v^{j}_{*}+ \frac{1}{2} \left( h^*_{tt} + 2 \,k^*_{j}\, v^{j}_{*} +{g}^{(0)}_{jk} \, v^{j}_{*}\,  v^{k}_{*} \right)\, \left( {v}^*_i +  {k}^*_i \right) \right] + {\cal O}(\aleph^{-4})
\end{split}
\end{equation}	
which can alternately be written as 
\begin{equation}
\begin{split}
u^\mu \partial_\mu &=\partial_t + \aleph^{-1} \, v_*^i \partial_i \\
&\quad + \frac{\aleph^{-2}}{2} \left( h^*_{tt} + 2 \,k^*_{j}\, v^{j}_{*} + {g}^{(0)}_{jk} \, v^{j}_{*}\,  v^{k}_{*} \right)\partial_t + \frac{\aleph^{-3}}{2} \left( h^*_{tt} + 2 \,k^*_{j}\, v^{j}_{*} + {g}^{(0)}_{jk} \, v^{j}_{*}\,  v^{k}_{*} \right)\, v_*^i\partial_i + {\cal O}(\aleph^{-4})\\
{u}_\mu dx^\mu &= -dt +{\aleph}^{-1}\, \left( {v}^*_i +  {k}^*_i \right) dx^i - \frac{1}{2}\,  {\aleph}^{-2} \, \left(- {h}^*_{tt} +  {g}^{(0)}_{jk} \, {v}^j_* \,  {v}^k_* \right)dt\\
&\quad + \aleph^{-3}\left[{h}^*_{ij}v^{j}_{*}+ \frac{1}{2} \left( h^*_{tt} + 2 \,k^*_{j}\, v^{j}_{*} + {g}^{(0)}_{jk} \, v^{j}_{*}\,  v^{k}_{*} \right)\, \left( {v}^*_i +  {k}^*_i \right) \right] dx^i + {\cal O}(\aleph^{-4})\\
\end{split}
\end{equation}

Now, we are ready to calculate the velocity gradients - with some foresight, we will use the fact 
that the BMW limit is also an incompressibility limit where $v_*^i$ is divergenceless (see below).
With this in mind, we can write the velocity gradients as
\begin{equation}
\begin{split}
{\theta} &= {\cal O}( {\aleph}^{-4}) \\
{a}_{\mu} dx^{\mu} &=  {\aleph}^{-3} \left[ \partial_{t} {v}_{i}^{*}+{v}_{*}^{j}  {\nabla}^{(0)}_{j}  {v}_{i}^{*} -f_i^* \right] dx^{i}  + {\cal O}( {\aleph}^{-4})  \\
{a}^{\mu} \partial_{\mu} &=  {\aleph}^{-3} \left[ \partial_{t} {v}^{i}_{*}+{v}_{*}^{j}  {\nabla}^{(0)}_{j}  {v}^{i}_{*} -f^i_* \right] \partial_{i}  + {\cal O}( {\aleph}^{-4})  \\
{\sigma}_{\mu \nu} dx^{\mu} dx^{\nu} &= {\aleph}^{-2}\,  {\nabla}^{(0)}_{(i}  {v}^{*}_{j)} \,dx^{i} dx^{j} 
 -2{\aleph}^{-3} \,  {v}^{j}_{*} \, {\nabla}^{(0)}_{(i}  {v}^{*}_{j)} \,dx^{i} dt + {\cal O}( {\aleph}^{-4}) \\ 
{\sigma}^{\mu \nu}  \partial_\mu\otimes\partial_\nu &=   {\aleph}^{-2}{\nabla}_{(0)}^{(i}  {v}_{*}^{j)}\partial_i\otimes\partial_j+  2 {\aleph}^{-3}   ({v}_{i}^{*}+k_i^*) {\nabla}_{(0)}^{(i}  {v}_{*}^{j)}\partial_t\otimes_s\partial_j  + {\cal O}( {\aleph}^{-4})\\
{\omega}_{\mu \nu} dx^{\mu}\wedge dx^{\nu} &= {\aleph}^{-2}\, {\nabla}^{(0)}_{[i}  {v}^{*}_{j]} \,dx^{i} \wedge dx^{j} 
- 2{\aleph}^{-3} {v}^{j}_{*} \, {\nabla}^{(0)}_{[i}  {v}^{*}_{j]} \,dx^{i}\wedge dt  + {\cal O}( {\aleph}^{-4}) \\
\end{split}
\end{equation}
where $\nabla^{(0)}_{i}$ is the covariant derivative compatible with $\hat{g}^{(0)}(x^{i})$ and $f_i^*$
is the `gravitational force' acting on the fluid 
\begin{equation}
f_i^* \equiv \frac{1}{2}\partial_i {h}^*_{tt} - \partial_t {k}^*_i +\left[\nabla^{(0)}_i k^*_j - \nabla^{(0)}_j k^*_i\right] {v}_*^j = \frac{1}{2}\partial_i {h}^*_{tt} - \partial_t {k}^*_i +q^*_{ij} {v}_*^j 
\label{fidefn}
\end{equation}	
where we have introduced $q^*_{ij} \equiv \nabla^{(0)}_i k^*_j - \nabla^{(0)}_j k^*_i$.

\subsection{Navier-Stokes equations on a curved geometry}
\label{s:bmwB}

Now, we turn to the scaling of the thermodynamic variables. We define the mass density $\rho_0$, the
pressure per mass density $p_*$ and  the kinematic viscosity $\nu_0$ by
\begin{equation}
\begin{split}
\rho_0 \equiv \epsilon_0 + p_0\ ,\quad p = p_0 + {\aleph}^{-2} \rho_0\ {p}_{*}\quad\text{and}\quad \eta_0 =\rho_0\,  \nu_0
\end{split}
\end{equation}
where as before the subscript $0$ indicates the background value. All other thermodynamic variables
have similar scalings, for example,
\begin{equation}
\varepsilon = \varepsilon_0 + \aleph^{-2}\rho_0 \, \varepsilon_* \quad\text{and}\quad b = b_0 + \aleph^{-2} \, \delta b_*
\label{bmwd2}
\end{equation}	
The BMW limit is taken as to be the scaling as $\aleph \to \infty$ and in this limit the conservation equation $\nabla_\mu T^{\mu\nu} = 0$ reduces to 
\begin{equation}
{\cal O}(\aleph^{-2}): \, \qquad \nabla_i^{(0)} \, v^i_*  = 0 
 \label{incombdy}
\end{equation}	
%
%
and then a non-relativistic forced Navier-Stokes equation:
\begin{equation}
\begin{split}
{\cal O}(\aleph^{-3}): \qquad \left[\partial_t +v^j_*\nabla^{(0)}_j\right] v_i^* - 2\, \nu_0\, \nabla^{(0)^j} \left(\nabla^{(0)}_{(i} v^{*}_{j)}\right)= f_i^* -  \nabla^{(0)}_i \left[p_*+\nu_0^2\, \frac{d\, (d-3)}{(d-1)\,(d-2)} \; R^{(0)}\right] 
\end{split}
\label{nsbdy1}
\end{equation}
where the kinematic viscosity $\nu_0$ is given as 
\begin{equation}
\nu_0 \equiv \frac{\eta_0}{\rho_0} = \frac{b_0}{d}
\label{}
\end{equation}	

Before proceeding we should explain the origin of this equation since it differs from that presented in \cite{Bhattacharyya:2008kq}. On the l.h.s of \eqref{nsbdy1} we see a familiar term corresponding to the convective derivative of the non-relativistic velocity. The usual Laplacian term is modified due to the background curvature into the second derivative piece multiplying $\nu_0$. Its origin can be traced back to the term $-2 \, \eta \, \sigma_{\mu\nu}$ in the relativistic stress tensor \eqref{Tbdy2}.  On the r.h.s. of \eqref{nsbdy1} we have collected all the forcing terms: there is the familiar pressure gradient term along with two other terms that arise from curvature. $f_i^*$ is the forcing term that arises from the fluctuating part of the metric as is clear from \eqref{fidefn}; this term has been accounted for in \cite{Bhattacharyya:2008kq}. However, we also should see a forcing of the fluid from the `background' curvature: the spatial part of the metric $g^{(0)}_{\mu\nu}$ is a curved spatial metric $g_{ij}^{(0)}$ and its effect on the fluid turns out to be at the order ${\cal O}(\aleph^{-3})$, just the same as the other terms in the equation. However, its origins in the relativistic stress tensor are a bit more involved; it does not arise from any of the terms written down in \eqref{Tbdy2} but  rather from a second order gradient term in the relativistic stress tensor $2\,\eta\, b\, C_{\mu\alpha\nu\beta}u^\alpha u^\beta$ i.e., a coupling between the fluid and background curvature \cite{Bhattacharyya:2008mz}.

Note that we have here specialized to relativistic conformal fluids which have holographic duals, so that the transport coefficient multiplying the tensor structure  $C_{\mu\alpha\nu\beta}u^\alpha u^\beta$ is fixed to be $2\,\eta\, b$. For a general fluid we can have a new transport coefficient here $\kappa \propto \eta \, b$ and the correspondingly we would replace the  coefficient of $R^{(0)}$ in \eqref{nsbdy1} with   $\nu_0^2 \to \frac{\kappa_0}{\rho_0\, d}$.  Also, we should note that that while other tensor structures involving curvatures couplings are allowed for non-conformal fluids, these will necessarily have non-vanishing trace and as a result will only show up at sub-leading order in the BMW scaling limit.

At the risk of being overly pedantic we reiterate the fact that if we wish to place consider the non-relativistic BMW scaling of a relativistic fluid, then we must necessarily work with higher order gradient terms in the relativistic fluid stress tensor. To obtain the correct non-relativistic equations up to the order where we encounter Navier-Stokes equations the relevant part of the relativistic stress tensor is given to be 
\begin{equation}
T_{\mu\nu}  = p\, g_{\mu\nu} + (\varepsilon  + p) \, u_\mu\,u_\nu -2\,\eta\, \sigma_{\mu\nu}+2\,\eta\, b\, C_{\mu\alpha\nu\beta}u^\alpha u^\beta
\label{}
\end{equation}	
which is a subset of the full second order stress tensor derived in \cite{Bhattacharyya:2008mz}
\begin{equation}
\label{enmom:eq}
\begin{split}
T_{\mu\nu}& =  p\, g_{\mu\nu} + (\varepsilon  + p) \, u_\mu\,u_\nu -2\,\eta\, \sigma_{\mu\nu}\\
&-2\,\eta \,\tau_\omega \, \left[u^{\lambda}\mathcal{D}_{\lambda}\sigma_{\mu \nu}+\omega_{\mu}{}^{\lambda}\sigma_{\lambda \nu}+\omega_\nu{}^\lambda \sigma_{\mu\lambda} \right]\\
&+2\,\eta\, b\left[u^{\lambda}\mathcal{D}_{\lambda}\sigma_{\mu \nu}+\sigma_{\mu}{}^{\lambda}\sigma_{\lambda \nu} -\frac{\sigma_{\alpha \beta}\sigma^{\alpha \beta}}{d-1}P_{\mu \nu}+ C_{\mu\alpha\nu\beta}u^\alpha u^\beta \right]+\ldots\\
\end{split}
\end{equation}
It is a simple exercise to verify that none of the other terms involved in the second order stress tensor \eqref{enmom:eq} contribute to the BMW scaled equations at ${\cal O}(\aleph^{-3})$.

\subsection{The bulk metric dual to a non-relativistic fluid on the boundary of AdS}
\label{s:bmwC}

One can also construct the gravitational solutions dual to such fluids as described in \cite{Bhattacharyya:2008kq}. To do so we simply need to apply the scaling of parameters described earlier as for e.g., in \eqref{bmwd1}, \eqref{bmwd2} to  the general fluid/gravity bulk metric dual to a relativistic fluid on the boundary of AdS. Such a metric correct to second order in the relativistic gradient expansion was originally derived in \cite{Bhattacharyya:2008mz} generalizing the original result of \cite{Bhattacharyya:2008jc}. It was believed that this in general would suffice to find the gravity dual for a non-relativistic fluid that satisfies the incompressible Navier-Stokes equations derived above \eqref{incombdy}, \eqref{nsbdy1}. In fact the original computation presented in  \cite{Bhattacharyya:2008kq} argued that it would actually suffice to consider the relativistic metric accurate to first order in gradients. Unfortunately, this turns out to be incorrect and one needs a subset of second order gradient terms along with one particular third order gradient term to solve Einstein's equations to order ${\cal O}(\aleph^{-3})$. Note that this is necessary because it is at ${\cal O}(\aleph^{-3})$ that we encounter the dynamical content of the boundary fluid equations, viz., the Navier-Stokes equation \eqref{nsbdy1}. 

The new ingredient in our analysis is that we need to worry about a particular third order term proportional to ${\cal D}_\mu {\cal R}$ in the fluid/gravity correspondence. The term in question turns out to be computable using the original algorithm outlined for constructing bulk metrics dual to boundary fluids in \cite{Bhattacharyya:2008jc} and thankfully involves a decoupled tensor structure that can be sourced independently.  Including this term, we find that for the non-relativistic fluid on the Dirichlet surface it suffices to consider the following truncation of the relativistic fluid/gravity metric to third order in boundary gradient expansion:
\begin{equation}\label{metric3trunc}
\begin{split}
ds^2&=-2 u_\mu dx^\mu \left( dr + r\ \mathcal{A}_\nu dx^\nu \right) \\
&+ \left[ r^2 g_{\mu\nu} +2u_{(\mu}\mathcal{S}_{\nu)\lambda}u^\lambda -\frac{2}{3r}\frac{u_{(\mu}\mathcal{D}_{\nu)}\mathcal{R}}{(d-1)(d-2)} \right]dx^\mu dx^\nu\\
&+r^2\left[ \frac{u_\mu u_\nu}{(br)^d} +2b F(br) \sigma_{\mu\nu}+4b^2 \frac{L(br)}{(br)^d}u_{(\mu}P_{\nu)}^{\lambda}\mathcal{D}_{\alpha}{\sigma^{\alpha}}_{\lambda}\right]dx^\mu dx^\nu \\
&-2(br)^2\left[ H_1(br) C_{\mu\alpha\nu\beta}u^\alpha u^\beta+\frac{b N(br)}{(br)^d} \frac{(d-3)u_{(\mu}\mathcal{D}_{\nu)}\mathcal{R}}{(d-1)(d-2)}\right] dx^\mu dx^\nu \\
& +\ldots\\
\end{split}
\end{equation}
where the functions appearing above and their large $r$ asymptotics are given to be:
\begin{equation}\label{funcDreq}
\begin{split}
f(br) &\equiv 1-\frac{1}{(br)^{d}} \\
F(br)&\equiv \int_{br}^{\infty}\frac{y^{d-1}-1}{y(y^{d}-1)}dy \\
&\approx  \frac{1}{br} -\frac{1}{d(br)^d}+ \frac{1}{(d+1)(br)^{d+1}}+\frac{\#}{(br)^{2d}}+\ldots\\
L(br) &\equiv \int_{br}^\infty\xi^{d-1}d\xi\int_{\xi}^\infty dy\ \frac{y-1}{y^3(y^d
-1)} \\
&\approx \frac{1}{(d+1)(br)}-\frac{1}{2(d+2)(br)^2}+\frac{1}{(d+1)(2d+1)(br)^{d+1}}\\
&\quad-\frac{1}{(d+1)(2d+4)(br)^{d+2}} +\frac{\#}{(br)^{2d+1}}+\ldots \\
H_1(br)&\equiv \int_{br}^{\infty}\frac{y^{d-2}-1}{y(y^{d}-1)}dy \\
&\approx \frac{1}{2(br)^2}-\frac{1}{d(br)^d}+ \frac{1}{(d+2)(br)^{d+2}}+\frac{\#}{(br)^{2d}}+\ldots\\ 
N(br) &\equiv \int_{br}^\infty\xi^{d-1}d\xi\int_{\xi}^\infty dy\ \frac{y^2-1}{y^4(y^d
-1)} \\
&\approx\frac{1}{(d+1)(br)} -\frac{1}{3(d+3)(br)^3}+\frac{1}{(d+1)(2d+1)(br)^{d+1}}\\
&\quad-\frac{1}{(d+3)(2d+3)(br)^{d+3}} +\frac{\#}{(br)^{2d+1}} +\ldots  \\ 
\end{split}
\end{equation}

There are various new curvature tensors and derivatives introduced above. ${\cal D}$ denotes the Weyl covariant derivative introduced in \cite{Loganayagam:2008is} which was used to present the bulk metric dual relevant for fluid/gravity correspondence in the case of curved boundaries in \cite{Bhattacharyya:2008mz}. ${\cal R}_{\mu\nu}$ is likewise a Weyl covariant Ricci tensor and ${\cal S}_{\mu\nu}$ is a Weyl-covariant Schouten tensor 
\begin{equation}
{\cal S}_{\mu\nu} = \frac{1}{d-2} \left( {\cal R}_{\mu\nu}  - \frac{1}{2\, (d-1)}\, g_{\mu\nu}\, {\cal R}\right)
\label{}
\end{equation}	
For further details of these objects and the complete form for the second order fluid/gravity metric accurate to second order in gradients we refer the reader to \cite{Bhattacharyya:2008mz}.

Once we have the relativistic metric at hand it is a simple matter to employ the scalings outlined earlier. We find that the bulk metric dual to an incompressible Navier-Stokes fluid living on a spatially curved geometry at the boundary of AdS takes the form 
\begin{equation}\label{bdybmwf}
\begin{split}
ds^2 
&= ds_0^2 +  {\aleph}^{-1} ds_1^2 +  {\aleph}^{-2} ds_2^2 +  {\aleph}^{-3} ds_3^2 + {\cal O}( {\aleph}^{-4}) \\
&\text{with}\\
ds_0^2 &= 2  \ dt\ dr + r^2\left[- f_0 dt^2 +  {g}^{(0)}_{ij}dx^i dx^j\right]\\
ds_1^2 &= -2  \left(  {v}^*_i +   {k}^*_i \right)\ dx^i\ dr + 2 r^2 \left[ {k}^*_i -\left(1- f_0\right)\left(  {v}^*_i +   {k}^*_i \right)\right] dx^i dt \\
ds_2^2 &= 2    \left[- \frac{1}{2} {h}^*_{tt} + \frac{1}{2}  {g}^{(0)}_{jk} \,  {v}^j_* \,   {v}^k_* \right]dt\ dr + r^2\left[ {h}^*_{tt}\, dt^2 +  {h}^*_{ij} \, dx^i \, dx^j \right]\\
&\quad +r^2 \left(1-f_0\right) \left[\left(-   {h}^*_{tt} +  {g}^{(0)}_{jk} \,  {v}^j_* \,   {v}^k_*+ {p}_* d\right)dt^2 \right.\\
&\qquad \left. + \left(  {v}^*_i +   {k}^*_i \right) \left(  {v}^*_j +   {k}^*_j \right)dx^i dx^j \right]+2\,r^2\, b_0  {F}_0 {\nabla}^{(0)}_{(i}  {v}^{*}_{j)} \,dx^{i} dx^{j}\\
&\qquad 
\red{-\frac{R^{(0)}}{(d-1)\,(d-2)} \, dt^2 - 2\, H_0 \left(\text{S}_{ij}^{(0)} - \frac{R^{(0)}\, g_{ij}^{(0)}}{2\, (d-1)\, (d-2)}\right) dx^i \,dx^ j }\\
ds_3^2 &= -2  \left[ {h}^*_{ij} {v}^{j}_{*}+  \left( \frac{1}{2} {h}^*_{tt} +  \, {k}^*_{j}\,  {v}^{j}_{*} +\frac{1}{2}  {g}^{(0)}_{jk} \,  {v}^{j}_{*}\,   {v}^{k}_{*}\right)\, \left(  {v}^*_i +   {k}^*_i \right) \right] dx^i dr \\
&\quad + 2r
\left[ \partial_{t}  {v}_{i}^{*}+ {v}_{*}^{j}   {\nabla}^{(0)}_{j}   {v}_{i}^{*} - {f}_i^* \right] dx^{i}dt
 -4r^2b_0  {F}_0 {v}_{*}^j {\nabla}^{(0)}_{(i}  {v}^{*}_{j)} \,dx^{i} dt 
\\
&\quad -2\,r^2 \left(1- f_0\right)\left[ {h}^*_{ij} {v}^{j}_{*}+ \left( {k}^*_{j}\,  {v}^{j}_{*} +  {g}^{(0)}_{jk} \,  {v}^{j}_{*}\,   {v}^{k}_{*}+ {p}_* d \right)\, \left(  {v}^*_i +   {k}^*_i \right) \right] dx^i dt \\
&\quad  
\red{- 2\,\frac{L_0}{(b_0r)^{d-2}} {\nabla}^2_{(0)}v^*_{i}   \,dx^i \,dt -2 \, \text{S}_{ij}^{(0)}\, v_*^j \, dx^i \,dt- \frac{1}{d-2}\, {\nabla}^j_{(0)}q^*_{ij} \, dx^i \,dt}\\
&\quad
\red{+ \frac{R^{(0)}}{(d-1)\,(d-2)}\, (v^*_i + k^*_i) \, dx^i \,dt 
+ 4\, H_0 \left(\text{S}_{ij}^{(0)} - \frac{R^{(0)}\, g_{ij}^{(0)} }{2\, (d-1)\, (d-2)} \right) v_*^j \, dx^i\, dt }\\
&\quad
\red{+ 2\, \frac{b_0\, N_0}{(b_0\,r)^{d-2}} \, \frac{d-3}{(d-1)\, (d-2) }\, \nabla^{(0)}_i R^{(0)} \, dx^i \,dt}
\end{split}
\end{equation}
where we have highlighted  the terms that were missed in the previous analysis for quick comparison. Note that when the background spatial metric $g^{(0)}_{ij}$ is Ricci flat, many of the terms vanish except for  two terms in the third order metric (which are proportional to $\nabla^2_{(0)} v_i^*$ and $\nabla^j_{(0)} q^*_{ij}$ respectively). Finally we should note that $\text{S}_{ij}^{(0)}$ is used to denote the spatial components  of the Schouten tensor of the full background metric $g^{(0)}_{\mu\nu}$; in particular, it should not be confused with the Schouten tensor of the spatial metric $g_{ij}^{(0)}$. The functions that enter into the metric above are:
\begin{equation}
\begin{split}
f_0 &\equiv 1-(b_0 r)^{-d}\ ,\quad  {p}_* \equiv -\frac{\delta b_*}{b_0} \quad\text{and}\quad {F}_0 \equiv \int_{b_0 r}^{\infty}\frac{y^{d-1}-1}{y(y^{d}-1)}dy\\ 
{f}_i^* &\equiv  \frac{1}{2}\partial_i  {h}^*_{tt} - \partial_t  {k}^*_i +\left[ {\nabla}^{(0}_i  {k}^*_j -  {\nabla}^{(0)}_j  {k}^*_i\right] {v}_*^j = \frac{1}{2}\partial_i {h}^*_{tt} - \partial_t {k}^*_i + {q}^*_{ij} {v}_*^j  \\
L_0 &\equiv \int_{b_0 r}^\infty\xi^{d-1}d\xi\int_{\xi}^\infty dy\ \frac{y-1}{y^3(y^d
-1)}\\ 
H_0 &\equiv (b_0 r)^2\, \int_{b_0r}^{\infty}\frac{y^{d-2}-1}{y(y^{d}-1)}dy \\
N_0 &\equiv \int_{b_0r}^\infty\xi^{d-1}d\xi\int_{\xi}^\infty dy\ \frac{y^2-1}{y^4(y^d
-1)} \\
\end{split}
\end{equation}

Finally, let us note that the Navier-Stokes equations themselves have an interesting scaling symmetry. Given any solution to \req{incombdy} and \req{nsbdy1} with $g^{(0)}_{ij} = \delta_{ij}$ we can consider replacing 
\begin{equation}
p_* \to \epsilon^2 \, p_{*\epsilon} \ , \qquad v^i_* \to \epsilon\, v^i_{*\epsilon} \ , \qquad f^i_* \to \epsilon^3\, f^i_{*\epsilon}
\end{equation}	
where again the functions entering the dynamics with subscript $*\epsilon$ have spatial gradients $\partial_i \sim \epsilon$ and temporal gradients $\partial_t \sim \epsilon^2$. This fact makes it possible to compound the Navier-Stokes scaling which effectively allows one to replace $\aleph \to \aleph^{w}$ for some $w \geq 1$. Essentially the incompressible Navier-Stokes system of equations is a fixed point set of this scaling symmetry, a fact that we have made use of in \sec{s:nh}.

\section{Bulk dual of the non-relativistic Dirichlet fluid}
\label{s:bmwR}

In this appendix we present without derivation the results for the non-relativistic scaling limit of  the Dirichlet problem, generalizing the result quoted in \sec{s:nrhyp2}. Physically the only new content is that we allow the metric on the Dirichlet surface $\Sigma_D$ to be endowed with an arbitrarily slowly varying spatial metric. Thus in contrast to \sec{s:nrhyp2} we are relaxing the constraint on  $g_{ij}^{(0)}$ introduced in \eqref{Dg0} being Ricci flat. 

To indicate the differences note that the presence of a non-Ricci flat spatial metric $g^{(0)}_{ij}$ implies that the non-relativistic metric gets contributions from various tensor structures which appear at the second order in the gradient expansion of the fluid/gravity correspondence. In terms of the metric written down in \cite{Bhattacharyya:2008mz} some of these are straightforward to see -- any term involving curvature tensors of boundary data (and thus via the Dirichlet constitutive relation the hypersurface $\Sigma_D$ curvatures) will contribute at this order. However, we also get contribution from spatial gradients of the hypersurface curvature, i.e., in the BMW scaling limit encounter terms of the form $\nabla^{(0)}_i R^{(0)}$. To guide the reader towards a derivation, we quote simply the relativistic tensor structures which are relevant and their scaling behavior under the Dirichlet BMW scaling.
\begin{equation}
\begin{split}
\hat{g}_{\mu\nu}\, dx^\mu\, dx^\nu &= -dt^2 + \hat{g}^{(0)}_{ij} \, dx^i\, dx^j + 2\, {\hat{\aleph}}^{-1}\, \hat{k}^*_{i}\,   dt\, dx^i +  {\hat{\aleph}}^{-2}\, \left(\hat{h}^*_{tt}\, dt^2 + \hat{h}^*_{ij} \, dx^i \, dx^j \right).\\
\hat{u}_\mu dx^\mu &= -dt +{\hat{\aleph}}^{-1}\, \left(\hat{v}^*_i +  \hat{k}^*_i \right) dx^i - \frac{1}{2}\,  {\hat{\aleph}}^{-2} \, \left(- \hat{h}^*_{tt} +  \hat{g}^{(0)}_{jk} \, \hat{v}^j_* \,  \hat{v}^k_* \right)dt\\
&\quad + \hat{\aleph}^{-3}\left[\hat{h}^*_{ij}\hat{v}^{j}_{*}+ \frac{1}{2} \left( \hat{h}^*_{tt} + 2 \,\hat{k}^*_{j}\, \hat{v}^{j}_{*} + \hat{g}^{(0)}_{jk} \, \hat{v}^{j}_{*}\,  \hat{v}^{k}_{*} \right)\, \left( \hat{v}^*_i +  \hat{k}^*_i \right) \right] dx^i + {\cal O}(\hat{\aleph}^{-4})\\
\hat{a}_{\mu} dx^{\mu} &=  {\hat{\aleph}}^{-3} \left[ \partial_{t} \hat{v}_{i}^{*}+\hat{v}_{*}^{j}  \hat{\nabla}^{(0)}_{j}  \hat{v}_{i}^{*} -\hat{f}_i^* \right] dx^{i}  + {\cal O}( {\hat{\aleph}}^{-4})  \\
\hat{\sigma}_{\mu \nu} dx^{\mu} dx^{\nu} &= {\hat{\aleph}}^{-2}\,  \hat{\nabla}^{(0)}_{(i}  \hat{v}^{*}_{j)} \,dx^{i} dx^{j} 
 -2{\hat{\aleph}}^{-3} \,  \hat{v}^{j}_{*} \, \hat{\nabla}^{(0)}_{(i}  \hat{v}^{*}_{j)} \,dx^{i} dt + {\cal O}( {\hat{\aleph}}^{-4}) \\ 
 \end{split}
 \label{genscalefA}
 \end{equation}
 as before along with new tensor structures:
 \begin{equation}
 \begin{split}
\hat{\mathcal{S}}_{\nu\lambda}\hat{u}^\lambda dx^\nu &={\hat{\aleph}}^{-2}\frac{\hat{R}^{(0)}}{2(d-1)(d-2)}dt+{\hat{\aleph}}^{-3}\left[\hat{\text{S}}^{(0)}_{ij}\hat{v}^j_*+ \frac{1}{2(d-2)}\hat{\nabla}^j_{(0)}\hat{q}^*_{ij}\right] dx^i + {\cal O}( {\hat{\aleph}}^{-4})\\
\hat{\mathcal{R}}_{\nu\lambda}\hat{u}^\lambda dx^\nu &={\hat{\aleph}}^{-3}\left[\hat{R}^{(0)}_{ij}\hat{v}^j_*+ \frac{1}{2}\hat{\nabla}^j_{(0)}\hat{q}^*_{ij}\right] dx^i + {\cal O}( {\hat{\aleph}}^{-4})\\
\hat{P}_{\nu}^{\lambda}\hat{\mathcal{D}}_{\alpha}{\hat{\sigma}^{\alpha}}_{\lambda}dx^\nu &=\frac{{\hat{\aleph}}^{-3}}{2}\hat{\nabla}^2_{(0)}\hat{v}^*_{i} dx^i + {\cal O}( {\hat{\aleph}}^{-4}) \\
\hat{C}_{\mu\alpha\nu\beta}\hat{u}^\alpha \hat{u}^\beta dx^\mu dx^\nu &= {\hat{\aleph}}^{-2} \left[\hat{\text{S}}^{(0)}_{ij}-\frac{\hat{R}^{(0)}}{2(d-1)(d-2)}\hat{g}^{(0)}_{ij}\right] dx^i dx^j\\
&\quad- \; 2\,{\hat{\aleph}}^{-3} \left[\hat{\text{S}}^{(0)}_{ij}-\frac{\hat{R}^{(0)}}{2(d-1)(d-2)}\hat{g}^{(0)}_{ij}\right] \hat{v}^j_* dx^i dt\\
\hat{\mathcal{D}}_\nu\hat{\mathcal{R}} dx^\nu &= {\hat{\aleph}}^{-3}\,\hat{\nabla}^j_{(0)}\hat{R}^{(0)}dx^j
\end{split}
\label{genscalefB}
\end{equation}
where
\begin{equation}\begin{split}
\hat{f}_i^* &\equiv \frac{1}{2}\partial_i \hat{h}^*_{tt} - \partial_t \hat{k}^*_i +\left[\hat{\nabla}^{(0)}_i \hat{k}^*_j - \hat{\nabla}^{(0)}_j \hat{k}^*_i\right] \hat{v}_*^j = \frac{1}{2}\partial_i \hat{h}^*_{tt} - \partial_t \hat{k}^*_i +\hat{q}^*_{ij} \hat{v}_*^j \\
\hat{q}^*_{ij} &\equiv \hat{\nabla}^{(0)}_i \hat{k}^*_j - \hat{\nabla}^{(0)}_j \hat{k}^*_i
\end{split}\end{equation}
and we use $\hat{\text{S}}^{(0)}_{ij}$ to denote the spatial (i.e. $ij$-) components of the Schouten tensor for $g^{(0)}_{\mu\nu}$ keep the expressions somewhat compact. 

\subsection{Boundary data for non-relativistic fluids on $\Sigma_D$}
\label{s:}

Given the scalings to obtain the non-relativistic fluid on $\Sigma_D$, it is possible to identify the relevant terms of the third order fluid/gravity metric that we need to retain to solve the Dirichlet problem. Per se the set of terms we need in the bulk metric is still given by \eqref{metric3trunc}, though as in the main text we want to use this information to solve for the Dirichlet constitutive relations. We first quote the results for the 
hypersurface stress tensor which defines the hypersurface velocity $\hat{u}^\mu$ before indicating the answers for the boundary velocity field and metric in terms of the hypersurface data.

We can parameterize the hypersurface stress tensor as in the main text; to obtain the correct non-relativistic equations on $\Sigma_D$ we need to retain some second order gradient terms involving hypersurface curvature tensors (analogously to the situation at the boundary as described in \sec{s:bmwB}). The relevant piece of the relativistic hypersurface stress tensor turns out to be:
\begin{equation}
\begin{split}
\hat{T}_{\mu\nu}&=(\hat{\varepsilon}+\hat{p})\,\hat{u}_\mu\hat{u}_\nu + \hat{p}\,\hat{g}_{\mu\nu}-2\,\hat{\eta} \,\hat{\sigma}_{\mu\nu}
+\hat{\kappa}_C \, \hat{C}_{\mu\alpha\nu\beta}\, \hat{u}^\alpha \hat{u}^\beta +\ldots\\
\end{split}
\end{equation}
where
\begin{equation}
\begin{split}
\hat{\varepsilon} &\equiv \frac{d-1}{8\pi G_{d+1}b^d}\frac{\hat{\alpha}}{\hat{\alpha}+1}\left[1- \frac{\hat{\alpha}\hat{R}^{(0)}}{2\,r_D^2\,(d-1)\,(d-2)} \right]\\
\hat{\varepsilon}+\hat{p} &\equiv \frac{d\hat{\alpha}}{16\pi G_{d+1}b^d}\left[1- \frac{\hat{\alpha}^2\,\hat{R}^{(0)} }{2\,r_D^2\,(d-1)\,(d-2)} \right]-\frac{1}{8\pi G_{d+1}b^d}\frac{\hat{\alpha}^2}{\hat{\alpha}+1}\frac{\hat{R}^{(0)}}{r_D^2(d-1)(d-2)}\\
\hat{\eta} &= \frac{1}{8\pi G_{d+1}b^{d-1}}\\
\hat{\kappa}_C &= \frac{1}{8\pi G_{d+1}b^{d-2}}\left[1-\frac{\hat{\alpha}^2}{\hat{\alpha}+1}\frac{(br_D)^{d-2} -1}{(br_D)^d}\right]\\
\end{split}
\end{equation}
Conservation of this stress tensor together with the scaling forms introduced in \eqref{genscalefA}, \eqref{genscalefB} leads to the incompressible Navier-Stokes equations on $\Sigma_D$.

The boundary velocity field and metric can be expressed in terms of the Dirichlet data as before. Before we write out the exressions in their gory detail, let us introduce some new parameters $\hat{\kappa}_L$, $\hat{\kappa}_N$ which depend on the location of $\Sigma_D$ as 
\begin{equation}
\begin{split}
\hat{\kappa}_L &\equiv \frac{1}{d}\left[
\xi(\xi^d-1)\frac{d}{d\xi}\left[\xi^{-d}L(\xi)\right]+
\frac{1}{\xi\left[1+\frac{d}{2}(\hat{\alpha}^2-1)\right]}+\frac{1}{\xi^2(d-2)}\right]_{\xi=br_D}  \\
\hat{\kappa}_N &\equiv \frac{1}{d}\left\{
(d-3)\left[\xi(\xi^d-1)\frac{d}{d\xi}\left[\xi^{-d}N(\xi)\right]+
\frac{1}{\xi\left[1+\frac{d}{2}(\hat{\alpha}^2-1)\right]}\right] \right.\\
&\left. -\frac{d-2}{2\, \xi^3\left[1+\frac{d}{2}(\hat{\alpha}^2-1)\right]}\right\}_{\xi=br_D}
\end{split}
\end{equation}
These are in turn limiting values of certain functions which appear in various guises in the result for the Dirichlet constitutive relations and the bulk metric and are collected once and for all below.
\begin{equation}
\begin{split}
\hat{F}(br) &\equiv \frac{1}{\hat{\alpha}}\left(F(br)-F_D\right) \\
\hat{H}_1(br) &\equiv \left(H_1(br)-H_{1D}\right)\\
\hat{\xi}_1(br) &\equiv \frac{\hat{\alpha}}{b}\left(\frac{1}{r}-\frac{1}{r_D}\right)+\frac{\hat{\alpha}}{br_D}\left[1-\hat{\alpha}^2f(br)\right]=\frac{\hat{\alpha}}{br}\left[1-\frac{r}{r_D}\hat{\alpha}^2f(br)\right]\\
\hat{M}_1(br) &\equiv \frac{\hat{\alpha}^2}{b^2}\left(\frac{1}{r^2}-\frac{1}{r_D^2}\right)+\frac{\hat{\alpha}^2}{b^2r_D^2}\left[1-\hat{\alpha}^2f(br)\right]=\frac{\hat{\alpha}^2}{(br)^2}\left[1-\frac{r^2}{r_D^2}\hat{\alpha}^2f(br)\right]\\
\hat{M}_2(br) &\equiv\frac{\hat{\alpha}^2}{b^2r_D^2(d-2)}\left[1+\frac{2}{d\hat{\alpha}(\hat{\alpha}+1)}\right]
\left[1-\hat{\alpha}^2f(br)\right] \\
\hat{L}_1(br) &\equiv \frac{L(br)}{(br)^d}-\frac{L_D}{(br_D)^d}+\kappa_L \left[1-\hat{\alpha}^2f(br)\right]\\
&\quad-\frac{\hat{\alpha}^2-1}{2b^2(d-2)}\left(\frac{1}{r^2}-\frac{1}{r_D^2}\right)-\frac{(\hat{\alpha}^2-1)}{2b\left[1+\frac{d}{2}(\hat{\alpha}^2-1)\right]}\left(\frac{1}{r}-\frac{1}{r_D}\right)\\
\hat{N}_1(br) &\equiv \hat{\alpha}(d-3)\left(\frac{N(br)}{(br)^d}-\frac{N_D}{(br_D)^d}\right)+\kappa_N\hat{\alpha}\left[1-\hat{\alpha}^2f(br)\right]\\
&+\frac{\hat{\alpha}}{3b^3}\left(\frac{1}{r^3}-\frac{1}{r_D^3}\right)-\frac{\hat{\alpha}\left[\frac{\hat{\alpha}^2}{r_D^2}+(d-3)b^2(\hat{\alpha}^2-1)\right]}{2b^3\left[1+\frac{d}{2}(\hat{\alpha}^2-1)\right]}\left(\frac{1}{r}-\frac{1}{r_D}\right)\\
\end{split}
\label{allhatfunx}
\end{equation}
where the functions entering the above expressions and their asymptotics have been previously been collected together in \eqref{funcDreq}.

The final result of this exercise leads to:
\begin{equation}\begin{split}
u_t  &=  -\hat{\alpha}_0 - \hat{\aleph}^{-2} \hat{\alpha}_0 \left[- \frac{1}{2}\hat{h}^*_{tt} + \frac{1}{2} \hat{g}^{(0)}_{jk} \, \hat{v}^j_* \,  \hat{v}^k_* +\hat{p}_* \frac{\frac{d}{2}\, (\hat{\alpha}_0^2 -1)}{1+\frac{d}{2}\, (\hat{\alpha}_0^2 -1)} - \frac{\hat{\alpha}_0^2}{r_D^2} \, \frac{\hat{R}^{(0)}}{2\,(d-1)\,(d-2)}
\right]  + {\cal O}(\hat{\aleph}^{-4})  \\
\end{split}\end{equation}	

\begin{equation}\begin{split}
u_i &= \hat{\aleph}^{-1}\, \hat{\alpha}_0\, \left(\hat{v}^*_i + \hat{k}^*_i \right)\\
&\quad +\hat{\aleph}^{-3}\,
\hat{\alpha}_0\left[\hat{h}^*_{ij}\hat{v}^{j}_{*}+  \left( \frac{1}{2}\hat{h}^*_{tt} +  \,\hat{k}^*_{j}\, \hat{v}^{j}_{*} +\frac{1}{2} \hat{g}^{(0)}_{jk} \, \hat{v}^{j}_{*}\,  \hat{v}^{k}_{*} +\hat{p}_* \frac{\frac{d}{2}\, (\hat{\alpha}_0^2 -1)}{1+\frac{d}{2}\, (\hat{\alpha}_0^2 -1)}\right)\, \left( \hat{v}^*_i +  \hat{k}^*_i \right) \right]\\
&\qquad -\hat{\aleph}^{-3}\, \frac{\hat{\alpha}_0^2}{r_D\left(1+\frac{d}{2}(\hat{\alpha}_0^2-1)\right)}\left[ \partial_{t} \hat{v}_{i}^{*}+\hat{v}_{*}^{j}  \hat{\nabla}^{(0)}_{j}  \hat{v}_{i}^{*}-\hat{f}_i^* \right] \\
&\qquad 
+\hat{\aleph}^{-3}\left[ b_0^2\, \hat{\kappa}_L\,\hat{\alpha}_0\,  \hat{\nabla}^2_{(0)}v^*_i -  b_0^3\, \hat{\kappa}_N \, \hat{\alpha}_0^2\, \frac{\nabla^{(0)}_i\, R^{(0)}}{(d-1)(d-2)} \right] \\
&\qquad
+ \hat{\aleph}^{-3}\, \frac{\hat{\alpha}_0^3}{r_D^2} \, \left[\hat{S}^{(0)}_{ij}\hat{v}^j_* - \frac{1}{d-2}\, \left(1 + \frac{2}{d\, \hat{\alpha}_0\, (\hat{\alpha}_0 +1) } \right) \hat{R}^{(0)}_{ij}\hat{v}^j_*
 - \frac{\hat{\nabla}^j_{(0)}\hat{q}^*_{ij}}{d\,(d-2)\, \hat{\alpha}_0\, (\hat{\alpha}_0 +1) }\right]\\
&\qquad 
+ {\cal O}(\hat{\aleph}^{-4}) 
\end{split}\end{equation}	

Further the Dirichlet constitutive relation for the boundary metric is  
\begin{equation}\begin{split}
g_{tt} &= -\hat{\alpha}_0^2 + \hat{\aleph}^{-2} \, \left[\hat{h}^*_{tt}\ +
\left(1-\hat{\alpha}_0^2 \right) \left(- \hat{h}^*_{tt} + \hat{g}^{(0)}_{jk} \, \hat{v}^j_* \,  \hat{v}^k_*+\hat{p}_* \frac{d\hat{\alpha}_0^2 }{1+\frac{d}{2}\, (\hat{\alpha}_0^2 -1)} \right) \right] \\
& \quad 
+ \hat{\aleph}^{-2} \, \left[\frac{\hat{\alpha}_0^4}{r_D^2}\,\frac{\hat{R}^{(0)}}{(d-1)\,(d-2)}\right]
+ {\cal O}(\hat{\aleph}^{-4}) \\
g_{ti} &= \hat{\aleph}^{-1}\left( \hat{k}^*_i + (\hat{\alpha}_0^2-1)\, (\hat{k}_i^* + \hat{v}_i^*) \right) 
- \hat{\aleph}^{-3}\frac{\hat{\alpha}_0^3}{r_D\left(1+\frac{d}{2}\, (\hat{\alpha}_0^2 -1)\right)}
\left[ \partial_{t} \hat{v}_{i}^{*}+\hat{v}_{*}^{j}  \hat{\nabla}^{(0)}_{j}  \hat{v}_{i}^{*} -\hat{f}_i^* \right] \\
&\quad -\hat{\aleph}^{-3} \left(1-\hat{\alpha}_0^2 \right)\left[\hat{h}^*_{ij}\hat{v}^{j}_{*}+ \left(\hat{k}^*_{j}\, \hat{v}^{j}_{*} + \hat{g}^{(0)}_{jk} \, \hat{v}^{j}_{*}\,  \hat{v}^{k}_{*}+\hat{p}_* \frac{d\hat{\alpha}_0^2 }{1+\frac{d}{2}\, (\hat{\alpha}_0^2 -1)} \right)\, \left( \hat{v}^*_i +  \hat{k}^*_i \right) \right]\\
&\quad +\;\hat{\aleph}^{-3} \left[\frac{2 \, b_0}{\hat{\alpha}_0}\, F(b_0\,r_D)\hat{v}_{*}^j\hat{\nabla}^{(0)}_{(i} \hat{v}^{*}_{j)} - 2\, b_0^2{H}_{1}(b_0 r_D) \left(\hat{\text{S}}^{(0)}_{ij}-\frac{\hat{R}^{(0)}}{2(d-1)(d-2)}\hat{g}^{(0)}_{ij}\right)v_*^j \right] \\
&\quad
+\;\hat{\aleph}^{-3} \, \frac{\hat{\alpha}_0^4}{2\,r_D^2}\left[2\,\hat{\text{S}}^{(0)}_{ij}\,\hat{v}^j_*+ \frac{1}{(d-2)}\hat{\nabla}^j_{(0)}\hat{q}^*_{ij} -\frac{\hat{R}^{(0)}}{(d-1)\,(d-2)} \left(\hat{v}^*_i +  \hat{k}^*_i \right) \right] \\
& \quad -\;\hat{\aleph}^{-3} \,
 \frac{\hat{\alpha}_0^2\,(\hat{\alpha}_0^2-1)}{2\,r_D^2\,(d-2)}\left[1+\frac{2}{d\,\hat{\alpha}_0\,(\hat{\alpha}_0+1)}\right] \left[2\,\hat{R}^{(0)}_{ij}\hat{v}^j_*+ \hat{\nabla}^j_{(0)}\hat{q}^*_{ij}\right] \\
 &\quad
+\;\hat{\aleph}^{-3} \, \left[\frac{1}{(d-1)\, (d-2)}\, b_0^3\,\hat{N}_1(\infty) \,\hat{\nabla}^j_{(0)}\hat{R}^{(0)}
-\, b_0^2\, \hat{L}_1(\infty) \, \hat{\nabla}^2_{(0)}\hat{v}^*_{i} \right]\\
&\quad
+ \;{\cal O}(\hat{\aleph}^{-4})\\
g_{ij} &= \hat{g}^{(0)}_{ij} + \hat{\aleph}^{-2} \left(\hat{h}^*_{ij} -(\hat{\alpha}_0^2 -1)\, (\hat{v}^*_i + \hat{k}^*_i) \,(\hat{v}^*_j + \hat{k}^*_j) - \frac{2 \, b_0}{\hat{\alpha}_0}\, F(b_0\,r_D)  \,\hat{\nabla}^{(0)}_{(i} \hat{v}^{*}_{j)} \right) \\
& \quad + \; \hat{\aleph}^{-2} \,b_0^2\,{H}_{1}(b_0 r_D) \left(2\,\hat{\text{S}}^{(0)}_{ij}-\frac{\hat{R}^{(0)}}{(d-1)(d-2)}\hat{g}^{(0)}_{ij}\right) \\
& \quad + \;{\cal O}(\hat{\aleph}^{-4})
\end{split}\end{equation}	
where 
\begin{equation}
\begin{split}
\hat{L}_1(\infty) &\equiv\frac{1}{d}\left[
\frac{b_0\, r_D}{\hat{\alpha}_0^2}\, L'(b_0\, r_D)+
\frac{1-\frac{d}{2}}{b_0\, r_D\left[1+\frac{d}{2}(\hat{\alpha}_0^2-1)\right]}+\frac{1-\frac{d}{2}}{b_0\, r_D^2(d-2)}\right] \left(1-\hat{\alpha}_0^2\right)\\
\hat{N}_1(\infty) &\equiv \frac{1}{d}\left\{
(d-3)\left[\frac{b_0\, r_D}{\hat{\alpha}_0^2}N'(b_0\, r_D)+
\frac{1-d/2}{b_0\, r_D\left[1+\frac{d}{2}(\hat{\alpha}_0^2-1)\right]}\right] \right.\\
&\left. -\frac{d-2}{2\, b_0\, r_D^3\left[1+\frac{d}{2}(\hat{\alpha}_0^2-1)\right]}\right\}\hat{\alpha}_0\left[1-\hat{\alpha}_0^2\right] -\frac{\hat{\alpha}_0}{3\, b_0\, r_D^3}+\frac{\hat{\alpha}_0^3}{2\, b_0\, r_D^3\left[1+\frac{d}{2}(\hat{\alpha}_0^2-1)\right]}\\
\end{split}
\end{equation}

\subsection{The bulk dual for arbitrarily spatially curved metric on $\Sigma_D$}
\label{s:}

The final result for the bulk metric dual to the non-relativistic fluid living on the Dirichlet hypersurface $\Sigma_D$ is simply obtained by plugging in the scaling form \eqref{genscalefA}, \eqref{genscalefB} into \req{metric3trunc}, having eliminated the boundary data in favor of the hypersurface data using the Dirichlet One obtains: 
\begin{equation}
ds^2 = ds_0^2 +  {\aleph}^{-1} ds_1^2 +  {\aleph}^{-2} ds_2^2 +  {\aleph}^{-3} ds_3^2 + {\cal O}( {\aleph}^{-4})
\label{}
\end{equation}	
with 
\begin{equation}\label{hypbmwf1aR012}
\begin{split}
ds_0^2 &= 
	2\,\hat{\alpha}_0\ dt\ dr + r^2\left(-\hat{\alpha}_0^2 \,f_0 \,dt^2 
	+ \hat{g}^{(0)}_{ij}\,dx^i dx^j\right)\\
ds_1^2 &= 
	-2\, \hat{\alpha}_0\left( \hat{v}^*_i +  \hat{k}^*_i \right)\ dx^i\ dr 
	+ 2\, r^2 \left[\hat{k}^*_i -\left(1-\hat{\alpha}_0^2\, f_0\right)
	\left( \hat{v}^*_i +  \hat{k}^*_i \right)\right] dx^i\, dt \\
ds_2^2 &= 
	2\, \hat{\alpha}_0 \left[- \frac{1}{2}\hat{h}^*_{tt} 
	+ \frac{1}{2}\, \hat{g}^{(0)}_{jk} \, \hat{v}^j_* \,  \hat{v}^k_* 
	+\hat{p}_* \frac{\frac{d}{2}\, (\hat{\alpha}_0^2 -1)}{1+\frac{d}{2}\, (\hat{\alpha}_0^2 -1)} 	\right]dt\ dr 
	+ r^2\left[\hat{h}^*_{tt}\, dt^2 + \hat{h}^*_{ij} \, dx^i \, dx^j \right]\\
&\quad 
	+r^2 \left(1-\hat{\alpha}_0^2\, f_0\right) \left[\left(- \hat{h}^*_{tt} 
	+ \hat{g}^{(0)}_{jk} \, \hat{v}^j_* \,  \hat{v}^k_*
	+\hat{p}_* \frac{d\hat{\alpha}_0^2 }{1+\frac{d}{2}\, (\hat{\alpha}_0^2 -1)}
	 \right)dt^2 \right.\\
&\qquad \left. \qquad 
	+ \left( \hat{v}^*_i +  \hat{k}^*_i \right) \left( \hat{v}^*_j 
	+  \hat{k}^*_j \right)dx^i dx^j \right]
	+2\,r^2\,b_0\, \hat{F}_0\,\hat{\nabla}^{(0)}_{(i} \hat{v}^{*}_{j)} \,dx^{i} dx^{j}\\
&\qquad 
- \frac{\hat{\alpha}_0^3}{r_D^2}	 \;\frac{R^{(0)}}{(d-1)(d-2)}\; dt \,dr
 -2\, b_0^2\, r^2\, \hat{H}_1(b_0r) \left[\hat{\text{S}}^{(0)}_{ij}-\frac{\hat{R}^{(0)}}{2(d-1)(d-2)}\hat{g}^{(0)}_{ij}\right] dx^i dx^j  \\ 
 & \qquad 
- \, (b_0r)^2\, \hat{M}_1(b_0r) \, \frac{R^{(0)}}{(d-1)(d-2)}\, dt^2 
 \end{split}
\end{equation}
 \begin{equation}\label{hypbmwf1aR3}
\begin{split}
ds_3^2 &= 
	-2\,\hat{\alpha}_0\left[\hat{h}^*_{ij}\hat{v}^{j}_{*}+ 
	 \left( \frac{1}{2}\hat{h}^*_{tt} +  \,\hat{k}^*_{j}\, \hat{v}^{j}_{*} 
	 +\frac{1}{2} \hat{g}^{(0)}_{jk} \, \hat{v}^{j}_{*}\,  \hat{v}^{k}_{*} 
	 +\hat{p}_* \frac{\frac{d}{2}\, (\hat{\alpha}_0^2 -1)}{1+\frac{d}{2}\, (\hat{\alpha}_0^2 -1)}	\right)\, \left( \hat{v}^*_i +  \hat{k}^*_i \right) \right] dx^i dr \\
&\quad  
         +\frac{2\hat{\alpha}_0^2}{r_D\left(1+\frac{d}{2}(\hat{\alpha}_0^2-1)\right)}\left[ \partial_{t} \hat{v}_{i}^{*}+\hat{v}_{*}^{j}  \hat{\nabla}^{(0)}_{j}  \hat{v}_{i}^{*} 
	-\hat{f}_i^* \right] dx^{i}dr\\
&\quad 
	+ 2r\,\frac{\hat{\alpha}_0(2\,\hat{\xi}_0-1)}{1+\frac{d}{2}\, (\hat{\alpha}_0^2 -1)}
	\left[ \partial_{t} \hat{v}_{i}^{*}+\hat{v}_{*}^{j}  \hat{\nabla}^{(0)}_{j}  \hat{v}_{i}^{*} 
	-\hat{f}_i^* \right] dx^{i}dt\\
&\quad 
	-2\,r^2 \left(1-\hat{\alpha}_0^2 \,f_0\right)\left[\hat{h}^*_{ij}\hat{v}^{j}_{*}
	+ \left(\hat{k}^*_{j}\, \hat{v}^{j}_{*} + \hat{g}^{(0)}_{jk} \, \hat{v}^{j}_{*}\,  \hat{v}^{k}_{*}
	+\hat{p}_* \frac{d\hat{\alpha}_0^2 }{1+\frac{d}{2}\, (\hat{\alpha}_0^2 -1)} \right)\, 
	\left( \hat{v}^*_i +  \hat{k}^*_i \right) \right] dx^i dt \\
&\qquad 
	-4\, r^2\, b_0\, \hat{F}_0\, \hat{v}_{*}^j\hat{\nabla}^{(0)}_{(i} \hat{v}^{*}_{j)} \,dx^{i} dt
-2\, b_0^2\, r^2\, \hat{L}_1	\nabla^2_{(0)} v_i^* \, dt \, dx^i +2\, b_0^3\, r^2\, \hat{N}_1
\, \frac{\nabla_i^{(0)} R^{(0)}}{(d-1) (d-2)} \, dx^i\, dt 
	\\
&\qquad	
-2\left[ b_0^2\, \kappa_L\,\hat{\alpha}_0\,  \hat{\nabla}^2_{(0)}v^*_i -  b_0^3\, \kappa_N \, \hat{\alpha}_0^2\, \frac{\nabla^{(0)}_i\, R^{(0)}}{(d-1)(d-2)} \right]  dx^i \, dr\\
&\qquad
-2\, \frac{\hat{\alpha}_0^3}{r_D^2} \, \left[\hat{\text{S}}^{(0)}_{ij}\hat{v}^j_* - \frac{1}{d-2}\, \left(1 + \frac{2}{d\, \hat{\alpha}_0\, (\hat{\alpha}_0 +1) } \right) \hat{R}^{(0)}_{ij}\hat{v}^j_*
 - \frac{\hat{\nabla}^j_{(0)}\hat{q}^*_{ij}}{d\,(d-2)\, \hat{\alpha}_0\, (\hat{\alpha}_0 +1) }\right]
 dx^i\, dr
 \\
&\qquad  
+4 \, b_0^2\, r^2\,\hat{H}_1(b_0r) \left[\hat{\text{S}}^{(0)}_{ij}-\frac{\hat{R}^{(0)}}{2(d-1)(d-2)}\hat{g}^{(0)}_{ij}\right] \hat{v}^j_* dx^i dt \\ 
&\qquad 
+ 2\, (b_0r)^2\, \hat{M}_1(b_0r) \, \left( (v_i^* + k_i^*)\, \frac{\hat{R}^{(0)}}{2(d-1)(d-2)} - 
\hat{\text{S}}^{(0)}_{ij}\hat{v}^j_* - \frac{1}{2(d-2)}\hat{\nabla}^j_{(0)}\hat{q}^*_{ij}\right)dx^i\, dt \\
&\qquad 
+ 2\, (b_0r)^2\, \hat{M}_2(b_0r) \,\left[\hat{R}^{(0)}_{ij}\hat{v}^j_*+ \frac{1}{2}\hat{\nabla}^j_{(0)}\hat{q}^*_{ij}\right] dx^i \, dt \\
&\qquad 
\end{split}
\end{equation}


\providecommand{\href}[2]{#2}\begingroup\raggedright\endgroup
\end{document}